\DeclareUnicodeCharacter{03B1}{\ensuremath{\alpha}}
\DeclareUnicodeCharacter{03B2}{\ensuremath{\alpha}}
\PassOptionsToPackage{unicode}{hyperref}
\PassOptionsToPackage{hyphens}{url}
\documentclass[]{elsarticle}
\usepackage{amsmath,amssymb}
\usepackage{iftex}
\ifPDFTeX
\usepackage[T1]{fontenc}
\usepackage[utf8]{inputenc}
\usepackage{textcomp}
\else
\usepackage{unicode-math}
\defaultfontfeatures{Scale=MatchLowercase}\defaultfontfeatures[\rmfamily]{Ligatures=TeX,Scale=1}\fi\usepackage{lmodern}\ifPDFTeX\else\fi\IfFileExists{upquote.sty}{\usepackage{upquote}}{}\IfFileExists{microtype.sty}{\usepackage[]{microtype}\UseMicrotypeSet[protrusion]{basicmath}}{}\usepackage{xcolor}\usepackage[left=2.5cm,right=2.5cm,top=2.5cm,bottom=2.5cm]{geometry}\usepackage{longtable,booktabs,array}\usepackage{calc}\usepackage{etoolbox}\makeatletter\patchcmd\longtable{\par}{\if@noskipsec\mbox{}\fi\par}{}{}\makeatother\IfFileExists{footnotehyper.sty}{\usepackage{footnotehyper}}{\usepackage{footnote}}\makesavenoteenv{longtable}\usepackage{graphicx}\makeatletter\def\maxwidth{\ifdim\Gin@nat@width>\linewidth\linewidth\else\Gin@nat@width\fi}\def\maxheight{\ifdim\Gin@nat@height>\textheight\textheight\else\Gin@nat@height\fi}\makeatother\setkeys{Gin}{width=\maxwidth,height=\maxheight,keepaspectratio}\makeatletter\def\fps@figure{htbp}\makeatother\newlength{\cslhangindent}\newlength{\csllabelwidth}\newlength{\cslentryspacingunit}\usepackage{calc}\usepackage[fontsize=11pt]{scrextend}\usepackage{setspace}\usepackage{placeins}\usepackage{soul}\usepackage{color}\usepackage{floatrow}\usepackage{float}\usepackage{fancyhdr}\usepackage{booktabs}\usepackage{longtable}\usepackage{array}\usepackage{multirow}\usepackage{wrapfig}\usepackage{float}\usepackage{colortbl}\usepackage{pdflscape}\usepackage{tabu}\usepackage{threeparttable}\usepackage{threeparttablex}\usepackage[normalem]{ulem}\usepackage{makecell}\usepackage{xcolor}\ifLuaTeX\usepackage{selnolig}\fi\IfFileExists{bookmark.sty}{\usepackage{bookmark}}{\usepackage{hyperref}}\IfFileExists{xurl.sty}{\usepackage{xurl}}{}\hypersetup{hidelinks,pdfcreator={LaTeX via pandoc}}\author{}\date{\vspace{-2.5em}}
\usepackage{natbib}
\setcitestyle{authoryear,open={(},close={)}}
\makeatletter
\def\ps@pprintTitle{%
     \let\@oddhead\@empty
     \let\@evenhead\@empty
     \def\@oddfoot{\footnotesize\itshape
       Preprint submitted to ...\hfill\today}%
     \let\@evenfoot\@oddfoot}
\makeatother
\begin{document}

\begin{frontmatter}

\title{Deep Reinforcement Learning Framework for Diversified Portfolio Management Across Global Equity Markets}

\vspace{1\baselineskip} % Adds one line space after the title

\author[WNEUW]{Kamil Kashif\fnref{1}}
   
\author[WNEUW2]{Robert Ślepaczuk\corref{cor1}\fnref{2}}
\ead{rslepaczuk@wne.uw.edu.pl} 
   
\affiliation[WNEUW]{Quantitative Finance Research Group, Faculty of Economic Sciences, University of Warsaw, ul. Dluga 44-50, 00-241, Warsaw, Poland}

\affiliation[WNEUW2]{Quantitative Finance Research Group, Department of Quantitative Finance and Machine Learning, Faculty of Economic Sciences, University of Warsaw, ul. Dluga 44-50, 00-241, Warsaw, Poland}

\cortext[cor1]{Corresponding author}

\fntext[1]{ORCID: https://orcid.org/0009-0004-0666-581X}
\fntext[2]{ORCID: https://orcid.org/0000-0001-5227-2014}

  \begin{abstract}
This study develops and evaluates a deep reinforcement learning framework for dynamic portfolio allocation across global equity markets. The Soft Actor--Critic algorithm is used to learn continuous portfolio weights within a Markov Decision Process, incorporating transaction costs, turnover penalties, and diversification constraints into the reward function. Five model configurations are compared, varying in reward formulation, policy structure (flat versus hierarchical Dirichlet), portfolio constraints, and temporal encoder (LSTM versus Transformer), and evaluated via walk-forward optimization across sixteen out-of-sample folds spanning 2003--2026 on the Nasdaq-100, Nikkei 225, and Euro Stoxx 50. Results show that RL strategies achieve competitive risk-adjusted performance primarily in the Euro Stoxx 50, where statistically significant abnormal returns are observed, but the central hypothesis is only partially confirmed: no strategy achieves statistically significant excess returns relative to Buy \& Hold under HAC-robust inference across all markets. Regime analysis reveals that RL adds the most value during periods of elevated uncertainty, while ensemble aggregation across markets improves risk-adjusted performance and confirms the benefits of geographic diversification.
  \end{abstract}

  
 \end{frontmatter}

\linespread{1.5}\selectfont

\hypertarget{introduction}{%
\section{Introduction}\label{introduction}}

Predicting and navigating financial markets remains a challenging task due to their inherent volatility, complex structure, and constantly evolving economic environment. Over the past decades, major events such as the 2008 global financial crisis and the COVID-19 pandemic have demonstrated how rapidly market conditions can change, making it difficult for traditional models to generalize across different regimes. As a result, researchers and practitioners have explored a wide range of approaches, from classical statistical models to advanced machine learning techniques, in an attempt to develop robust and profitable investment strategies. In recent years, reinforcement learning (RL) has emerged as a promising framework for financial decision-making. Unlike traditional supervised learning approaches that focus on predicting future prices or returns, RL directly models the portfolio allocation problem as a sequential decision-making task. This allows the agent to learn trading strategies by interacting with the market environment and optimizing a long-term objective.

The main objective of this study is to develop and evaluate reinforcement learning-based portfolio allocation strategies under realistic market conditions. We formulate the portfolio allocation problem as a Markov Decision Process and employ the Soft Actor--Critic (SAC) algorithm to learn dynamic asset allocation policies. The framework incorporates transaction costs, turnover penalties, and diversification constraints directly into the learning objective. In addition, a walk-forward optimization (WFO) scheme with adaptive retraining is used to simulate a realistic deployment setting.

A central aspect of this research is the systematic comparison of different design choices within the RL framework. Specifically, we analyze how variations in reward formulation (absolute versus benchmark-relative), policy structure (flat versus hierarchical allocation), portfolio constraints (fully invested versus flexible exposure), and temporal encoding (LSTM versus Transformer) influence the performance and robustness of the learned strategies.

The central hypothesis of this study is that reinforcement learning-based portfolio allocation strategies can achieve robust and economically meaningful performance when trained under realistic constraints and evaluation procedures, and that their effectiveness is strongly influenced by architectural and economic design choices. Furthermore, it is hypothesised that the relative performance of reinforcement learning strategies varies across macroeconomic regimes, with active allocation adding greater value during periods of elevated uncertainty and lower trend persistence.

The following research questions are addressed:

\begin{itemize}
\item RQ1. How does the choice of reward formulation (absolute vs.\ benchmark-relative) affect the performance and stability of learned strategies?
\item RQ2. What is the impact of different policy structures (flat vs.\ hierarchical allocation) on portfolio performance and diversification?
\item RQ3. How do portfolio constraints (fully invested vs.\ flexible exposure) influence trading behavior and risk-adjusted returns?
\item RQ4. Does the inclusion of different temporal encoders (LSTM vs.\ Transformer) improve the model’s ability to capture market dynamics and enhance performance?
\item RQ5. Does combining strategy-specific signals across multiple markets through an ensemble approach improve risk-adjusted performance compared to single-market strategies and the benchmark?
\item RQ6. Does the performance of reinforcement learning strategies vary systematically across macroeconomic regimes, and in which market environments do RL strategies add the most value relative to passive and classical benchmarks?
\end{itemize}

To evaluate the proposed reinforcement learning strategies, we consider a cross-sectional equity universe based on three major stock indices: the \textit{Nasdaq 100}, the \textit{Nikkei 225}, and the \textit{Euro Stoxx 50}, representing U.S., Asian, and European markets, respectively. This selection allows for a comprehensive analysis across different economic regions and market conditions. The full sample spans the period from \textit{2 January 2003} to \textit{13 March 2026} with \textit{daily} frequency, covering multiple market regimes, including periods of high volatility and structural change.

The empirical framework is evaluated using a walk-forward optimization (WFO) scheme, where each iteration consists of a \textit{5-year training} period, followed by \textit{1 year of validation} and \textit{1 year of testing} set. This setup allows the model to adapt to evolving market conditions while ensuring a robust and realistic evaluation of performance.

This study contributes to the existing literature in several ways. First, it provides a unified multi-market evaluation framework applying identical walk-forward optimization and assessment procedures across three economically distinct equity markets, enabling direct cross-market comparison absent from prior single-market studies. Second, it introduces a hierarchical Dirichlet policy structure that separates the equity-cash allocation from individual asset selection, extending flat policy architectures used in related work. Third, it proposes an adaptive retraining criterion that selectively updates the model based on rolling validation performance, reducing computational cost while maintaining deployment realism. Fourth, it conducts a systematic regime decomposition across three macroeconomic periods, providing evidence on the conditions under which reinforcement learning allocation adds value relative to passive and classical benchmarks.

The remainder of this thesis is structured as follows. Section~\ref{literature-review} reviews the relevant literature on algorithmic trading, machine learning, and reinforcement learning in finance. Section~\ref{data} describes the data sources and preprocessing steps. Section~\ref{Theoretical_Background} presents the theoretical background underlying the reinforcement learning framework. Section~\ref{methodology} outlines the methodology, including the state representation, action space, reward function, model architecture, and validation procedure. Section~\ref{empirical_results} reports the empirical results of the proposed strategies. Section~\ref{regimeness} presents the sub-period macroeconomic regime analysis. Section~\ref{ensemble} presents the ensemble total fund perspective. Section~\ref{discussion} discusses the main findings and their implications. Finally, Section~\ref{conclusions} concludes the thesis and outlines directions for future research.

\hypertarget{literature-review}{%
\section{Literature Review}\label{literature-review}}

Historically, algorithmic trading has been known with rule-based execution, econometric modeling, and supervised machine learning aimed at avoiding human biases in market entry and exit. However, a significant paradigm shift is occurring as researchers move beyond static price forecasting. Recent literature has begun to explore the robust capabilities of Reinforcement Learning (RL). Unlike traditional statistical models that focus on time predictions, RL frameworks, specifically when integrated with deep architectures, offer a comprehensive approach to sequential decision-making, portfolio allocation, and risk optimization.

\hypertarget{algorithmic-trading-and-portfolio-optimization}{%
\subsection{Portfolio Optimization}\label{algorithmic-trading-and-portfolio-optimization}}

Let's go back a couple of steps and recall ourselves how it started and was up until now.

The foundational framework for modern portfolio optimization started with the work of \citet{Markowitz_1952}, who introduced mean-variance optimization (MVO) where we balance expected returns against risk via covariance matrices, enabling efficient frontiers for diversed allocations. While, this is applicable in terms of mathematics, its applications in real-world showed instability. For example, \citet{Black_Literman} in their paper critiqued that MVO often leads to highly concentrated portfolios hypersensitive to estimation errors

Most recently, the literature has shifted toward integrating predictive modeling into the optimization loop. \citet{slusarczyk2025optimal} modernized MVO by forecasting returns using hybrid time series and machine learning models such as ARIMA-GARCH and the XGBoost models. The paper demonstrates that combining forecasting models with traditional optimization can improve portfolio outcomes compared to purely historical approaches.

\citet{chaweewanchon2022markowitz}  propose a hybrid portfolio optimization framework that integrates machine learning-based stock prediction with the classical mean-variance (MV) model. Their approach employs a CNN-BiLSTM architecture with robust statistical preprocessing to improve prediction accuracy of financial time-series data, followed by a stock preselection step based on predicted returns. The selected assets are then used within the MV framework to construct optimal portfolios. Empirical results on SET50 data demonstrate that incorporating predictive stock selection significantly improves portfolio performance in terms of return, risk, and Sharpe ratio compared to traditional and benchmark models.

\citet{lopez2025enhancing} investigate the incorporation of machine learning techniques into the Markowitz portfolio selection framework, aiming to enhance key aspects of portfolio management such as alpha generation, risk management, and the optimization of risk measures including conditional value at risk. Their work highlights how machine learning enables more dynamic decision-making by leveraging complex financial datasets, while also addressing the practical considerations and challenges of applying these methods in real-world portfolio management.

A particularly relevant benchmark in the context of portfolio optimization is the equal-weight ($1/N$) strategy, which allocates capital uniformly across all available assets without estimation or optimization. \citet{DE_MIGEUL} demonstrate that this naive diversification strategy is remarkably difficult to outperform out-of-sample, consistently matching or exceeding the performance of sophisticated mean-variance optimized portfolios across a wide range of datasets and evaluation periods. Their findings highlight that estimation error in expected returns and covariances severely limits the practical advantages of complex optimization models, establishing the equal-weight portfolio as a natural and stringent benchmark for any active portfolio allocation strategy.

\hypertarget{machine-learning-in-finance}{%
\subsection{Machine and Deep Learning in Finance}\label{machine-learning-in-finance}}

The application of statistical learning to financial markets represents a shift from testing economic hypotheses to maximizing predictive accuracy. While traditional econometrics assumes linear relationships, financial time series are non-linear. To navigate this complexity, researchers have increasingly turned to supervised learning frameworks.

Support Vector Machines (SVM) were among the first early-stage ML tools to gain prominence for their ability to handle high-dimensional data by identifying optimal hyperplanes for classification. \citet{Kim2003} demonstrated that SVMs could outperform traditional Artificial Neural Networks (ANN) in predicting the direction of the daily KOSPI index, largely due to their ability to minimize structural risk rather than just empirical error. Similarly, Random Forests (RF) have been known for their robustness against the outliers common in financial data. \citet{Khaidem2016} utilized RF to capture complex market patterns, finding that the ensemble based approach of decision trees effectively mitigates the risk faced with investing the stock market. \citet{Grudniewicz_Slepaczuk_2023} explored the application of ML in algorithmic trading across global stock markets, where they concluded that Linear SVM outperforms Bayesian GLM and the benchmark.

However, still the challenge remains the noise within financial signals and the high probability of backtest overfitting. \citet{Bailey2014} provided a rigorous mathematical framework to quantify the probability of backtest overfitting using a model-free method by proposing a framework that measure the probability of that an optimal strategy on the training set outpeforms the average on the test. \citet{LopezDePrado2018} also argues that most machine leanring funds often fail in finance, proposing a among other a walk-forward backtesing and a deflated sharpe ratio to ensure that reported performance is statistically legitimate.

While early machine learning handled cross-sectional data effectively, financial time series require architectures capable of modeling high-dimensional features over time. The literature has evolved from basic sequential models to sophisticated attention frameworks.

The integration of Recurrent Neural Networks (RNN) marked the first major attempt to give financial models memory by allowing information to persist across time steps. \citet{Giles2001} provided an early benchmark showing that RNNs could capture the noisy, non-stationary dynamics of stock prices however data pre-processing is vital. They proposed a zero-crossing rate, which is a ratio of changes in the sign noting that this approach outperforms the traditional appraoches for the S\&P 500 equity index. \citet{Bieganowski_Slepaczuk_2024}  utilized supervised autoencoders with recurrent structures to enhance feature extraction, proving that even simpler recurrent architectures can perform when combined with robust labeling techniques like the Triple Barrier Method.

In order to solve the vanishing gradient problem, the Long Short-Term Memory (LSTM) model comes in which introduces a gating mechanism that decides which information to keep or discard. \citet{Fischer2018} established the gold standard by proving LSTMs significantly outperform traditional deep neural networks in predicting S\&P 500 constituents, specifically by capturing long range interactions between historical lags. \citet{Krynska_Slepaczuk_2022} cconducted a granular analysis of various LSTM architectures across daily and intraday both wiht regression and classificaiton problem, concluding that the classificiton approach outperforms the benchmark. \citet{Kashif_Slepaczuk_2024} developed a hybrid LSTM-ARIMA model, illustrating that the outperformance of an LSTM is maximized when it is paired with a classical econometric component to handle linear residuals of ARIMA.

The current frontier in the literature is the Transformer architecture, which replaces recurrence with Self-Attention, allowing the model to focus on the most important historical events regardless of how far back they occurred. \citet{Ding2025} argued that while LSTMs process data linearly, Transformers' parallelization and attention mechanisms combined allows them to significantly outperform them individually. \citet{Stefaniuk_Slepaczuk_2025} explored the application of the Informer architecture specifically on high-frequency Bitcoin data, demonstrating that attention when designed appropriately with the correct loss function can outperform the benchmark.

\hypertarget{deep-and-reinforcement-learning-for-trading}{%
\subsection{Reinforcement and Deep Reinforcement Learning in Finance}\label{deep-and-reinforcement-learning-for-trading}}

The integration of Reinforcement Learning (RL) into financial systems represents a transition from static price forecasting to an active, agentic decision-making. Unlike supervised learning models that minimize prediction error, RL frameworks optimize for an object which is typciall long-term cumulative performance, or risk-adjusted returns.

This concept of modern financial reinforcement learning is surveyed by \citet{Hambly_2023}, who contrast RL with classical stochastic control approaches that rely heavily on explicit model assumptions. Instead, RL frameworks enable agents to learn optimal decision-making policies directly through interaction with the environment, using observed state transitions and rewards. The survey highlights that, particularly in model-free settings, RL can effectively handle complex and high-dimensional financial problems without requiring a fully specified model of market dynamics. Furthermore, RL naturally integrates prediction and decision-making, making it well suited for tasks such as optimal execution, portfolio optimization, and smart order routing.

The Direct Reinforcement (DR) approach was introduced by \citet{MOODY} as a method for optimizing trading strategies by viewing investment decision-making as a stochastic control problem. They propose the Recurrent Reinforcement Learning (RRL) algorithm, which eliminates the need to construct forecasting models and instead directly learns investment policies. Unlike traditional reinforcement learning methods such as Q-learning and TD-learning, which rely on value function estimation, the DR approach simplifies the problem representation and avoids Bellman’s curse of dimensionality. The method optimizes risk-adjusted returns using the differential Sharpe ratio while accounting for transaction costs. Empirical results on financial data, including applications to the S\&P 500 and T-Bills, show that RRL-based strategies achieve better trading performance than value-function-based approaches.

\citet{Deng2016} propose a recurrent deep neural network framework for real-time financial signal representation and trading, combining deep learning and reinforcement learning components. In this approach, the deep learning module captures dynamic market conditions and extracts informative features, while the reinforcement learning component interacts with these representations to make trading decisions in an unknown environment. The model is implemented as a deep and recurrent neural network, and a task-aware backpropagation through time method is introduced to address the vanishing gradient problem during training. The proposed system is evaluated on both stock and commodity futures markets, where its robustness is demonstrated under a range of testing conditions.

\citet{Buehler_2019} propose a framework for hedging derivative portfolios using deep reinforcement learning in the presence of market frictions such as transaction costs, liquidity constraints, and risk limits. Their approach applies reinforcement learning to non-linear reward structures based on convex risk measures and learns trading strategies directly without relying on specific assumptions about market dynamics. The framework is designed to scale efficiently to high-dimensional settings and generalizes across different hedging instruments. Empirical results, including experiments on the S\&P 500 and a synthetic market based on the Heston model, show that the method outperforms standard complete-market hedging approaches under transaction costs.

\citet{maringer2012regime} propose a regime-switching extension of the recurrent reinforcement learning (RRL) model, referred to as RSRRL, to better capture the complexity of financial time series. The framework introduces regime-dependent dynamics, where switching between regimes is driven by a volatility-based indicator. The authors present two variants of the model—a threshold version and a smooth transition version—and evaluate their performance in automated trading and portfolio management tasks. Out-of-sample results generally favour the RSRRL models over the standard RRL approach, although concerns regarding robustness remain, particularly in the presence of transaction costs.

\citet{Liu_2025} propose the FinRL-Meta framework, which provides a universe of near real-market environments for data-driven deep reinforcement learning in finance. The framework separates financial data processing from the design of trading strategies and offers a unified data engineering pipeline for accessing, cleaning, and processing large-scale financial data. It includes hundreds of market environments for various trading tasks and supports efficient training through multiprocessing using GPU resources. By enabling diverse DRL agents to be trained across these environments, the framework addresses challenges related to data quality, simulation complexity, and scalability in financial reinforcement learning.

\citet{Meng2026} propose a deep reinforcement learning framework enhanced by cluster embedding (CE) and zero-shot prediction for financial trading. The approach groups feature channels with intrinsic similarities and leverages clustering information instead of individual channel features to improve representation learning. Zero-shot prediction is achieved by assigning unseen samples to appropriate clusters, enabling predictions without additional training. The framework also integrates predicted and observed OHLCV data to form the reinforcement learning state space. Experimental results on multiple real-world financial datasets show improvements in predictive accuracy and trading performance, including a cumulative return of 137.94\% on the S\&P 500.

Deep hedging literature was recently extended by \citet{bracha2025application}, who apply a Twin Delayed Deep Deterministic Policy Gradient (TD3) agent to at-the-money S\&P 500 options. Using a walk-forward procedure, the model is evaluated over a long out-of-sample period and benchmarked against Black–Scholes delta hedging. The results show that the DRL agent can outperform traditional hedging methods, particularly in volatile or high-cost environments. The study further demonstrates that higher risk-awareness penalties can deteriorate performance, while longer volatility estimation windows lead to more stable results.

\citet{Zhang_2020} introduce a deep reinforcement learning framework for option replication and hedging under realistic market frictions, including discrete trading, round lotting, and nonlinear transaction costs. The authors employ several DRL methods, including proximal policy optimization (PPO), to learn hedging strategies that optimize a trade-off between replication error and trading costs. Using simulated market environments, they show that the learned strategies achieve similar or better performance compared to Black–Scholes delta hedging, with PPO exhibiting superior efficiency in terms of profit and loss, training time, and data requirements.

\citet{kabbani2022deep} propose a deep reinforcement learning framework for automated stock trading, formulating the problem as a partially observed Markov decision process (POMDP) that integrates both portfolio allocation and price prediction. The authors employ the Twin Delayed Deep Deterministic Policy Gradient (TD3) algorithm to learn trading strategies in a continuous action space, incorporating technical indicators and news sentiment scores into the state representation. Using historical market data, their approach achieves strong performance, reporting a Sharpe ratio of 2.68 on unseen test data, and demonstrating that enriched state representations significantly improve trading outcomes.

\citet{rani2025deep} propose a deep reinforcement learning framework for optimal trade execution in high-frequency trading that explicitly incorporates market impact into the decision-making process. The model integrates market impact directly into the action space and uses high-dimensional inputs such as limit order book states, trading volumes, and price dynamics. The agent is trained to maximize cumulative returns while accounting for execution costs and both transient and permanent market impact. The learned policies adapt order type and quantity based on market conditions. Experiments using simulated and historical limit order book data show that the DRL agent achieves a favorable balance between profitability and slippage reduction compared to traditional rule-based and optimization-based methods.

\citet{yang2020deep} propose an ensemble deep reinforcement learning strategy for automated stock trading that combines multiple actor-critic algorithms, including PPO, A2C, and DDPG. The ensemble integrates the strengths of these individual models to improve robustness across different market conditions. The approach is evaluated on a portfolio of 30 Dow Jones stocks and compared against individual DRL agents as well as traditional benchmarks such as the Dow Jones Industrial Average and a minimum-variance portfolio strategy. The results show that the ensemble method achieves higher risk-adjusted returns, as measured by the Sharpe ratio, than both the individual models and the baseline strategies.

\citet{Ohyun2025} formulate portfolio optimization as a constrained Markov decision process (CMDP) within a risk-averse reinforcement learning framework. To enforce risk-related constraints required by investors and regulators, the authors employ an augmented Lagrangian multiplier (ALM) method, enabling the agent to incorporate constraint satisfaction directly into the decision-making process. The proposed approach demonstrates no constraint violations during testing and outperforms other risk-averse reinforcement learning methods, highlighting its effectiveness for constrained portfolio optimization.

\citet{park2022portfolio} apply a Deep Q-Network (DQN) framework with experience replay to portfolio optimization, where an agent interacts with financial data to dynamically allocate capital and maximize returns while managing risk. The model learns to predict optimal portfolio weights over time based on historical price data, refining its strategy through past experiences. Empirical evaluation on real-world data shows that the proposed approach achieves superior performance compared to traditional methods, as measured by Sharpe ratio and cumulative returns.

\citet{SOLEYMANI2020113456} proposed DeepBreath, a deep reinforcement learning framework for portfolio management that combines a restricted stacked autoencoder for feature extraction with a convolutional neural network to learn investment policies. The model employs both offline and online learning to adapt to market dynamics, while blockchain is used to mitigate settlement risk. Empirical results show that the approach outperforms existing strategies in terms of return while minimizing risk.

\citet{JIANG2024101016} proposed a model-free deep reinforcement learning framework for portfolio selection that incorporates transaction costs and investor risk aversion into an extended mean–variance reward function. Using a twin-delayed deep deterministic policy gradient (TD3) algorithm, the approach constructs optimal portfolios in high-dimensional markets and demonstrates superior performance compared to traditional and DRL-based benchmarks.

\citet{sterling2026deep} investigate the application of deep reinforcement learning to dynamic portfolio optimization in financial markets, emphasizing its role as an adaptive alternative to traditional static allocation methods in environments characterized by non-linear dependencies and stochastic regimes. The study adopts a system-level perspective, analyzing architectural design choices, reward function construction, and state-space representation, while also addressing broader considerations such as algorithmic governance, systemic risk, and the infrastructure required for deploying DRL in financial systems.

\citet{cheng2024multiagent} propose a multi-agent deep reinforcement learning framework for multi-asset trading and portfolio management, where individual agents operate on separate assets and their experiences are integrated to generate trading strategies and portfolio allocations. The framework combines a trading action module and a portfolio management module, leveraging the TD3 algorithm to improve training stability and decision quality. Experimental results on stock market data show that the approach outperforms single-agent reinforcement learning methods and yields more stable returns.

\citet{analytics2030031} propose a hierarchical model-based deep reinforcement learning framework for single-asset trading, where a high-level agent dynamically selects among specialized low-level agents trained on clustered market regimes. By integrating predictive models into the decision process, the framework bridges model-based and model-free reinforcement learning. Empirical results on cryptocurrency data show improved performance and support risk-aware optimization via CVaR-based rewards.

\citet{jrfm16030201} propose a fuzzy ensemble deep reinforcement learning framework for stock portfolio management, where fuzzy logic is used to represent market trends through continuous degrees of bullish, bearish, and oscillatory behavior. The approach combines multiple reinforcement learning algorithms, including A2C, PPO, DDPG, ACKTR, and TRPO, by averaging their action outputs to form trading decisions. Using S\&P 100 constituent stocks and historical market data, the method demonstrates improved performance compared to benchmark strategies and individual reinforcement learning models.

\citet{shavandi2022multi} propose a multi-agent deep reinforcement learning framework for algorithmic trading, where multiple agents, each specialized in a distinct trading timeframe, collaboratively learn trading strategies. The framework employs a hierarchical feedback mechanism that transfers information from higher- to lower-timeframe agents, enabling the integration of multi-scale market dynamics. Evaluated on historical EUR/USD data, the approach outperforms single-agent models and benchmark trading strategies across multiple return- and risk-based performance metrics.

\citet{wu2020adaptive} propose adaptive stock trading strategies based on deep reinforcement learning, integrating Gated Recurrent Units (GRU) for feature extraction from financial time-series data. They introduce two frameworks: a critic-only Gated Deep Q-Network (GDQN) and an actor–critic Gated Deterministic Policy Gradient (GDPG). Evaluated on stocks from U.S., U.K., and Chinese markets under both trending and volatile conditions, the proposed methods achieve superior returns and improved risk-adjusted performance compared to benchmark strategies, with GDPG demonstrating greater stability.

\hypertarget{non-trading-reinforcement-learning}{%
\subsection{Reinforcement Learning Beyond Finance}\label{non-trading-reinforcement-learning}}

The reinforcement learning techniques employed in this study originate from foundational advances in broader decision-making domains. \citet{mnih2015human} introduced the Deep Q-Network (DQN), establishing that deep neural networks can learn control policies directly from high-dimensional inputs without handcrafted features. \citet{schulman2017proximal} subsequently introduced Proximal Policy Optimization (PPO), addressing training instability in policy gradient methods through a clipped surrogate objective that constrains policy updates. The Soft Actor-Critic algorithm employed in this study builds on these foundations, extending the actor-critic framework with maximum-entropy reinforcement learning to encourage exploration and improve stability in continuous action spaces \citep{Haarnoja2018}.

These methodological advances are directly relevant to portfolio allocation, where agents must operate in high-dimensional, non-stationary environments and optimize long-term risk-adjusted objectives under realistic constraints.

\hypertarget{literature-summary}{%
\subsection{Summary}\label{literature-summary}}

The literature reviewed in this chapter shows a progressive transformation in quantitative finance, moving from static optimization frameworks toward adaptive, data-driven decision-making systems. Early approaches to portfolio construction, were in mean-variance optimization, provided a mathematical foundation for balancing risk and return. However, their practical limitations, particularly sensitivity to estimation error and instability in non-stationary environments motivated the development of more robust and dynamic methodologies. Subsequent advancements incorporated regime based optimization, and predictive modeling, makeing a shift from purely historical estimation toward forward-looking portfolio construction.

The introduction of machine learning further accelerated this shift by enabling models to capture non-linear relationships and complex interactions within financial data. Nevertheless, the literature consistently highlights fundamental challenges, including the low signal to noise ratio of financial time series, susceptibility to overfitting, and the evolving nature of market dynamics. While deep learning architectures, such as recurrent networks and attention based models, have improved the ability to model temporal dependencies and extract latent features, they remain inherently dependent on forecasting future returns, which limits their robustness in highly stochastic environments.

Reinforcement learning represents a conceptual departure from these approaches by reframing financial decision making as a sequential optimization problem rather than a prediction task. Instead of estimating future prices, reinforcement learning agents directly learn policies that maximize long-term, risk adjusted performance under realistic constraints such as transaction costs and market frictions. The literature demonstrates that, when combined with deep neural networks, these methods are capable of operating in high dimensional state spaces and adapting to changing market conditions. At the same time, research emphasizes that their performance is highly sensitive to reward specification, model stability, and the handling of non-stationarity, indicating that careful design and evaluation are essential.

In summary, the literature suggests that while traditional and supervised learning approaches provide valuable foundations, they are insufficient to fully address the dynamic and sequential nature of financial decision making. This motivates the adoption of deep reinforcement learning as a unified framework for portfolio allocation, capable of integrating prediction, execution, and risk management within a single adaptive system.

\hypertarget{data}{%
\section{Data Engineering}\label{data}}

This study considers three major equity indices to represent different global markets and provide diversification across regions. Specifically, the \textit{Nasdaq-100 equity index}, the \textit{Nikkei 225 equity index}, and the \textit{Euro Stoxx 50 equity index} are included in the analysis. These indices capture the dynamics of  U.S., Asian, and European equity markets, respectively, allowing for a comprehensive assessment of cross-market behavior. 

\hypertarget{data_description}{%
\subsection{Data Description}\label{data_description}}

The empirical analysis is based on a multi-source financial dataset constructed for the purpose of training and evaluating a reinforcement learning framework for dynamic portfolio allocation across multiple equity indices. The dataset integrates asset-level price information, index membership dynamics, and benchmark market series.

Daily equity price data for all constituents were obtained using the \textit{yfinance} API. These data include standard market variables and form the primary input for the modeling framework. Historical information on index composition, including constituent additions and deletions over time, was retrieved from the \textit{Bloomberg Terminal Anywhere} subscription. This information was used to reconstruct time varying index membership and to address survivorship bias by ensuring that only assets available at each point in time are considered.

In addition to constituent-level data, benchmark index proxies were incorporated to represent the performance of each investment universe. For the NASDAQ-100, the Invesco QQQ Trust ETF (\textit{QQQ}) was used as a tradable benchmark. For the Euro Stoxx 50, the SPDR Euro Stoxx 50 ETF (\textit{FEZ}) was employed due to its long and consistent price history. For the Nikkei 225, the iShares MSCI Japan ETF (\textit{EWJ}) is used as a tradable benchmark proxy. 

The full sample spans the period from \textit{2 January 2003} to \textit{13 March 2026}. Following data collection, all sources were aligned within a unified preprocessing pipeline to ensure temporal consistency across price series, benchmark data, and index membership records.

\begin{figure}
{\centering \includegraphics[width=0.95\linewidth,height=0.8\textheight]{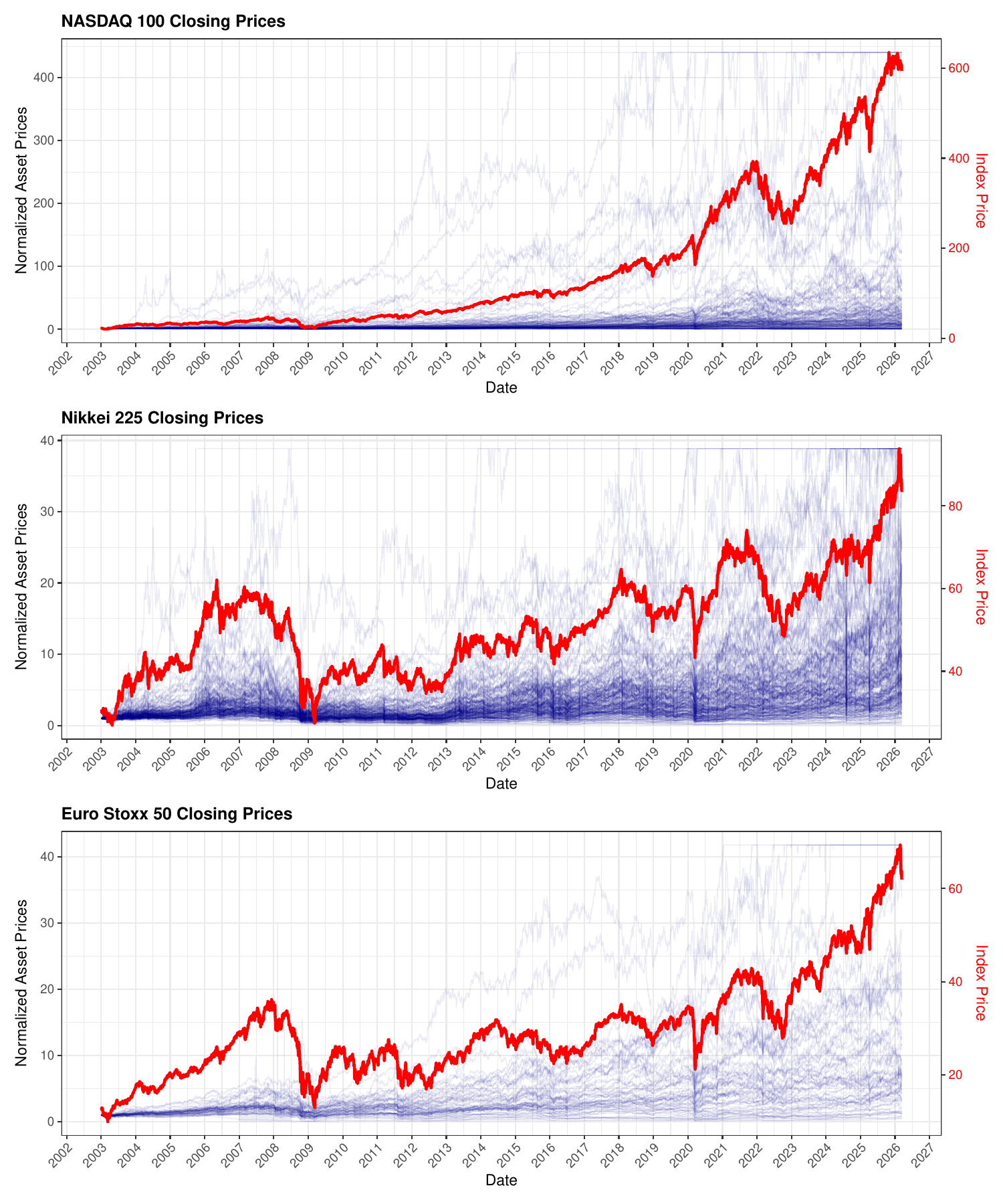} }
\caption{Normalized price paths of assets in the two investment universes}\label{fig:PRICE_PATHS}
\end{figure}

Figures~\ref{fig:PRICE_PATHS} present the normalized price paths of the assets included in the investment universe, with each series scaled to one at its first available observation. Tables~\ref{tab:index_summary} and~\ref{tab:return_summary} report descriptive statistics of the equity indices at the price level and daily log-return level, respectively.

\vspace{-0.4cm}

\noindent\linespread{0.65}\selectfont {\scriptsize \textbf{Note}: \textit{The figure displays the price paths of all constituent assets considered in the study. Each series is normalized to unity at its initial observation. The red line in each panel represents the corresponding benchmark ETF in its original (non-normalized) price scale.}}

\linespread{1.5}\selectfont

\newpage

\begin{table}[htbp]
\centering
\caption{Descriptive statistics of the equity indices used in the study.}
\label{tab:index_summary}
\begin{tabular}{lrrrrrrrr}
\toprule
Index & Count & Mean & Std & Min & 25\% & 50\% & 75\% & Max \\
\midrule
NASDAQ 100 & 5836 & 154 & 154 & 20 & 38 & 87 & 226 & 635 \\
Nikkei 225 & 5676 & 51 & 12 & 27 & 41 & 51 & 59 & 94 \\
Euro Stoxx 50 & 5836 & 30 & 11 & 10 & 23 & 28 & 33 & 69 \\
\bottomrule
\end{tabular}
\end{table}

\vspace{-0.4cm}

\noindent\linespread{0.65}\selectfont {\scriptsize \textbf{Note}: \textit{The descriptive statistics for NASDAQ 100, NIKKEI 225, EUROSTOXX 50 are calculated on the closing price in the period from 2003-01-02 until 2026-03-13}}

\linespread{1.5}\selectfont

\begin{table}[htbp]
\centering
\caption{Descriptive statistics of daily log-returns for the equity indices.}
\label{tab:return_summary}
\begin{tabular}{lrrrrrrr}
\toprule
Index & Mean & Std & Skewness & Ex. Kurtosis & Min & Max & JB p-value \\
\midrule
NASDAQ 100    & 0.0005 & 0.0136 & $-$0.2378 & 7.1260  & $-$0.1276 & 0.1148 & $<$0.001 \\
Nikkei 225    & 0.0001 & 0.0128 & $-$0.1225 & 8.6171 & $-$0.1078 & 0.1406 & $<$0.001 \\
Euro Stoxx 50 & 0.0002 & 0.0159 & $-$0.3300 & 9.8994  & $-$0.1331 & 0.1616 & $<$0.001 \\
\bottomrule
\end{tabular}
\end{table}

\vspace{-0.4cm}

\noindent\linespread{0.65}\selectfont
{\scriptsize \textbf{Note}: \textit{Statistics are computed on daily log-returns 
over the full sample period 2003-01-02 to 2026-03-13. Ex.\ Kurtosis denotes 
excess kurtosis relative to the normal distribution. The Jarque--Bera test 
rejects normality at the 1\% significance level for all three series, 
confirming the fat-tailed and negatively skewed nature of daily equity returns.}}

\linespread{1.5}\selectfont

\hypertarget{Data-Preperation}{%
\subsection{Data Preparation}\label{Data-Preperation}}

This section describes the procedures applied to transform the raw financial data into a consistent and model ready dataset. Particular attention is given to handling time varying index membership, ensuring alignment across data sources, and constructing a tradable universe that accurately reflects real world investment constraints.

\subsubsection{Constituents Membership and Survivorship Bias}
\label{sec:survivorship}

\begin{figure}
{\centering \includegraphics[width=0.95\linewidth,height=0.5\textheight]{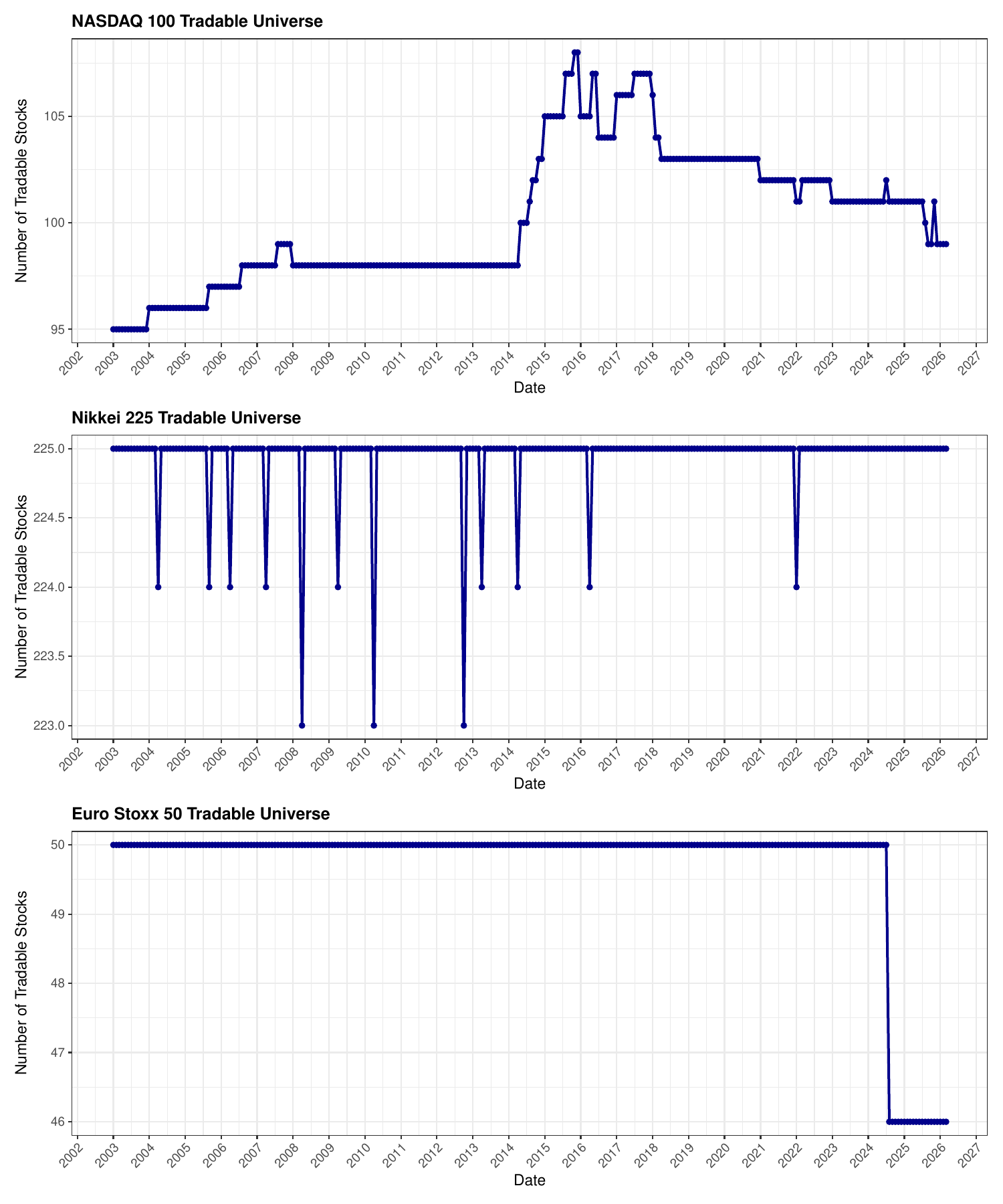} }
\caption{Evolution of the number of constiunts over time}\label{fig:tradable_universe}
\end{figure}

A key challenge in empirical financial studies is the presence of survivorship bias, which arises when only currently listed assets are considered, ignoring those that were removed from the index over time. This leads to an upward bias in performance estimates, as poorly performing or delisted assets are systematically excluded. To address this issue, historical index membership information was obtained from the Bloomberg Terminal. The dataset includes all constituent additions and deletions throughout the sample period, allowing for the reconstruction of the true, time-varying composition of each index. Figure~\ref{fig:tradable_universe} illustrates the evolution of the number of constituents over time, highlighting the dynamic nature of the investment universe.

\vspace{-0.4cm}

\noindent\linespread{0.65}\selectfont {\scriptsize \textbf{Note}: \textit{This figure presents how the number of constituents change over time during the period 2003-01-2 until 2026-03-13}}

\linespread{1.5}\selectfont

\subsubsection{Tradable Universe Construction}

Following data alignment, asset price data were merged with the historical index membership matrix described in Section~\ref{sec:survivorship}. This matrix was expanded to the daily frequency and combined with the price dataset to construct a tradability mask. The resulting mask indicates whether a given asset was both a member of the index and associated with a valid price observation at each point in time. Formally, the tradability condition ensures that only assets satisfying both membership and data availability constraints are considered in the investment universe. This approach guarantees that the model operates on a realistic and time-consistent set of tradable assets, effectively eliminating survivorship bias and preventing look-ahead errors.

Following the construction of the tradability mask, the two data sources Bloomberg membership records and yfinance price series are synchronized onto a common daily date index. Specifically, the Bloomberg membership matrix is expanded to daily frequency by forward-filling constituency status between recorded addition and deletion events, ensuring that each asset's tradability is correctly reflected on every trading day. Price data obtained from yfinance are aligned to this index, with any missing price observations handled via forward-fill to propagate the most recent available closing price. Assets with no valid price observation at a given date are excluded from the tradable universe regardless of their membership status, as enforced by the tradability mask. Following this alignment and filtering procedure, the effective date ranges for the three indices are 2003-01-02 to 2026-03-13 for both the NASDAQ-100 and EURO STOXX 50, and 2003-01-02 to 2026-03-13 for the Nikkei 225, with minor differences in the number of valid trading days across markets due to differing national holiday calendars.

\subsection{Summary}

\begin{figure}
{\centering \includegraphics[width=0.6\linewidth,height=0.4\textheight]{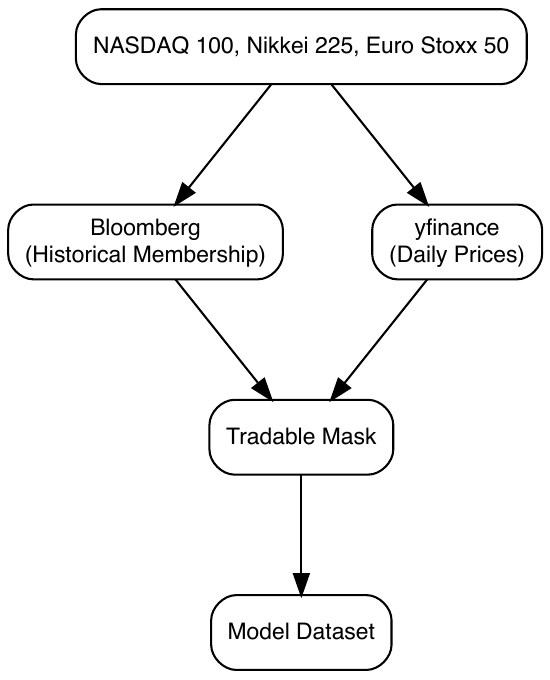} }
\caption{Data construction pipeline for the reinforcement learning framework}\label{fig:data_pipeline}
\end{figure}

To summarize the data construction process, Figure~\ref{fig:data_pipeline} presents the full pipeline from raw data acquisition to the final model-ready dataset.

The process begins with the extraction of historical index membership information from Bloomberg and price data from yfinance. These datasets are then aligned temporally and combined to construct the tradable universe through a masking procedure. Finally, the processed data serve as input to the reinforcement learning framework described in the following section.

\vspace{-0.4cm}

\begin{center}
\linespread{0.65}\selectfont 
{\scriptsize \textbf{Note}: \textit{The figure illustrates the process of constructing the model-ready dataset.}}
\end{center}

\linespread{1.5}\selectfont

\newpage

\hypertarget{Theoretical_Background}{%
\section{Theoretical Background}\label{Theoretical_Background}}

Financial decision-making inherently involves sequential choices under uncertainty, where actions taken at a given point in time influence future outcomes and available opportunities. In the context of portfolio allocation, an investor repeatedly observes market conditions, selects portfolio weights, and realizes returns that depend on both current decisions and future market dynamics.

This sequential structure naturally aligns with the framework of reinforcement learning (RL), where an agent interacts with an environment by observing states, taking actions, and receiving rewards. Unlike traditional optimization approaches that rely on static assumptions or one-step forecasts, RL provides a dynamic formulation that directly optimizes long-term performance.

This section introduces the theoretical foundations underlying the methodology adopted in this study. First, the portfolio allocation problem is framed as a sequential decision-making process within a Markov Decision Process (MDP). Next, the core concepts of reinforcement learning and value-based optimization are presented. Finally, the extension to deep reinforcement learning and the role of temporal representations in financial time series are discussed.

\hypertarget{MDP}{%
\subsection{Markov Decision Processes}\label{MDP}}

Portfolio allocation can be naturally formulated as a sequential decision problem. At each time step \(t\), an agent observes the current market information and selects an action, which corresponds to a portfolio allocation. The outcome of this decision is uncertain and depends on future market movements, making the problem inherently stochastic.

A standard framework to model such problems is the Markov Decision Process (MDP), defined by the tuple \((\mathcal{S}, \mathcal{A}, \mathcal{P}, r, \gamma)\), where:
\begin{itemize}
    \item \(\mathcal{S}\) is the set of states, representing the information available at time \(t\),
    \item \(\mathcal{A}\) is the set of actions, corresponding to possible portfolio allocations,
    \item \(\mathcal{P}(s_{t+1} \mid s_t, a_t)\) denotes the transition dynamics of the environment,
    \item \(r(s_t, a_t)\) is the reward function,
    \item \(\gamma \in (0,1)\) is the discount factor.
\end{itemize}

At each time step, the agent observes a state \(s_t \in \mathcal{S}\), selects an action \(a_t \in \mathcal{A}\), and receives a reward \(r_t\), after which the system transitions to a new state \(s_{t+1}\). A key assumption in this framework is the Markov property, which states that the future evolution of the system depends only on the current state and action, and not on the full history:

\begin{equation}
\mathbb{P}(s_{t+1} \mid s_t, a_t, s_{t-1}, a_{t-1}, \dots) 
= 
\mathbb{P}(s_{t+1} \mid s_t, a_t).
\end{equation}

In financial applications, this assumption is an approximation, as market dynamics often depend on past information. In practice, this limitation is addressed by constructing state representations that include recent observations, allowing the agent to incorporate relevant historical information into its decisions.

\hypertarget{Reinforcement-Learning}{%
\subsection{Reinforcement Learning}\label{Reinforcement-Learning}}

Within the MDP framework, the objective of the agent is to learn a decision rule, or policy, that maximizes long-term performance. A policy is defined as a mapping from states to actions, denoted by \(\pi(a \mid s)\), which specifies the probability of taking action \(a\) when the system is in state \(s\). The quality of a policy is evaluated through the expected cumulative reward, referred to as the return. Starting from time \(t\), the return is defined as:

\begin{equation}
G_t = \sum_{k=0}^{\infty} \gamma^k r_{t+k+1},
\end{equation}

where \(\gamma \in (0,1)\) is the discount factor, controlling the relative importance of immediate versus future rewards.

To assess the desirability of states and actions, reinforcement learning introduces value functions. The state-value function (\(V^\pi\)) measures the expected return starting from a given state under policy \(\pi\), while the action-value function (\(Q^\pi\)) evaluates the expected return when taking a specific action \(a\) in state \(s\):

\begin{equation}
V^\pi(s) = \mathbb{E}_\pi \left[ G_t \mid s_t = s \right],
\end{equation}
\begin{equation}
Q^\pi(s,a) = \mathbb{E}_\pi \left[ G_t \mid s_t = s, a_t = a \right].
\end{equation}

The goal of reinforcement learning is to find an optimal policy \(\pi^*\) that maximizes the expected return. In practice, directly optimizing over all possible policies is infeasible in complex environments. Instead, modern approaches rely on parameterized policies and value functions, which can be learned from data through interaction with the environment.

A common class of methods is the actor–critic framework, where two components are learned simultaneously: the actor, which represents the policy, and the critic, which estimates value functions. This combination enables stable learning by using value estimates to guide policy updates.

\hypertarget{Deep-Reinforcement-Learning}{%
\subsection{Deep Reinforcement Learning}\label{Deep-Reinforcement-Learning}}

In practical applications such as portfolio allocation, the state space is typically high-dimensional and complex, consisting of multiple assets and market indicators observed over time. In such settings, classical reinforcement learning methods that rely on tabular representations become infeasible. 

Deep reinforcement learning addresses this limitation by using neural networks as function approximators for both the policy and value functions. Instead of explicitly storing values for each state-action pair, the agent learns parameterized functions that generalize across similar states. This allows the model to operate in environments with large state spaces and rich feature representations.

However, combining reinforcement learning with function approximation introduces several challenges. First, the data used for learning is sequential and highly correlated, which can lead to unstable updates. Second, the target values used for training evolve over time as the policy improves, creating a moving optimization objective. To mitigate these issues, modern algorithms employ techniques such as experience replay, target networks, and stochastic optimization.

Another important aspect in financial applications is that decisions depend not only on the current observation but also on recent patterns in the data. To capture such dependencies, the state representation is typically constructed to include a window of past observations, allowing the model to incorporate relevant historical information.

\hypertarget{RL_SD}{%
\subsection{Representation Learning for Sequential Data}\label{RL_SD}}

Financial markets exhibit strong dependence on past observations, where recent price movements, volatility patterns, and market conditions provide important context for decision-making. As a result, using only current features may lead to a loss of relevant information. To address this, modern approaches incorporate models that can extract meaningful representations from sequences of past observations.

Recurrent neural networks (RNNs), and in particular Long Short-Term Memory (LSTM) networks, are designed to capture such dependencies by maintaining an internal state that evolves as new data becomes available. This allows the model to summarize relevant information from previous observations and use it to inform current decisions. Compared to standard feed-forward networks, LSTM architectures are more effective at handling longer dependencies and mitigating issues related to vanishing gradients.

More recently, attention-based models such as Transformers have been proposed as an alternative approach. Instead of processing data sequentially, these models use attention mechanisms to identify and weight the most relevant observations within a given history window. This enables the model to capture relationships across different time steps without relying on recurrent structures.

\begin{figure}[htbp]
\centering
\includegraphics[width=0.75\linewidth]{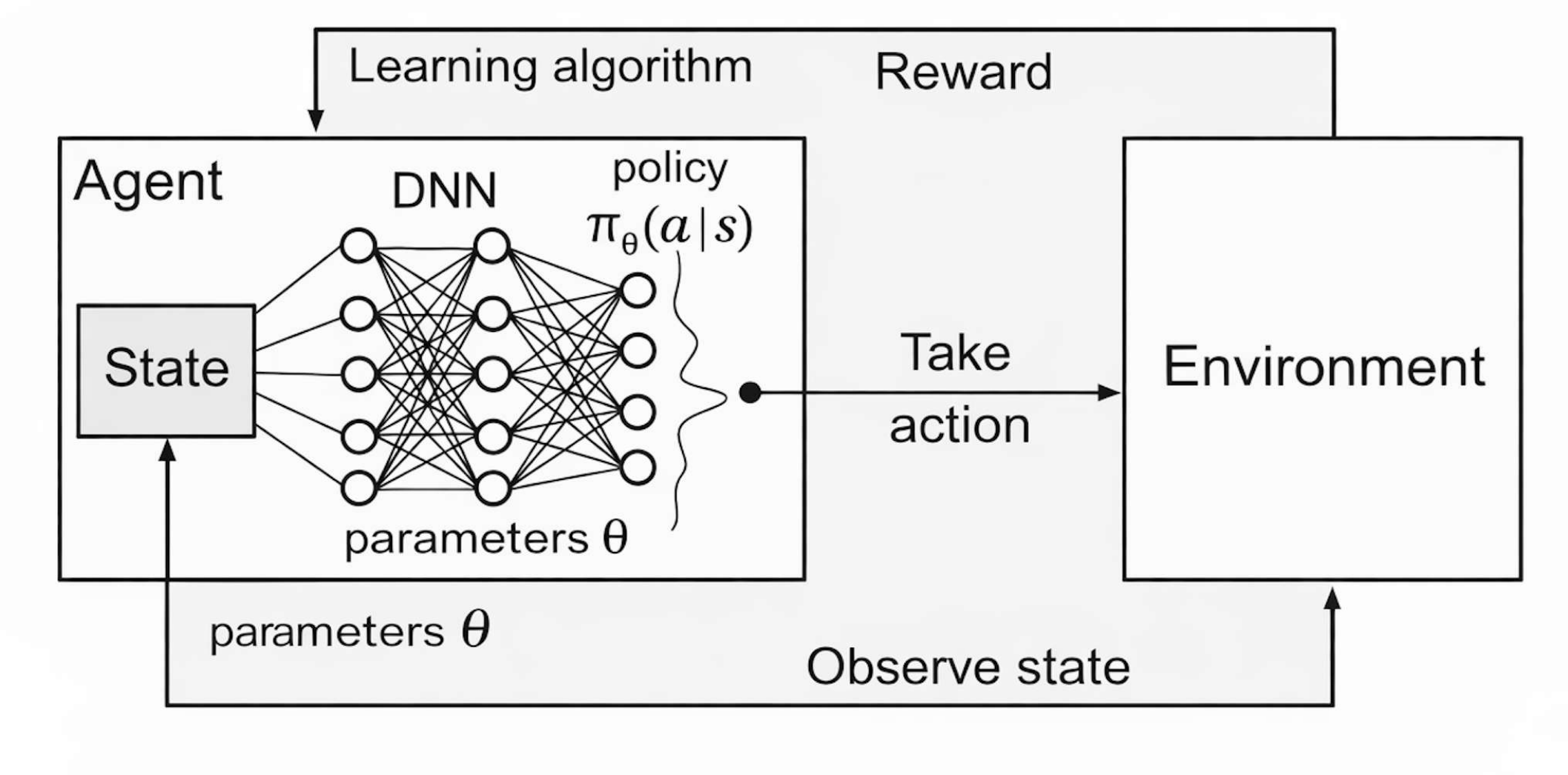}
\caption{Illustration of the reinforcement learning agent}
\label{fig:rl_basic}
\end{figure}

Figure~\ref{fig:rl_basic} provides a schematic illustration of this interaction, highlighting the sequential loop of state observation, action selection, reward feedback, and state transition that characterizes reinforcement learning problems.

\vspace{-0.4cm}

\linespread{0.65}\selectfont
{\noindent\scriptsize \textbf{Note}: \textit{The diagram illustrates the interaction between an agent and its environment in a reinforcement learning setting. At each time step, the agent observes the current state \(s_t\), selects an action \(a_t\) according to its policy \(\pi_\theta(a \mid s)\), and receives a reward signal. The environment transitions to a new state \(s_{t+1}\), forming a sequential decision-making loop. \newline Source: Author's own illustration.}}

\linespread{1.5}\selectfont

\hypertarget{PA_SCP}{%
\subsection{Portfolio Allocation as a Sequential Control Problem}\label{PA_SCP}}

The portfolio allocation problem can be naturally framed as a sequential control task within the reinforcement learning framework. At each time step \(t\), the agent observes the current state of the market and selects a portfolio allocation across a set of tradable assets. The action is represented by a vector of portfolio weights:

\begin{equation}
\mathbf{w}_t = (w_{1,t}, w_{2,t}, \dots, w_{N_t,t}),
\end{equation}
where \(N_t\) denotes the number of available assets at time \(t\). The weights satisfy the budget constraint:
\begin{equation}
\sum_{i=1}^{N_t} w_{i,t} = 1,
\end{equation}
and, in the long-only setting, \(w_{i,t} \geq 0\) for all \(i\).

Given a portfolio allocation, the realized portfolio return over the next period is given by:
\begin{equation}
R_{p,t+1} = \sum_{i=1}^{N_t} w_{i,t} R_{i,t+1},
\end{equation}
where \(R_{i,t+1}\) denotes the return of asset \(i\).

In practice, portfolio rebalancing incurs transaction costs, which depend on changes in portfolio weights. These costs introduce a trade-off between responsiveness to new information and excessive turnover. As a result, the objective is not only to maximize returns, but to optimize performance net of trading frictions.

Within the reinforcement learning framework, this objective is captured through the reward function, which reflects realized portfolio performance adjusted for costs and risk considerations. The agent therefore learns a policy that maps observed market conditions to portfolio allocations, with the aim of maximizing long-term cumulative reward.

This formulation provides a flexible and dynamic alternative to classical portfolio optimization approaches, allowing the model to adapt to changing market conditions and incorporate complex, non-linear relationships between assets and market variables.

\hypertarget{methodology}{%
\section{Methodology}\label{methodology}}

The aim of this section is to present the methodological framework used to construct and evaluate reinforcement learning based portfolio allocation strategies. Building on the theoretical foundations introduced in the previous section, the portfolio selection problem is formulated as a sequential decision-making task, where an agent dynamically allocates capital across a set of tradable assets over time.

The proposed framework combines a structured state representation derived from asset-level and market-level features with a reinforcement learning algorithm designed to optimize portfolio performance under realistic market conditions. In particular, the model accounts for transaction costs, portfolio rebalancing, and varying market environments through a walk-forward optimization scheme.

The remainder of this section is organized as follows. First, the state representation, action space, and reward function are introduced. This is followed by a description of the model architecture and the considered configuration variants. The training and validation procedure, including walk-forward optimization and adaptive retraining, is then presented. Finally, the performance metrics used in the empirical analysis are defined.

\hypertarget{State_Representation}{%
\subsection{State Representation}\label{State_Representation}}

The reinforcement learning agent operates on a structured state representation composed of both asset-level and global market features. The objective of the feature set is to capture cross-sectional differences between assets as well as the prevailing market environment, enabling the agent to make allocation decisions based on both relative and absolute signals. A set of features were constructed for each asset \(i\) at time \(t\). These features are grouped into four categories.

Before the state is passed to the reinforcement learning agent, a cross-sectional asset selection step is applied whenever a top-\(k\) constraint is imposed. Specifically, the agent is restricted to the \(k\) tradable assets with the highest 120-trading-day momentum, where \(k \in \{20,30\}\) depending on the configuration. The momentum measure is defined as
\begin{equation}
m_{i,t}^{(120)} = \frac{P_{i,t}}{P_{i,t-120}} - 1,
\end{equation}
where \(P_{i,t}\) denotes the daily closing price of asset \(i\) at time \(t\). The ranking is computed only among assets that are tradable and have valid price observations at time \(t\). This selection rule is exogenous to the reinforcement learning agent and is used to reduce the dimensionality of the action space while focusing the model on assets with the strongest recent price performance.

\subsubsection{Momentum features}
 
These capture short and medium term price dynamics through log returns computed over multiple horizons:

\begin{equation}
r^{(X)}_{i,t} = \log\left(\frac{P_{i,t}}{P_{i,t-X}}\right),
\end{equation}
where \(X \in \{1,5,20,60\}\) trading days.

\subsubsection{Volatility features}

These describe recent return variability using rolling standard deviations of daily log returns:

\begin{equation}
\sigma^{(X)}_{i,t} =
\sqrt{
\frac{1}{X}
\sum_{j=1}^{X}
\left(r_{i,t-j} - \bar{r}_{i,t}\right)^2
},
\end{equation}
where \(X \in \{5,20\}\).

\subsubsection{Technical indicators features}

These were included to capture non linear price dynamics and mean reversion effects. These consist of the relative strength index (RSI), the moving average convergence divergence (MACD) histogram, Bollinger Band position (\%B), the distance from the rolling 20-day high, and a mean-reversion signal based on deviations from the 20-day moving average.

\subsubsection{Market-relative features}

These include a rolling beta estimate computed over a 60-day window with respect to the market proxy, for example QQQ in case of NASDAQ 100, allowing the agent to account for systematic exposure. And an absolute performance feature was included in the form of a 20-day log return series, treated separately from the cross-sectional signals.

\subsubsection{Global features}

These include the level of the volatility index (VIX) and its 5-day change. Aggregate market conditions are captured through the cross-sectional average return of tradable assets and its 5-day rolling volatility. Additional features include market breadth, defined as the proportion of assets with positive medium-term returns, as well as normalized returns of the market proxy, for example QQQ in case of NASDAQ 100, over 5-day and 20-day horizons.

\subsubsection{Dynamic asset universe}
In order to reduce the dimensionality of the action space and focus on the most relevant investment opportunities, the set of tradable assets is dynamically filtered at each time step. Specifically, the agent operates on a subset of assets selected based on cross-sectional momentum. At each time \(t\), only the top-\(k\) assets ranked by their 120-day momentum are included in the investable universe. Momentum is defined as:
\begin{equation}
\text{momentum}_{i,t} = \frac{P_{i,t}}{P_{i,t-120}} - 1.
\end{equation}
This selection procedure ensures that the agent concentrates on assets exhibiting strong recent performance while maintaining a manageable action space. Since $k$ is fixed across all time steps and folds, the top-$k$ selection procedure guarantees a constant state vector dimension regardless of fluctuations in the number of tradable constituents $N_t$ over time. This ensures that the neural network encoder receives a fixed-size input at every step, eliminating the need for dynamic padding or variable-length architectures to handle the time-varying index composition illustrated in Figure~\ref{fig:tradable_universe}.

\hypertarget{Action_Space}{%
\subsection{Action Space}\label{Action_Space}}

The action of the agent at each time step \(t\) corresponds to the selection of portfolio weights across the available assets. Formally, the action is defined as:
\begin{equation}
a_t = w_t = (w_{1,t}, \dots, w_{N_t,t}, w^c_t),
\end{equation}
where \(w_{i,t}\) denotes the allocation to asset \(i\), \(w^c_t\) represents the allocation to cash, and \(N_t\) is the number of tradable assets at time \(t\). In configurations with a top-\(k\) restriction, \(N_t\) refers to the number of selected assets after the momentum-based pre-selection step, rather than the full set of tradable index constituents.

The portfolio weights are constrained such that:
\begin{equation}
\sum_{i=1}^{N_t} w_{i,t} + w^c_t = 1.
\end{equation}

To ensure valid portfolio allocations, the policy network outputs parameters of a Dirichlet distribution, from which the weight vector is sampled. This guarantees that all weights are non-negative and sum to one.

Two action formulations are considered. In the first, a \textit{flat} allocation is used, where a single Dirichlet distribution directly determines the weights across all assets (and cash). In the second, a \textit{hierarchical} structure is employed, where the policy first determines the overall allocation to equities relative to cash, and subsequently distributes the equity portion across individual assets using a Dirichlet distribution.

Depending on the configuration, the environment enforces different constraints. In some setups, the portfolio is restricted to fully invested long-only positions, while in others, allocations may include cash and allow for more flexible exposure. These variations enable the analysis of different portfolio construction regimes within a unified reinforcement learning framework.

\hypertarget{Reward_Function}{%
\subsection{Reward Function}\label{Reward_Function}}

The reward function determines the objective that the reinforcement learning agent seeks to optimize. In this framework, it is designed to balance portfolio performance with realistic trading considerations, including transaction costs, trading activity, and diversification.

The reward function is designed to reflect the trade-offs faced by real-world portfolio managers. In practice, maximizing raw returns alone leads to unstable and highly concentrated strategies. Therefore, the reward combines three components: (i) portfolio performance, (ii) trading frictions, and (iii) diversification. This ensures that the agent learns economically meaningful strategies that balance profitability, stability, and implementability.

At each time step, portfolio performance is measured using the log return of the portfolio value. Let \(r^{\text{net}}_{p,t}\) denote the portfolio return net of transaction costs. A logarithmic transformation is applied to stabilize the learning signal:
\begin{equation}
\tilde{r}_{p,t} = \log(1 + r^{\text{net}}_{p,t}).
\end{equation}

In addition to returns, two penalty terms are incorporated. First, a turnover penalty discourages excessive trading, where turnover is defined as the $L^1$ distance between consecutive portfolio weight vectors. Second, a concentration penalty encourages diversification by penalizing portfolios that are overly concentrated in a small number of assets. This is measured using the Herfindahl index. The Herfindahl–Hirschman Index \((HHI)\) is a standard measure of concentration commonly used in economics and finance. In the context of portfolio allocation, it quantifies how capital is distributed across assets. A higher HHI indicates that the portfolio is concentrated in fewer assets, while a lower value corresponds to a more diversified allocation. \(HHI\) is defined below:

\begin{equation}
\text{HHI}_t = \sum_{i=1}^{N_t} w_{i,t}^2,
\end{equation}
with the minimum value given by \(\text{HHI}^{\min}_t = 1 / N_t\), corresponding to an equal-weight allocation.

Two reward formulations are considered. In the first, the agent maximizes absolute portfolio performance:
\begin{equation}
r_t = 1000 \cdot \tilde{r}_{p,t}
- \lambda_{\text{TO}} \cdot \text{TO}_t \cdot 100
- \lambda_{\text{conc}} \cdot (\text{HHI}_t - \text{HHI}^{\min}_t) \cdot 100,
\end{equation}
where \(\text{TO}_t\) denotes portfolio turnover, \(\lambda_{\text{TO}}\) is the turnover penalty coefficient, and \(\lambda_{\text{conc}}\) controls the strength of the concentration penalty.

In the second formulation, the agent maximizes performance relative to a benchmark:
\begin{equation}
r_t = 1000 \cdot \left( \tilde{r}_{p,t} - \tilde{r}_{b,t} \right)
- \lambda_{\text{TO}} \cdot \text{TO}_t \cdot 100
- \lambda_{\text{conc}} \cdot (\text{HHI}_t - \text{HHI}^{\min}_t) \cdot 100,
\end{equation}
where \(\tilde{r}_{b,t}\) denotes the benchmark log return.

Transaction costs are incorporated directly in \(r^{\text{net}}_{p,t}\), while the turnover penalty provides an additional regularization mechanism to discourage excessive rebalancing beyond the direct cost of trading.

\begin{figure}

{\centering \includegraphics[width=0.8\linewidth,height=0.4\textheight]{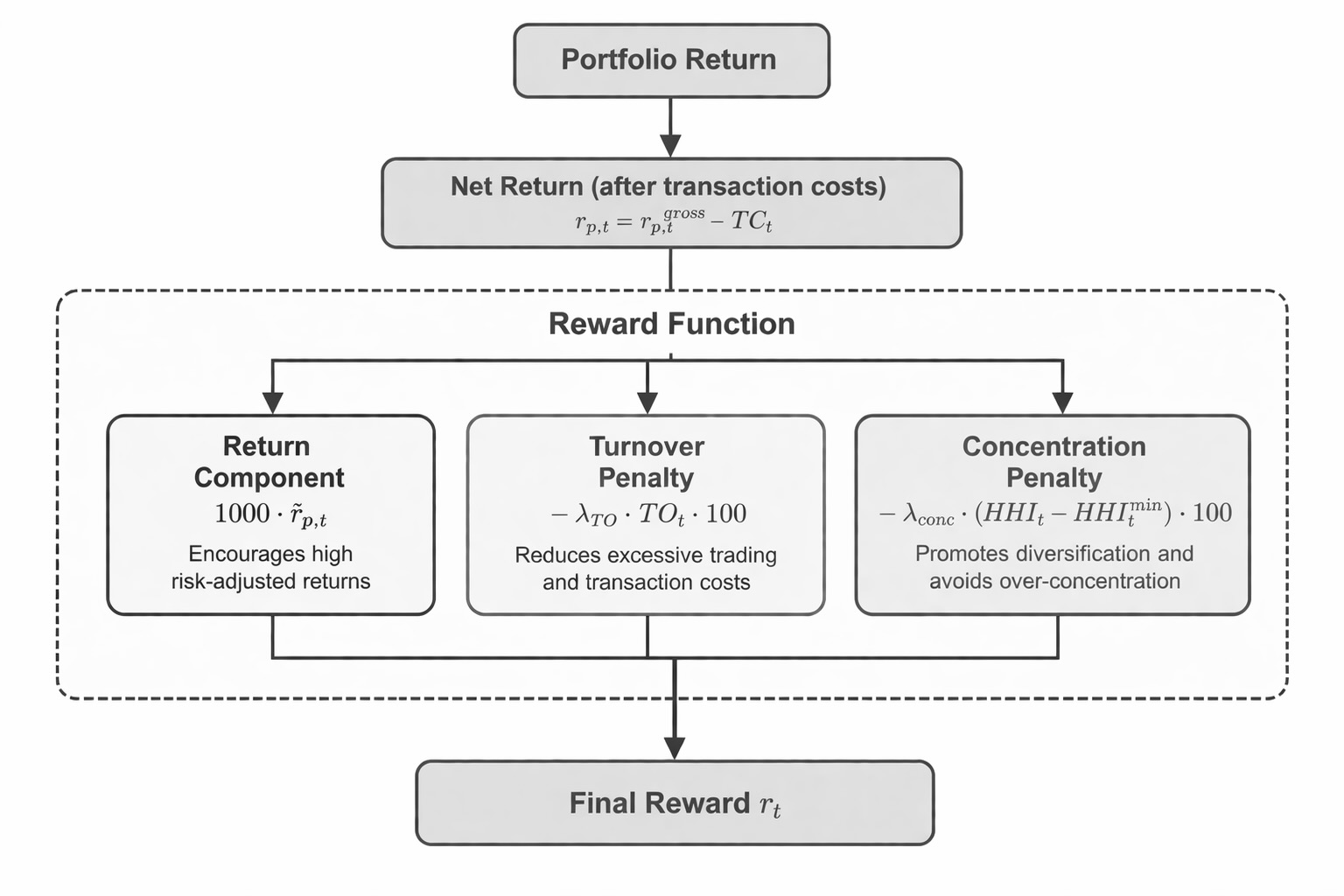} 

}

\caption{Decomposition of the Reinforcement Learning Reward Function}\label{fig:reward}
\end{figure}

The scaling factors applied to the reward components are introduced to balance their relative magnitudes and stabilize training. In reinforcement learning, poorly scaled rewards can lead to unstable gradients or slow convergence. The chosen scaling ensures that return, turnover, and concentration terms contribute meaningfully to the learning signal. Figure~\ref{fig:reward} summarizes the reward functions.

\vspace{-0.4cm}

\noindent\linespread{0.65}\selectfont {\scriptsize \textbf{Note}: \textit{The diagram illustrates the structure of the reward function used in the reinforcement learning framework. \textbf{Source}: Author's own illustration.}}

\linespread{1.5}\selectfont

\hypertarget{Model_Architecture}{%
\subsection{Model Architecture}\label{Model_Architecture}}

The portfolio allocation problem is addressed using a deep reinforcement learning framework based on the Soft Actor--Critic (SAC) algorithm. The model combines a stochastic policy with function approximation through neural networks, enabling it to operate in a high-dimensional and continuous action space.

\subsubsection{Soft Actor--Critic}
SAC is an off-policy actor--critic algorithm that learns a stochastic policy while maximizing both expected return and policy entropy. This entropy regularization encourages exploration and leads to more stable learning. The agent consists of a policy network (actor) and two action-value networks (critics), which are trained using experience replay and soft target updates. Target networks are updated via soft updates controlled by the smoothing parameter $\tau \in (0,1)$, whereby the target network parameters are shifted toward the current network parameters at each step: $\theta^{-} \leftarrow \tau\theta + (1-\tau)\theta^{-}$. The entropy coefficient $\alpha$ is held fixed at 0.2 throughout all experiments rather than using automatic tuning, as adaptive $\alpha$ introduced training instability across walk-forward folds due to the non-stationarity of financial return series and the relatively short validation windows. A fixed value ensures a consistent and reproducible exploration-exploitation trade-off across all markets and configurations.

\subsubsection{State encoding}
To process the structured state representation, a neural encoder is employed to transform asset-level and global features into a compact representation. Asset-level features are first encoded individually, and their historical information is summarized using a sequence model. In the baseline configuration, this is achieved using a Long Short-Term Memory (LSTM) network applied to each asset. The resulting embeddings are then combined using a cross-sectional attention mechanism, allowing the model to capture relationships across assets.

In one configuration, the LSTM encoder is replaced by a Transformer-based architecture. This alternative uses self-attention to model dependencies across the input sequence and allows for a more flexible representation of historical patterns.

\subsubsection{Policy network}
The policy network maps the encoded state representation to a portfolio allocation. To ensure valid portfolio weights, the policy outputs parameters of a Dirichlet distribution, from which the allocation vector is sampled.

Two policy structures are considered. In the \textit{flat} formulation, a single Dirichlet distribution directly determines the allocation across all assets and cash. In the \textit{hierarchical} formulation, the policy first determines the overall allocation to equities relative to cash, and then distributes the equity allocation across individual assets using a Dirichlet distribution. This structure allows the model to separate high-level allocation decisions from asset selection.

\subsubsection{Critic networks}
The critic consists of two independent neural networks that estimate the action-value function. Both critics take as input the encoded state representation together with the proposed action and output an estimate of the expected return. The use of twin critics mitigates overestimation bias and improves training stability.

\hypertarget{Model-Configurations}{%
\subsection{Model Configurations}\label{Model-Configurations}}

This study evaluates five reinforcement learning configurations, each representing a distinct combination of model architecture, policy structure, reward specification, and portfolio constraints. While all configurations are based on the Soft Actor--Critic framework and share a common feature representation as mentioned in section~\ref{State_Representation}, they differ in key design choices that affect portfolio construction and learning dynamics.

Each row in Table~\ref{tab:config_overview} represents a distinct model configuration, defined by a specific combination of encoder, policy structure, reward formulation, and portfolio constraints.

\begin{table}[H]
\centering
\small
\setlength{\tabcolsep}{3pt}   % default ~6pt → tighter
\renewcommand{\arraystretch}{1.1}
\begin{tabular}{lccccc}
\toprule
\textbf{Configuration} & \textbf{Encoder} & \textbf{Policy} & \textbf{Reward} & \textbf{Constraints} & \textbf{Top-\(k\)} \\
\midrule
LSTM\_1 & LSTM & Flat Dirichlet & Log return & Cash + flexible exposure & 20 / 30 \\
LSTM\_2 & LSTM & Hierarchical & Log return & Cash + flexible exposure & 20 / 30 \\
LSTM\_NC\_1 & LSTM & Flat Dirichlet & Benchmark-relative & Fully invested (no cash) & 10 / 20 \\
LSTM\_NC\_2 & LSTM & Flat Dirichlet & Benchmark-relative & Fully invested (no cash) & 20 \\
TRANSFORMERS & Transformer & Flat Dirichlet & Log return & Cash + flexible exposure & 20 / 30 \\
\bottomrule
\end{tabular}
\caption{Overview of reinforcement learning configurations.}
\label{tab:config_overview}
\end{table}
\begin{center}
\vspace{-0.4cm}
\linespread{0.65}\selectfont
{\scriptsize \textbf{Note}: \textit{All configurations are based on the Soft Actor--Critic (SAC) algorithm. “Cash + flexible exposure” indicates that the agent can dynamically allocate capital between risky assets and cash, allowing for partial investment and varying market exposure. In contrast, “fully invested (no cash)” denotes that the portfolio is entirely allocated to risky assets at all times.}}
\end{center}

\linespread{1.5}\selectfont

Table~\ref{tab:rl_params} reports the key reinforcement learning hyperparameters shared across configurations. These parameters govern the optimization process, including learning rates, replay buffer size, and update dynamics of the SAC algorithm.

\begin{table}[H]
\centering
\small
\setlength{\tabcolsep}{3pt}
\renewcommand{\arraystretch}{0.90}

\begin{tabular}{lcc}
\toprule
\textbf{Parameter} & \textbf{Value} \\
\midrule
Learning rate (actor) & $3 \times 10^{-4}$ \\
Learning rate (critic) & $5 \times 10^{-4}$ \\
Learning rate (entropy) & $3 \times 10^{-4}$ \\
Discount factor $\gamma$ & 0.99 \\
Target smoothing $\tau$ & 0.005 \\
Gradient steps & 2 \\
Batch size & 128 \\
Replay buffer size & 20{,}000 \\
Warm-up steps & 500 \\
Entropy coefficient $\alpha$ & 0.2 (fixed) \\
\bottomrule
\end{tabular}
\caption{Reinforcement learning and optimization hyperparameters.}
\label{tab:rl_params}
\end{table}
\vspace{-0.7cm}
\linespread{0.65}\selectfont 
\begin{center}
{\scriptsize \textbf{Note}: \textit{This table reports the common reinforcement learning hyperparameters used across all configurations.}}
\end{center}

\linespread{1.5}\selectfont

Table~\ref{tab:model_params} summarizes the sequence encoder architectures used in the model. The configuration of LSTM differ in hidden size, while the configuration of Transformer uses a two-layer self-attention encoder.

\begin{table}[H]
\centering
\small
\begin{tabular}{lccc}
\toprule
\textbf{Encoder Type} & \textbf{Hidden Size} & \textbf{Layers} \\
\midrule
LSTM & 64 / 128 & 1 \\
Transformer & 128 & 2 \\
\bottomrule
\end{tabular}
\caption{Neural network architecture parameters across configurations.}
\label{tab:model_params}
\end{table}
\vspace{-0.8cm}
\linespread{0.65}\selectfont 
\begin{center}
{\scriptsize \textbf{Note}: \textit{The table summarizes the architecture of the sequence encoders used in the model.}}
\end{center}

\linespread{1.5}\selectfont

Table~\ref{tab:env_params} presents the environment and training parameters affecting portfolio dynamics and learning. The transaction costs and penalty terms directly affect the reward function, while the lookback window defines the length of historical input provided to the model.

Furthermore, the transaction cost assumption of 2 basis points per unit of turnover is consistent with the lower bound of equity commissions available to institutional and high-volume traders on Interactive Brokers (IBKR) under the tiered pricing schedule, which ranges from 0.05 to 0.35 bps per share depending on monthly volume.

\begin{table}[H]
\centering
\small
\begin{tabular}{lcc}
\toprule
\textbf{Parameter} & \textbf{Value} \\
\midrule
Lookback window & 60 days \\
Transaction cost & 2 bps \\
Turnover penalty & 0.003 \\
Concentration penalty & 0.0 / 0.1 / 0.5 (config-dependent) \\
Epochs & 50 \\
Early stopping patience & 8 \\
\bottomrule
\end{tabular}
\caption{Environment configuration parameters.}
\label{tab:env_params}
\end{table}
\vspace{-0.8cm}
\linespread{0.65}\selectfont 
\begin{center}
{\noindent\scriptsize \textbf{Note}: \textit{This table reports the environment and training-related parameters.}}
\end{center}

\linespread{1.5}\selectfont

\hypertarget{Training-and-Validation-Procedure}{%
\subsection{Training and Validation Procedure}\label{Training-and-Validation-Procedure}}

The training and validation procedure is designed to ensure a realistic and robust evaluation of the proposed reinforcement learning strategies. Given the non-stationary nature of financial markets, the framework combines a walk-forward optimization scheme with adaptive retraining and systematic hyperparameter selection. This approach allows the model to be evaluated in a strictly out-of-sample setting while dynamically adjusting to changing market conditions.

\hypertarget{walk-forward-optimization}{%
\subsubsection{Walk Forward Optimization}\label{walk-forward-optimization}}

Overfitting is a major concern in machine learning models, particularly in financial time-series applications. Standard cross-validation techniques such as \emph{k}-fold validation are generally not suitable for financial data because they ignore the temporal ordering of observations. As a result, model evaluation based on such methods may lead to overly optimistic performance estimates and poorly generalizing strategies. 

To address this issue, a walk-forward optimization (WFO) framework is employed. As noted by \citet{cartaetal9355141}, walk-forward optimization is one of the most widely used validation techniques in financial machine learning and algorithmic trading research. The WFO procedure simulates a realistic trading environment by repeatedly training the model on historical data and evaluating it on subsequent unseen observations. This approach reduces the risk of overfitting and allows the model to adapt to evolving market conditions.

Two variants of WFO are commonly used: anchored and non-anchored. In the anchored approach, each walk shares a common starting point and the training window expands over time. In contrast, the non-anchored approach uses rolling windows with a fixed length, where both the starting and ending points shift forward over time. In this study, the non-anchored variant is adopted because it maintains a constant training horizon and better reflects a realistic portfolio management setting.

\begin{figure}

{\centering \includegraphics[width=1\linewidth,height=0.6\textheight]{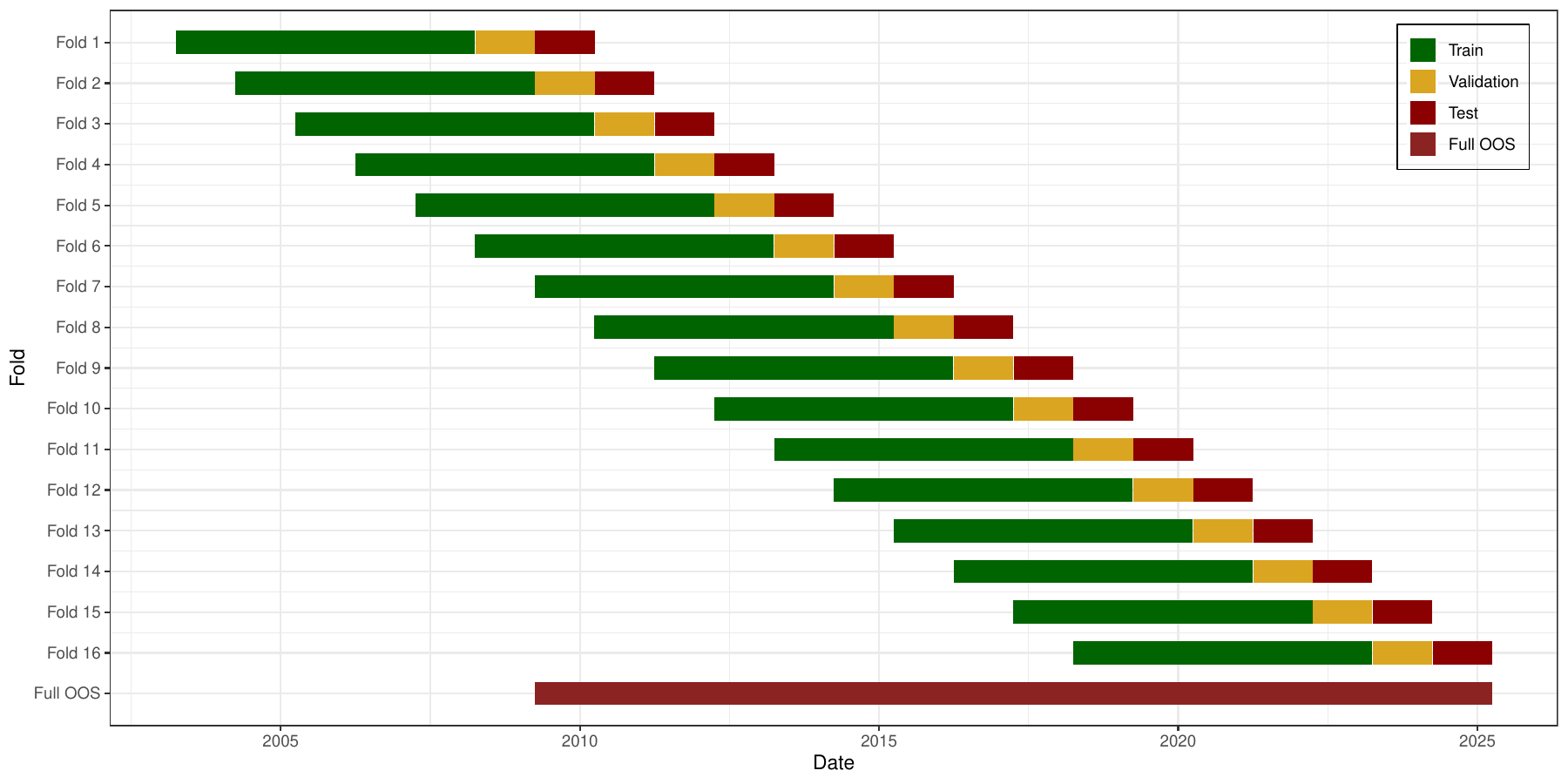} 

}

\caption{Walk-Forward Optimization Scheme (5-Year Training, 1-Year Validation, 1-Year Test)}\label{fig:WFO}
\end{figure}

The walk-forward scheme used in this study consists of a training window of 5 trading years, followed by validation and testing windows of one trading year each. The model is trained on the historical training window, reinforcement learning agent approach is selected using the both the training window and validation window as described in the subsequent sections, and final performance is evaluated on the out-of-sample testing period. The walk-forward procedure is illustrated in Figure~\ref{fig:WFO}.

It should be noted that the WFO architecture itself including the 5-year training window, 1-year validation period, and adaptive retraining criterion (presented in Section~\ref{Adaptive-Retraining-Strategy}) represents a set of design choices that were not subject to systematic optimization across alternatives. These parameters were selected based on established conventions in the financial machine learning literature \citep{cartaetal9355141} and practical considerations regarding the minimum  training horizon required for stable SAC convergence. As noted by \citet{Bailey2014}, ex-post selection of walk-forward parameters can itself constitute a form of meta-overfitting, producing apparently robust out-of-sample results that may not generalise to different time horizons.

\vspace{-0.4cm}

\noindent\linespread{0.65}\selectfont {\scriptsize \textbf{Note}: \textit{Green bars represent the training dataset, yellow bars the validation dataset, and red bars the out-of-sample testing dataset. Dark-red bars indicate the cumulative out-of-sample period. The plot is constructed using the final feature matrix provided to the reinforcement learning agent. The training window corresponds to five trading years, while the validation and testing windows correspond to one trading year each.}}

\linespread{1.5}\selectfont

\hypertarget{Adaptive-Retraining-Strategy}{%
\subsubsection{Adaptive Retraining Strategy}\label{Adaptive-Retraining-Strategy}}

In addition to the walk-forward structure, an adaptive retraining mechanism is employed to determine when the model should be re-estimated. Instead of retraining at every step, the procedure selectively updates the model based on its recent validation performance, thereby reducing computational cost while maintaining a realistic deployment setting.

At the first fold, the model is always trained. For each subsequent fold $k$, the current model is evaluated on the validation window using the Sharpe ratio computed from daily net returns. Based on this evaluation, a retraining decision is made.

Let $S_k$ denote the validation Sharpe ratio at fold $k$. A performance threshold is defined using recent validation results:
\begin{equation}
\label{eq:retrain_threshold}
\theta_k = \operatorname{median}(S_{k-m}, \dots, S_{k-1}) 
- \frac{1}{2}\operatorname{std}(S_{k-m}, \dots, S_{k-1}),
\end{equation}
where the window includes up to the last $m=5$ validation observations, provided that at least three past values are available. For the initial folds where fewer than three prior validation observations exist (i.e., $k < 3$), the threshold criterion in Equation~\ref{eq:retrain_threshold} cannot be computed and the model is always retrained regardless of validation performance, ensuring a well-defined cold-start behaviour.

Retraining is triggered if any of the following conditions are satisfied:
\begin{itemize}
\item $S_k < 0$,
\item $S_k < \theta_k$,
\item the number of consecutive folds without retraining exceeds a predefined limit (set to 3 folds).
\end{itemize}

If none of these conditions are met, the previously trained model is carried forward and applied to the next test period.

The validation Sharpe ratio is computed as:
\begin{equation}
\label{eq:val_sharpe}
S_k = \frac{\bar{r}_k}{\sigma_k} \sqrt{A},
\end{equation}
where $\bar{r}_k$ and $\sigma_k$ denote the mean and standard deviation of daily net portfolio returns on the validation set, and $A$ is the annualization factor ($A=252$).

\hypertarget{Hyperparameter-Selection}{%
\subsubsection{Hyperparameter-Selection}\label{Hyperparameter-Selection}}

Within each walk-forward training window, model hyperparameters are selected through a systematic evaluation of predefined candidate configurations. For each configuration, the agent is trained on the training dataset and subsequently evaluated on the corresponding validation period.

Performance is assessed using the Sharpe ratio computed over validation returns. To obtain a more robust estimate of performance, returns are aggregated at the monthly level, and the Sharpe ratio is computed for each month. This aggregation reduces the influence of short-term noise and provides a more stable measure of performance across different market conditions. These values are then summarized to guide model selection. 

The selection procedure follows a tiered approach. First, configurations that demonstrate consistently strong performance across both training and validation data are preferred. Among these, models with similar performance between training and validation are prioritized, as this indicates better generalization. If no configuration satisfies these criteria, the model with the highest validation performance is selected.

This procedure ensures that the selected hyperparameters balance performance and robustness, thereby reducing the likelihood of overfitting. In particular, the selection process avoids configurations that exhibit highly unstable performance, such as achieving strong results in a single period while performing poorly over the remainder of the validation window.

\hypertarget{Computational-Setup}{%
\subsection{Computational Setup}\label{Computational-Setup}}

All experiments were conducted on a cloud-based environment equipped with an NVIDIA L4 GPU (24GB VRAM) using a G2-standard instance configuration with 30GB of system memory. 

The computational requirements of the models differed depending on the architecture. Training a single LSTM based reinforcement learning agent required approximately 14 hours per walk-forward cycle, while the Transformer based model required approximately 23 hours due to its higher computational complexity and self attention mechanisms.

\hypertarget{performance-metrics}{%
\subsection{Performance Metrics}\label{performance-metrics}}

To evaluate the performance and robustness of the proposed trading strategies, a set of metrics commonly used in the financial literature is employed, following \citet{michankow2022lstm} and \citet{PairTrading}. These metrics provide a comprehensive assessment of profitability, risk, drawdowns, and trading activity.

Annualized Return Compounded (ARC) measures overall profitability, while Annualized Standard Deviation (ASD) captures return variability as a proxy for risk. Risk-adjusted performance is evaluated using both the Sharpe Ratio (SR) and Information Ratio (IR1), with the Modified Information Ratio (IR2) placing additional emphasis on drawdown control. Maximum Drawdown (MD) and Maximum Loss Duration (MLD) quantify the severity and persistence of losses. 

In addition, portfolio turnover is reported to assess trading intensity and the economic feasibility of the strategies, particularly in the presence of transaction costs. The detailed definitions of each metric are presented below.

\paragraph{Annualized Return Compounded (ARC)}
The Annualized Return Compounded shows the rate of return that
was annualized for the given strategy during the period of
\((0,...,T)\). It is expressed as a percentage.
\begin{equation}
\label{eq:arc}
ARC = (\space\prod_{t=1}^{N} \space (1+R_t)\space)^{\frac{252}{N}} - 1 \times 100\%
\end{equation}
where: \newline \(R_t\) - the percentage rate of return \newline
\(N\) - the sample size \newline \(R_t\) is calculated as follows:
\begin{equation}
\label{eq:rt}
R_t = \frac{P_t - P_{t-1}}{P_{t-1}}
\end{equation}
where: \newline \(P_t\) - the price at the point \(t\) \newline

\paragraph{Annualized Standard Deviation (ASD)}
The Annualized Standard Deviation is a risk measure.
\begin{equation}
\label{eq:asd}
ASD = \sqrt{252} \space\times\space \sqrt{\frac{1}{N-1}\sum_{t=1}^{N}\space(R_t\space - \bar{R})^2} \times 100\%
\end{equation}
where: \newline \(R_t\) - the percentage rate of return \newline
\(\bar{R}\) - the mean rate of return. \newline \(N\) - the sample size \newline \(\bar{R}\) is calculated in the following way:
\begin{equation}
\label{eq:rbar}
\bar{R} = \frac{1}{N} \sum_{t=1}^{N} R_t
\end{equation}

\paragraph{Maximum Drawdown (MD)}
The Maximum Drawdown gives us the maximum percentage drawdown
throughout the investment and is calculated as follows:
\begin{equation}
\label{eq:md}
MD(T) = \mathop{\mathrm{max}}_{s \in [0,T]} 
\left(\mathop{\mathrm{max}}_{t \in [0,s]}(R_{i,T} - R_{i,s})\right) \times 100\%
\end{equation}

\paragraph{Maximum Loss Duration (MLD)}
The Maximum Loss Duration represents the longest period, expressed in years, between two consecutive local equity curve maxima \citep{michankow2022lstm}. It is calculated as follows:
\begin{equation}
\label{eq:mld}
MLD = \max\left(\frac{m_j - m_i}{S}\right)
\end{equation}
where $m_i$ and $m_j$ denote the dates of two consecutive local maxima of the equity curve, and $S = 252$ is the number of trading days per year, such that MLD is expressed in years.

\paragraph{Sharpe Ratio (SR)}
The Sharpe ratio measures risk-adjusted performance by evaluating excess return per unit of total volatility. Let $(R_t)_{t=1}^{N}$ denote the sample of simple portfolio returns, with mean return $\bar{R} = \frac{1}{N} \sum_{t=1}^{N} R_t$. The Sharpe ratio is defined as:

\begin{equation}
\label{eq:sharpe}
SR = \frac{\bar{R}}{\sigma_R} \sqrt{252}
\end{equation}
where: \newline $\sigma_R$ - the standard deviation of returns

\paragraph{Information Ratio \((IR1)\)}
The Information Ratio describes the risk-adjusted return
metric based on the relation between ARC to its ASD and is calculated as
follows:
\begin{equation}
\label{eq:ir-star}
IR^{*} = \frac{ARC}{ASD}
\end{equation}

\paragraph{Modified Information Ratio \((IR2)\)}
The Modified Information Ratio is another more complex and
comprehensive risk-adjusted return metric which we regard as the
\textbf{most important} metric for the evaluation of strategies in this
research and is calculated as follows:
\begin{equation}
\label{eq:ir-starstar}
IR^{**} = IR^{*}\space\times\space ARC\space\times\space\frac{sign(ARC)}{MD}
\end{equation}

\paragraph{Extended Information Ratio (\(IR3\))}
In addition to \(IR^{**}\), an extended risk-adjusted performance metric is 
proposed that simultaneously penalizes volatility, maximum drawdown, and 
maximum loss duration:
\begin{equation}
\label{eq:ir3}
IR^{***} = \frac{ARC^3}{ASD \times MD \times MLD}
\end{equation}

\paragraph{Turnover (ADT)}
Portfolio turnover measures the extent of rebalancing activity over time and reflects the trading intensity of the strategy. Let $w_{i,t}$ denote the portfolio weight of asset $i$ at time $t$, and $w^{c}_{t}$ the corresponding cash weight. Given previous weights $(w^{-}_{i,t}, w^{c,-}_{t})$, turnover is defined as:

\begin{equation}
\label{eq:turnover}
TO_t = \sum_{i=1}^{N_t} |w_{i,t} - w^{-}_{i,t}| + |w^{c}_{t} - w^{c,-}_{t}|
\end{equation}

Turnover corresponds to the $L^1$ distance between consecutive portfolio allocations and captures the magnitude of trading required to rebalance the portfolio. Transaction costs are modeled as proportional to turnover, and average turnover is reported as the time-averaged percentage across the evaluation period.

\hypertarget{sec:benchmarks}{%
\subsection{Benchmark Strategies}\label{sec:benchmarks}}

To evaluate the performance of the proposed reinforcement learning strategies, four benchmark portfolios are constructed and maintained throughout the out-of-sample evaluation period.

\paragraph{Buy \& Hold strategy}  represents the primary benchmark against which all reinforcement learning models are compared. It consists of a passive investment in the market index ETF the Invesco QQQ Trust (QQQ) for the NASDAQ-100, the SPDR Euro Stoxx 50 ETF (FEZ) for the EURO STOXX 50, and the iShares MSCI Japan ETF (EWJ) for the Nikkei 225 held continuously without rebalancing throughout the evaluation period. This benchmark is considered the most stringent point of comparison, as it requires no active management, incurs no transaction costs, and captures the full market return. Outperforming a passive Buy \& Hold strategy on a risk-adjusted basis is widely regarded as a high bar in empirical portfolio management research \citep{DE_MIGEUL}, and constitutes the central performance criterion of this study.

\paragraph{Momentum Top-20 strategy}  selects the top 20 assets by 120-day price momentum at each rebalancing date and allocates capital equally among them. This benchmark captures the well-documented momentum premium and serves as a comparison point for the momentum pre-selection mechanism embedded in the RL framework, isolating whether the RL allocation policy adds value beyond simple momentum-based selection.

\paragraph{Equal Weight monthly strategy}  allocates capital uniformly across the same top-$k$ momentum-selected universe used by the RL agent, rebalanced at monthly frequency. Following \citet{DE_MIGEUL}, this naive diversification benchmark is included to assess whether the RL policy adds value over and above the momentum pre-filter alone. If the equal-weight portfolio achieves comparable performance to the RL strategies, it would suggest that the momentum selection step, rather than the learned allocation policy, is the primary driver of returns.

\paragraph {Markowitz Minimum Variance portfolio} is constructed within the same walk-forward optimization framework as the reinforcement learning strategies, using a 5-year rolling training window to estimate the covariance matrix. At each rebalancing date, portfolio weights are obtained numerically by solving a constrained minimum-variance problem subject to long-only, fully-invested, and maximum position size constraints, and the portfolio is rebalanced monthly (every 21 trading days). This benchmark provides a direct comparison between the proposed RL framework and a classical optimization-based approach under identical data constraints and evaluation conditions.

\hypertarget{Methodology-Summary}{%
\subsection{Methodology Summary}\label{Methodology-Summary}}

The overall research framework can be summarized as a sequential pipeline that integrates data processing, feature construction, model training, and evaluation within a unified reinforcement learning setting. sThe process begins with the collection and preprocessing of financial data, followed by the construction of asset-level and global features as described in Section~\ref{State_Representation}. These features form the input state representation for the reinforcement learning agent. The agent interacts with the environment by iteratively observing the state, selecting portfolio allocations, and receiving rewards based on portfolio performance. The model is trained using the Soft Actor--Critic algorithm under a walk-forward optimization framework, combined with adaptive retraining and systematic hyperparameter selection.

\begin{figure}
{\centering \includegraphics[height=0.4\textheight]{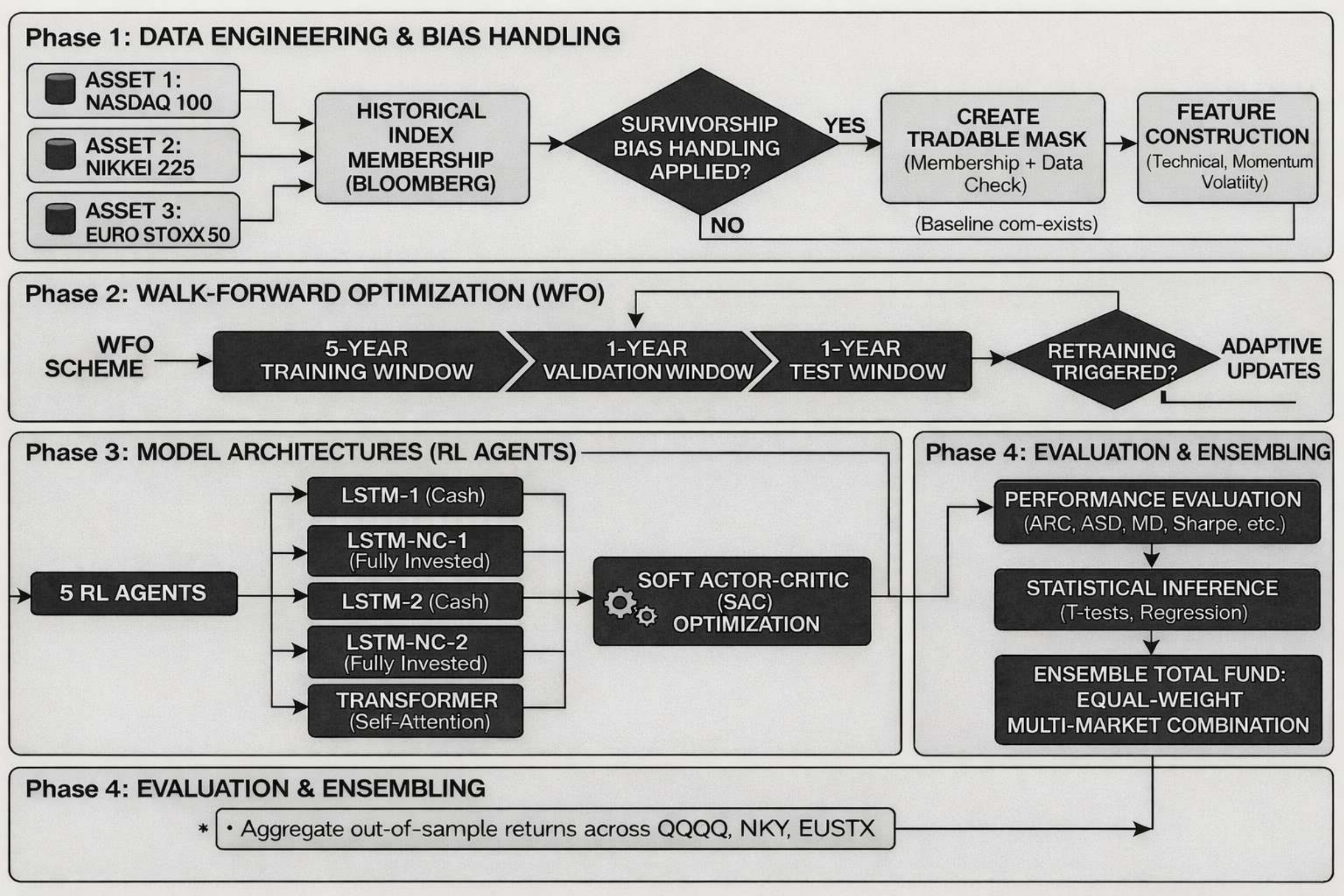} }
\caption{Overview of the reinforcement learning portfolio allocation research methodology}
\label{fig:method_pipeline}
\end{figure}

For each evaluation period, the trained model generates portfolio allocations, from which performance metrics are computed. This pipeline ensures that all results are obtained in a strictly out-of-sample setting and reflect realistic trading conditions. A schematic overview of the full methodology is presented in Figure~\ref{fig:method_pipeline}.

From an economic perspective, the proposed framework can be viewed as a dynamic portfolio allocation mechanism that adapts to changing market conditions. The reinforcement learning agent learns to balance expected returns, risk, and transaction costs through repeated interaction with the market environment. The incorporation of turnover and concentration penalties further aligns the learned strategies with realistic portfolio management objectives, such as diversification and cost efficiency.

\vspace{-0.4cm}

\linespread{0.65}\selectfont 
{\noindent\scriptsize \textbf{Note}: \textit{The figure illustrates the end-to-end workflow, from data preprocessing and feature construction to reinforcement learning training and out-of-sample evaluation.}}
\noindent\linespread{0.65}\selectfont
{\scriptsize \textbf{Source}: \textit{Authors’ own illustration.}}

\linespread{1.5}\selectfont

\hypertarget{empirical_results}{%
\section{Empirical results}\label{empirical_results}}

This section presents the empirical evaluation of the proposed reinforcement learning-based portfolio allocation strategies across different equity markets. For each market, the results are reported for three reinforcement learning configurations: LSTM with cash allowed, LSTM without cash, and a Transformer-based model. These approaches are benchmarked against four strategies described in Section~\ref{sec:benchmarks}: a passive Buy \& Hold strategy, a Momentum Top-20 portfolio, an Equal-Weight Monthly portfolio, and a Markowitz Minimum Variance portfolio.

The primary criterion for model selection in this study is the Modified Information Ratio (IR2), which captures risk-adjusted performance while explicitly penalizing drawdowns. This metric is considered the most comprehensive measure of strategy quality and is therefore used to identify the best-performing model within each setting. The Buy \& Hold strategy remains the primary benchmark for comparison, as it represents the most competitive passive alternative requiring no active management or transaction costs. Results are presented sequentially for each market to assess both within-market performance and cross-market consistency.

\hypertarget{er_QQQ}{%
\subsection{NASDAQ 100}\label{er_QQQ}}

\begin{figure}[H]
\centering

{\scriptsize \textbf\textbf{Panel A: RL with LSTM and Cash Allowed}\par}
\vspace{0.2cm}
\includegraphics[width=\textwidth, height=0.25\textheight, trim={0 0 0 0.65cm}, clip]{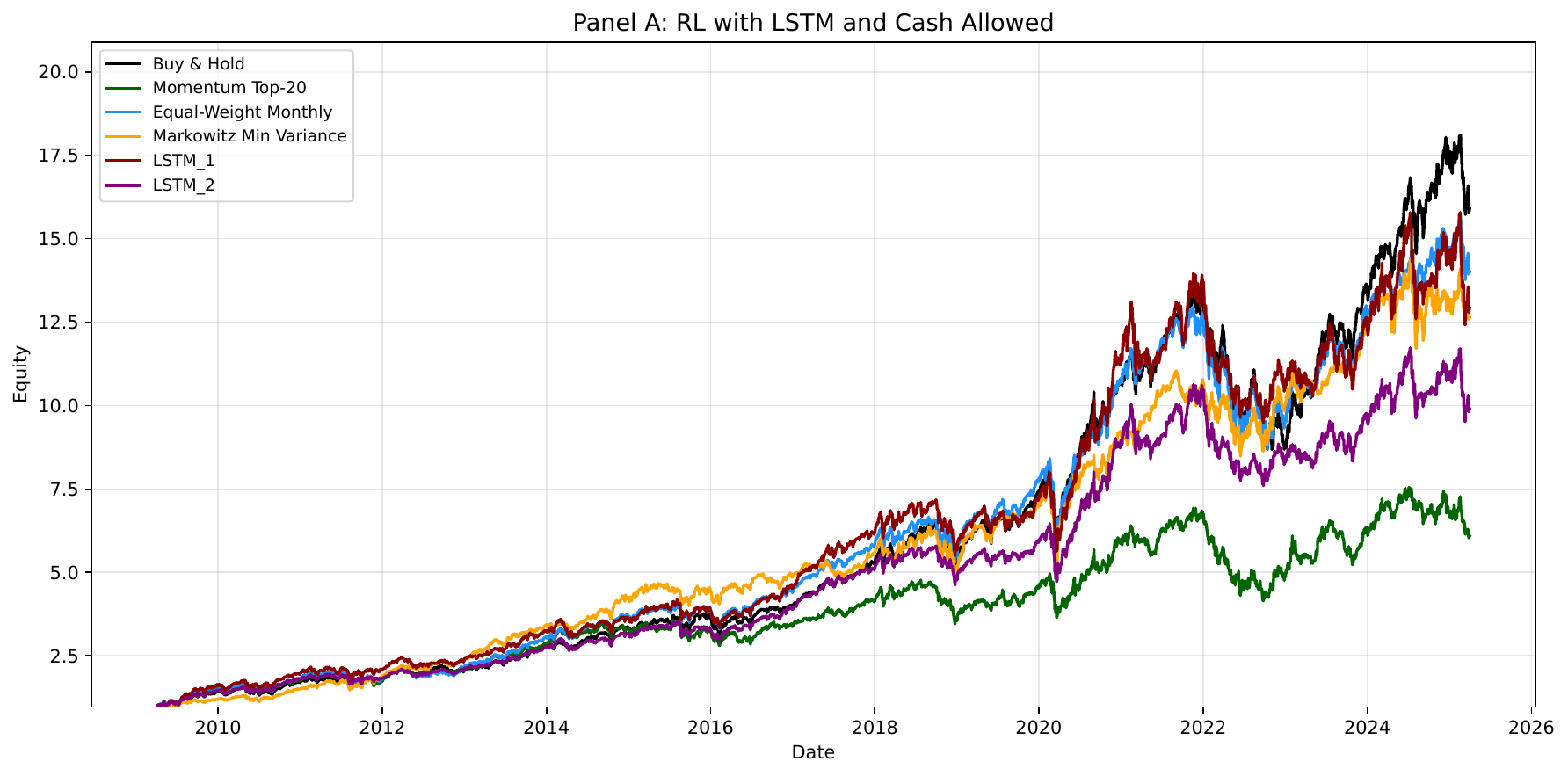}

{\scriptsize \textbf\textbf{Panel B: RL with LSTM and Cash Not-Allowed}\par}
\vspace{0.2cm}
\includegraphics[width=\textwidth, height=0.25\textheight, trim={0 0 0 0.65cm}, clip]{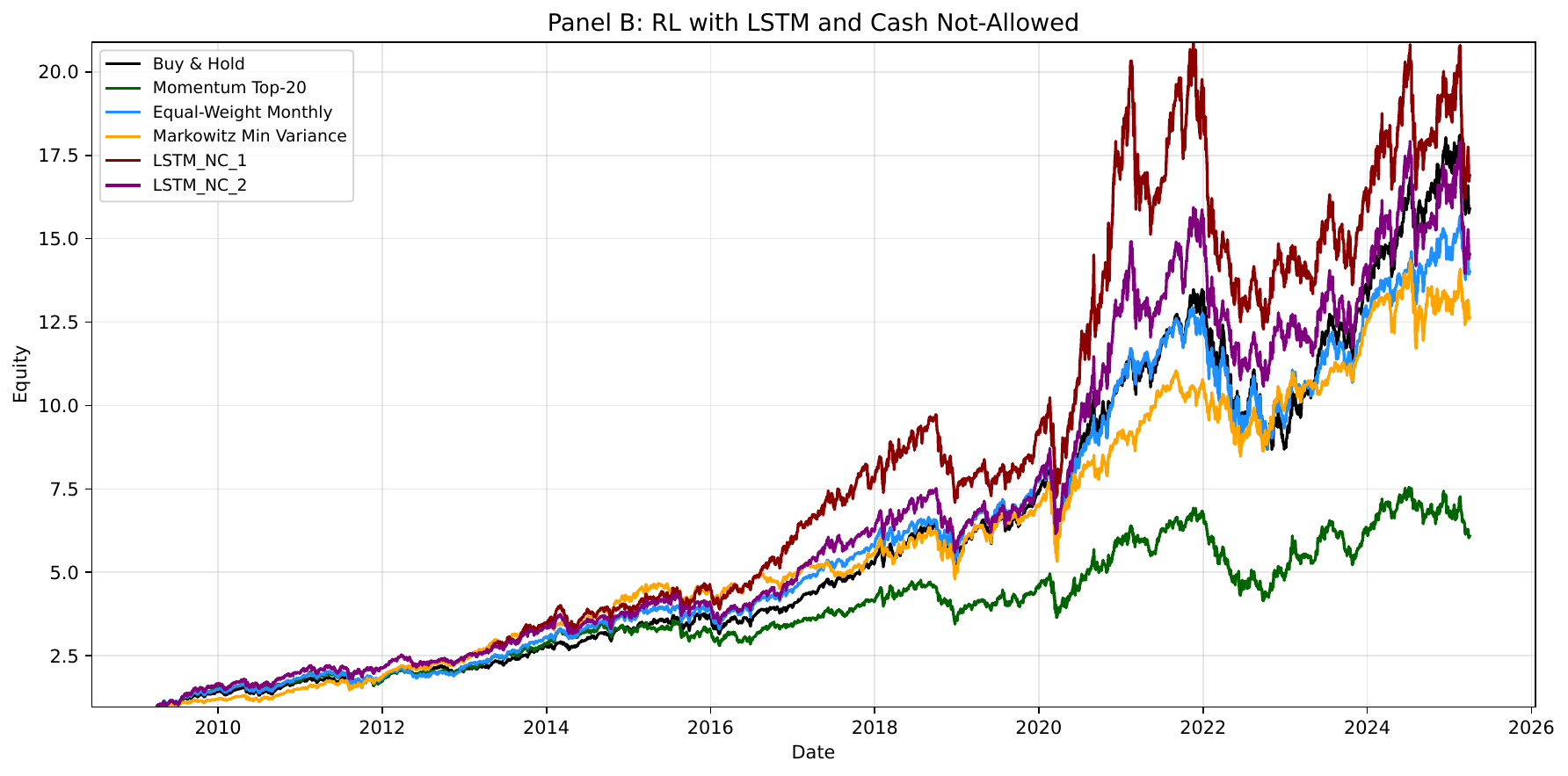}

{\scriptsize \textbf\textbf{Panel C: RL with Transformers}\par}
\vspace{0.2cm}
\includegraphics[width=\textwidth, height=0.25\textheight, trim={0 0 0 0.65cm}, clip]{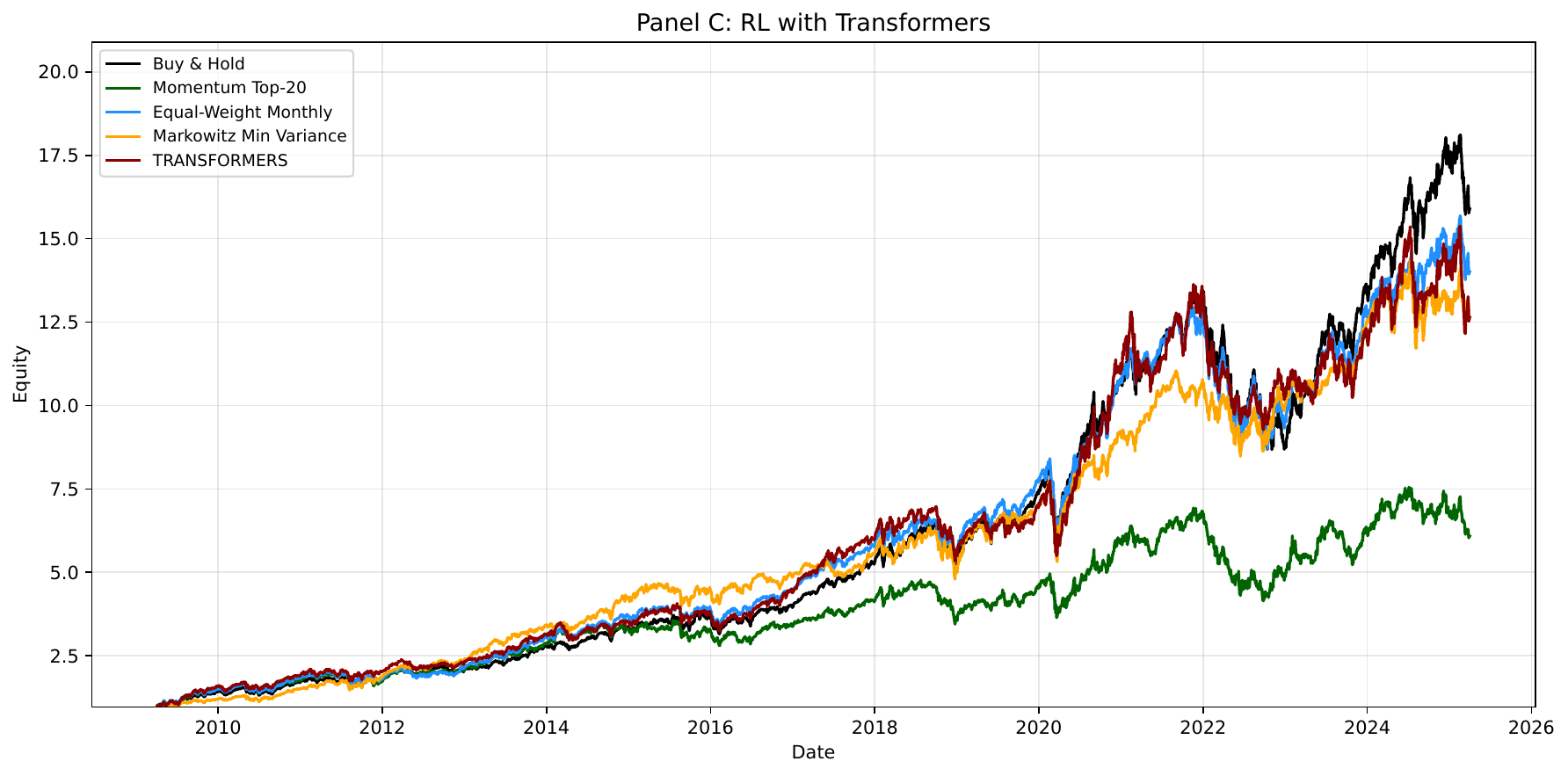}

\caption{Empirical results for the NASDAQ-100 across the three reinforcement learning configurations.}
\label{fig:qqq_empirical_results}
\end{figure}

Figure~\ref{fig:qqq_empirical_results} presents the equity curves for all strategies across the three configurations for the NASDAQ-100 equity index. Across all panels, the Buy \& Hold benchmark and the Equal-Weight Monthly portfolio exhibit the strongest long-term growth, with RL strategies consistently tracking below the passive benchmarks throughout the evaluation period. In Panel B, LSTM\_NC\_1 achieves the highest absolute return among RL strategies but at the cost of substantially higher volatility and drawdowns. The Transformer model in Panel C follows a similar trajectory to the LSTM models without a clear performance advantage.

Table~\ref{tab:perf_metric_qqq} reports the corresponding performance metrics. Based on the IR2 criterion, the Buy \& Hold benchmark achieves the highest value across all panels (IR2 = 0.52), followed closely by the Equal-Weight Monthly portfolio (IR2 = 0.49). Among the RL strategies, LSTM\_2 in Panel A achieves the strongest risk-adjusted performance (IR2 = 0.46) with the lowest volatility and drawdown across RL configurations (ASD = 18.67\%, MD = 28.77\%), reflecting the benefit of the hierarchical policy in managing downside risk. The Momentum Top-20 strategy underperforms all other strategies on IR2 despite its high turnover.

Overall, no RL strategy consistently outperforms the passive benchmarks on the NASDAQ-100 in terms of risk-adjusted performance. The strong and persistent upward trend of this index over the evaluation period systematically favors passive allocation strategies, limiting the value of active reinforcement learning allocation.

\vspace{-0.4cm}

\linespread{0.61}\selectfont
{\noindent\scriptsize \textbf{Note}: \textit{The figure presents the empirical results for the NASDAQ-100 under the three model configurations considered in the study. Panel A corresponds to the LSTM-based model with cash allowed (LSTM\_1: flat Dirichlet policy; LSTM\_2: hierarchical policy), Panel B to the LSTM-based model without cash and benchmark-relative reward (LSTM\_NC\_1: flat Dirichlet, top-$k$=10/20; LSTM\_NC\_2: flat Dirichlet, top-$k$=20), and Panel C to the Transformer-based model with flat Dirichlet policy and cash allowed (TRANSFORMERS). Trading begins on 2009-04-01 for the NASDAQ-100.}}

\linespread{1.5}\selectfont

\newpage

\begin{table}[htbp]
\centering
\scriptsize
\setlength{\tabcolsep}{3pt}
\renewcommand{\arraystretch}{0.95}

\caption{Performance Metrics for NASDAQ-100 Strategies}
\label{tab:perf_metric_qqq}
\resizebox{\textwidth}{!}{%
\begin{tabular}{lcccccccccc}
\toprule
\textbf{Strategy} & \textbf{AR (\%)} & \textbf{ARC (\%)} & \textbf{ASD (\%)} & \textbf{MD (\%)} & \textbf{MLD} & \textbf{IR1} & \textbf{IR2} & \textbf{IR3} & \textbf{Sharpe} & \textbf{ADT} \\
\midrule
\multicolumn{11}{l}{\textbf{Panel A: RL with LSTM and Cash Allowed}} \\
\midrule
Buy \& Hold & \textbf{1581.9} & \textbf{19.27} & 20.41 & 35.12 & 1.960 & \textbf{0.9441} & \textbf{0.5180} & \textbf{5.0956} & \textbf{0.9689} & 0.000 \\
Momentum Top-20 & 540.5 & 12.27 & 21.65 & 39.55 & 2.242 & 0.5667 & 0.1758 & 0.9626 & 0.6461 & 36.159 \\
Equal-Weight Monthly & 1338.2 & 18.13 & 20.47 & 32.55 & 2.071 & 0.8857 & 0.4933 & 4.3184 & 0.9187 & 0.326 \\
Markowitz Min Variance & 1235.2 & 16.65 & 19.18 & 31.00 & \textbf{1.667} & 0.8681 & 0.4662 & 4.6550 & 0.8978 & \textbf{0.214} \\
LSTM\_1 & 1234.8 & 17.61 & 21.63 & 32.34 & 2.258 & 0.8141 & 0.4433 & 3.4589 & 0.8622 & 13.782 \\
LSTM\_2 & 919.9 & 15.64 & \textbf{18.67} & \textbf{28.77} & 2.262 & 0.8377 & 0.4554 & 3.1500 & 0.8759 & 13.951 \\
\addlinespace
\multicolumn{11}{l}{\textbf{Panel B: RL with LSTM and Cash Not-Allowed}} \\
\midrule
Buy \& Hold & 1581.9 & 19.27 & 20.41 & 35.12 & 1.960 & \textbf{0.9441} & \textbf{0.5180} & \textbf{5.0956} & \textbf{0.9689} & 0.000 \\
Momentum Top-20 & 540.5 & 12.27 & 21.65 & 39.55 & 2.242 & 0.5667 & 0.1758 & 0.9626 & 0.6461 & 36.159 \\
Equal-Weight Monthly & 1338.2 & 18.13 & 20.47 & 32.55 & 2.071 & 0.8857 & 0.4933 & 4.3184 & 0.9187 & 0.326 \\
Markowitz Min Variance & 1235.2 & 16.65 & \textbf{19.18} & \textbf{31.00} & \textbf{1.667} & 0.8681 & 0.4662 & 4.6550 & 0.8978 & \textbf{0.214} \\
LSTM\_NC\_1 & \textbf{1666.4} & \textbf{19.70} & 24.20 & 41.23 & 2.595 & 0.8140 & 0.3890 & 2.9512 & 0.8678 & 16.326 \\
LSTM\_NC\_2 & 1407.9 & 18.51 & 22.63 & 33.95 & 2.262 & 0.8179 & 0.4460 & 3.6509 & 0.8676 & 14.782 \\
\addlinespace
\multicolumn{11}{l}{\textbf{Panel C: RL with Transformers}} \\
\midrule
Buy \& Hold & \textbf{1581.9} & \textbf{19.27} & 20.41 & 35.12 & 1.960 & \textbf{0.9441} & \textbf{0.5180} & \textbf{5.0956} & \textbf{0.9689} & 0.000 \\
Momentum Top-20 & 540.5 & 12.27 & 21.65 & 39.55 & 2.242 & 0.5667 & 0.1758 & 0.9626 & 0.6461 & 36.159 \\
Equal-Weight Monthly & 1338.2 & 18.13 & 20.47 & 32.55 & 2.071 & 0.8857 & 0.4933 & 4.3184 & 0.9187 & 0.326 \\
Markowitz Min Variance & 1235.2 & 16.65 & \textbf{19.18} & \textbf{31.00} & \textbf{1.667} & 0.8681 & 0.4662 & 4.6550 & 0.8978 & \textbf{0.214} \\
TRANSFORMERS & 1206.9 & 17.46 & 21.43 & 32.27 & 2.262 & 0.8147 & 0.4408 & 3.4014 & 0.8619 & 14.248 \\
\addlinespace
\bottomrule
\end{tabular}%
}
\end{table}

\vspace{-0.4cm}
\noindent\linespread{0.65}\selectfont
{\scriptsize \textbf{Note}: \textit{All strategies are evaluated on the NASDAQ-100, with trading commencing on 2009-04-06. All reported results are rounded to four significant figures. Strategy labels refer to the following configurations --- LSTM\_1: LSTM encoder, flat Dirichlet policy, log-return reward, cash allowed; LSTM\_2: LSTM encoder, hierarchical policy, log-return reward, cash allowed; LSTM\_NC\_1: LSTM encoder, flat Dirichlet policy, benchmark-relative reward, fully invested; LSTM\_NC\_2: LSTM encoder, flat Dirichlet policy, benchmark-relative reward, fully invested; TRANSFORMERS: Transformer encoder, flat Dirichlet policy, log-return reward, cash allowed. AR denotes the total cumulative return over the full evaluation period; ARC denotes the annualized compounded return and is the primary return criterion used throughout this study. IR2 is the primary evaluation metric for risk-adjusted performance.}}

\linespread{1.5}\selectfont

\hypertarget{er_NKSY}{%
\subsection{NIKKEI 225}\label{er_NKSY}}

\begin{figure}[H]
\centering

{\scriptsize \textbf\textbf{Panel A: RL with LSTM and Cash Allowed}\par}
\vspace{0.3cm}
\includegraphics[width=\textwidth, height=0.25\textheight, trim={0 0 0 0.65cm}, clip]{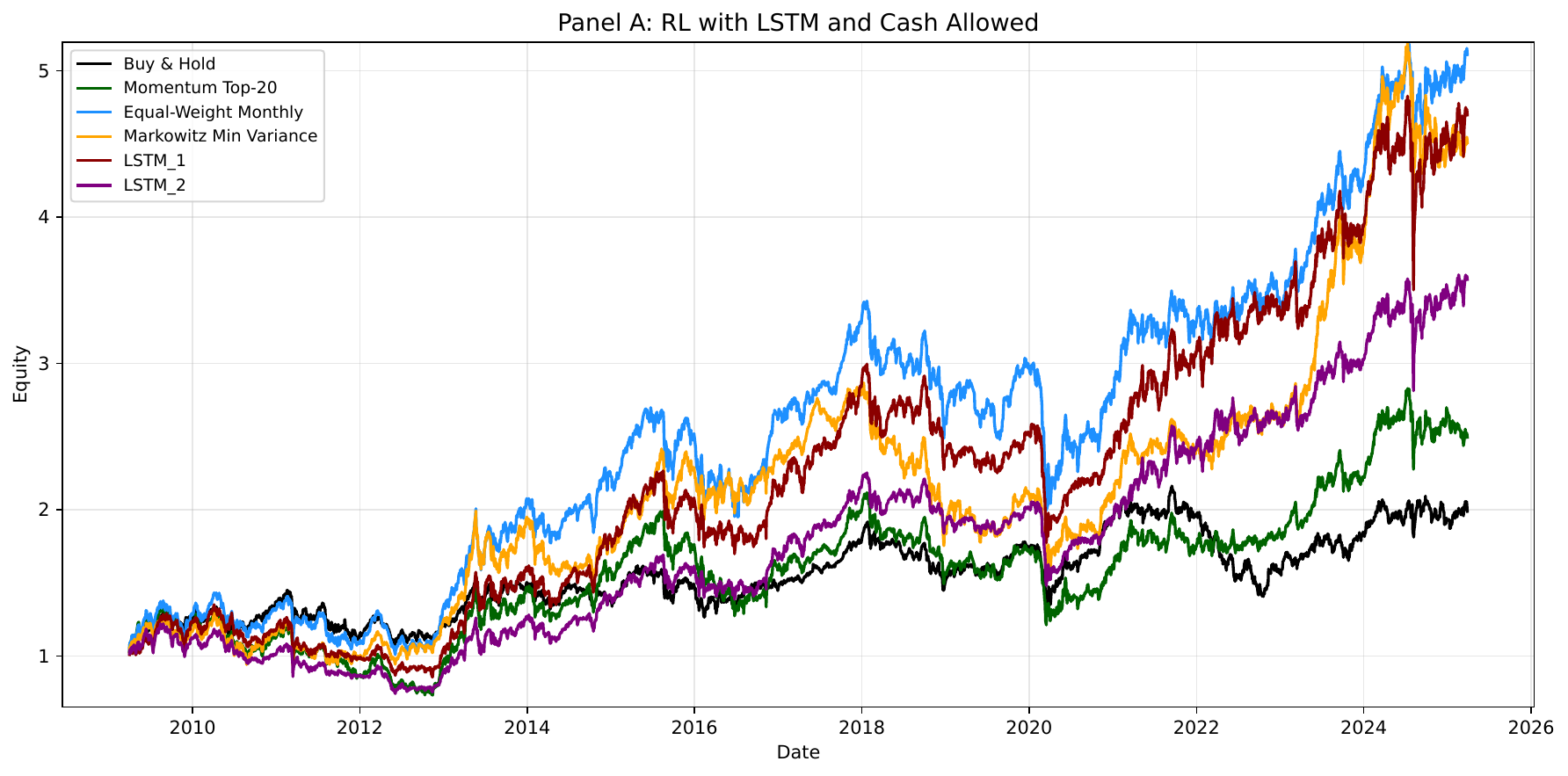}

{\scriptsize \textbf\textbf{Panel B: RL with LSTM and Cash Not-Allowed}\par}
\vspace{0.3cm}
\includegraphics[width=\textwidth, height=0.25\textheight, trim={0 0 0 0.65cm}, clip]{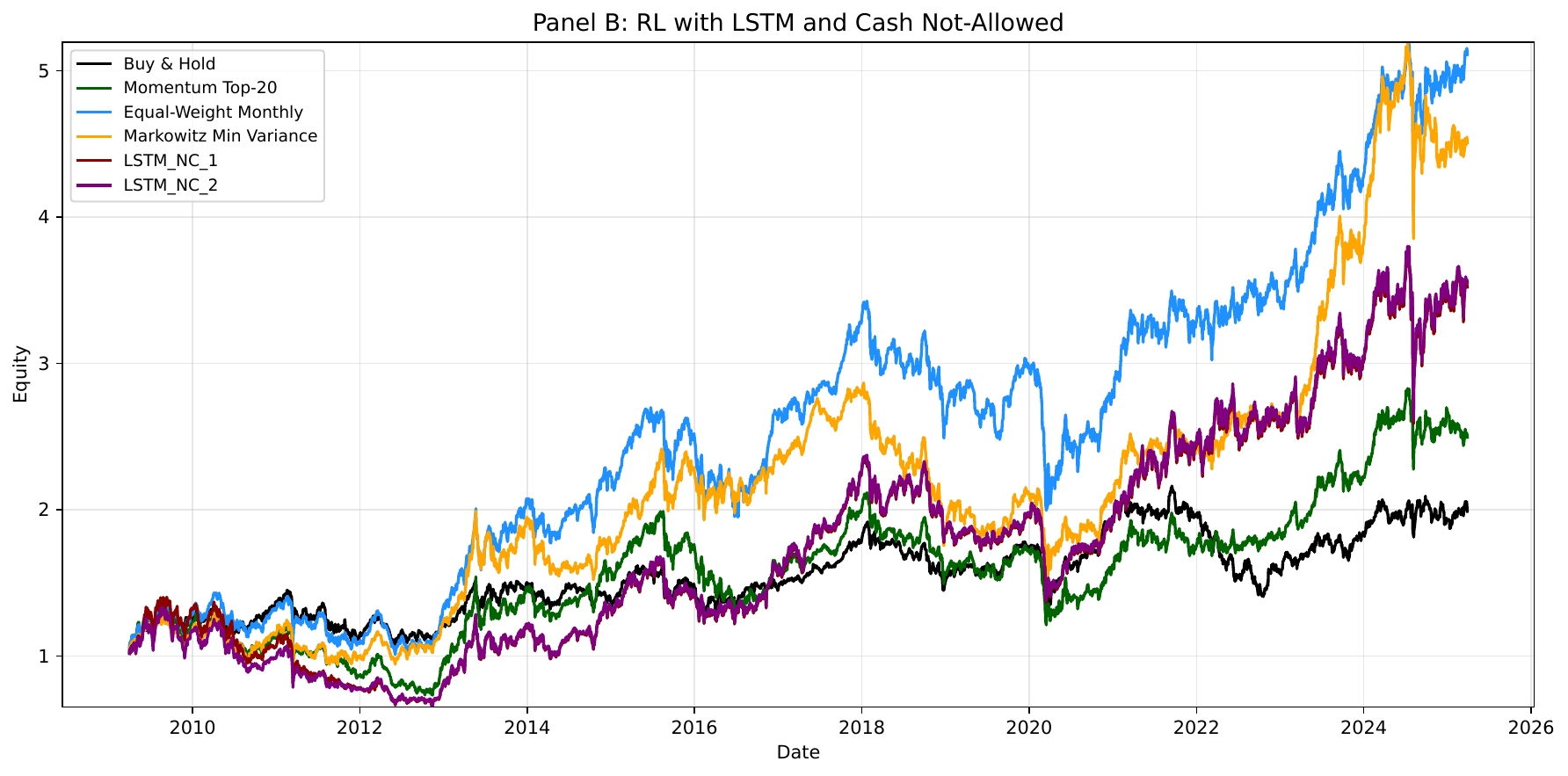}

{\scriptsize \textbf\textbf{Panel C: RL with Transformers}\par}
\vspace{0.3cm}
\includegraphics[width=\textwidth, height=0.25\textheight, trim={0 0 0 0.65cm}, clip]{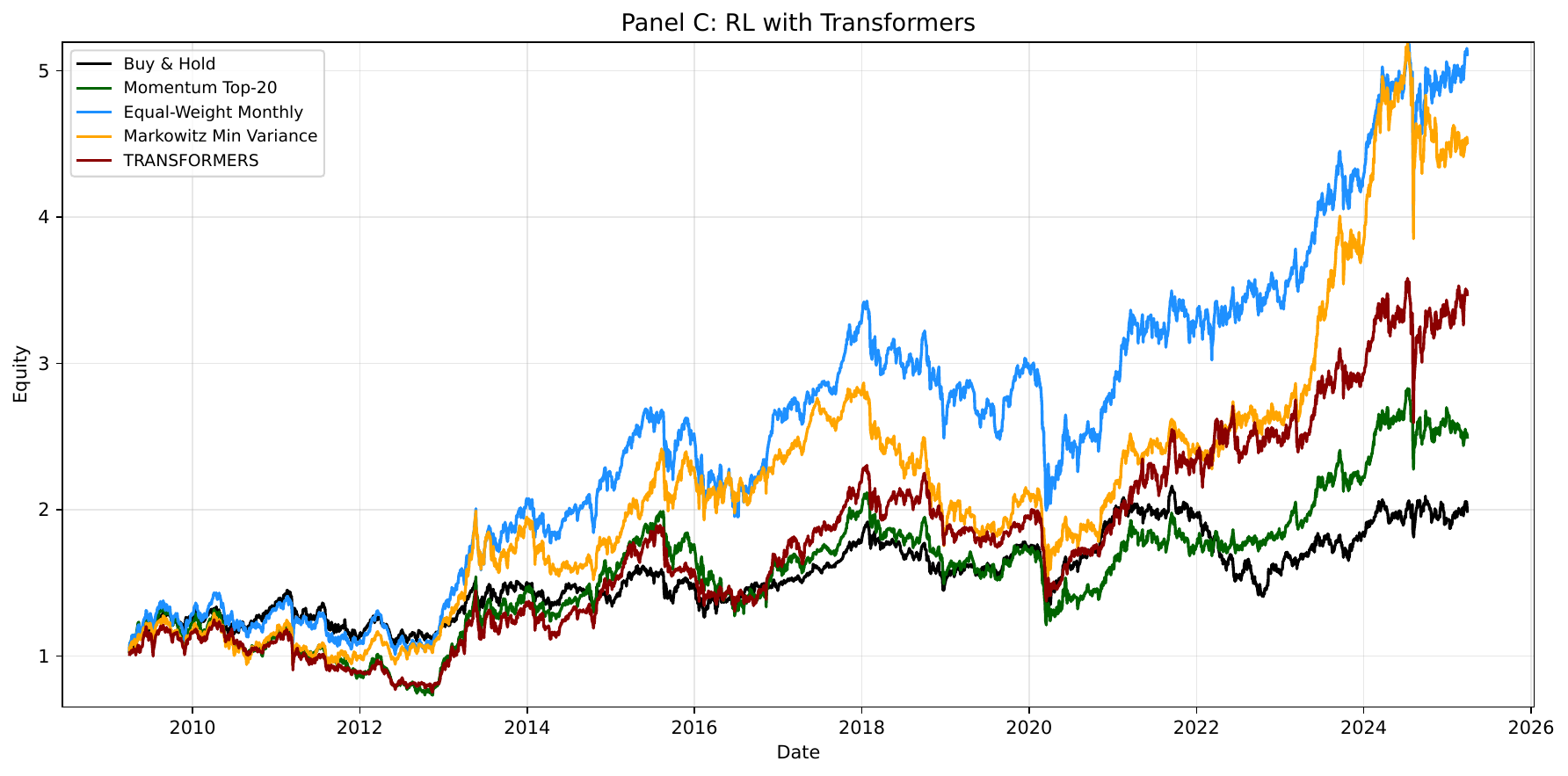}

\caption{Empirical results for the NIKKEI 225 across the three reinforcement learning configurations.}
\label{fig:nky_empirical_results}
\end{figure}

Figure~\ref{fig:nky_empirical_results} presents the equity curves for all strategies across the three configurations for the Nikkei 225 equity index. Across all panels, the Markowitz Minimum Variance and Equal-Weight Monthly portfolios exhibit the strongest long-term growth, substantially outperforming both the Buy \& Hold benchmark and all RL strategies. The Buy \& Hold benchmark delivers the weakest performance, reflecting the challenging nature of the Japanese equity market over the evaluation period. In Panel A, LSTM\_1 tracks closest to the leading benchmarks among RL strategies. In Panel B, the fully invested configurations trail substantially behind the passive benchmarks. Panel C shows the Transformer model achieving intermediate performance between the two LSTM configurations.

Table~\ref{tab:perf_metric_nky} reports the corresponding performance metrics. The Markowitz Minimum Variance portfolio achieves the highest ARC (13.17\%) and IR1 (0.61), while the Equal-Weight Monthly portfolio leads on IR2 (0.19). Among the RL strategies, LSTM\_1 in Panel A achieves the strongest risk-adjusted performance (IR2 = 0.15) and the highest IR3 (2.50), with the lowest MLD (0.690) reflecting effective drawdown recovery. The Buy \& Hold benchmark substantially underperforms all other strategies, driven by its extremely high MLD of 11.345 indicating a recovery period of over eleven years following the 2008 financial crisis.

Overall, the Nikkei 225 results reveal that simple diversification strategies such as Equal-Weight Monthly and Markowitz Minimum Variance add substantial value over passive index exposure in this market. While RL strategies outperform the Buy \& Hold benchmark, they do not consistently match the performance of the classical benchmarks.

\vspace{-0.4cm}

\linespread{0.61}\selectfont
{\noindent\scriptsize \textbf{Note}: \textit{The figure presents the empirical results for the Nikkei 225 under the three model configurations considered in the study. Panel A corresponds to the LSTM-based model with cash allowed (LSTM\_1: flat Dirichlet policy; LSTM\_2: hierarchical policy), Panel B to the LSTM-based model without cash and benchmark-relative reward (LSTM\_NC\_1: flat Dirichlet, top-$k$=10/20; LSTM\_NC\_2: flat Dirichlet, top-$k$=20), and Panel C to the Transformer-based model with flat Dirichlet policy and cash allowed (TRANSFORMERS). Trading begins on 2009-04-06 for the Nikkei 225.}}

\linespread{1.5}\selectfont

\newpage

\begin{table}[htbp]
\centering
\scriptsize
\setlength{\tabcolsep}{3pt}
\renewcommand{\arraystretch}{0.95}

\caption{Performance Metrics for Nikkei 225 Strategies}
\label{tab:perf_metric_nky}
\resizebox{\textwidth}{!}{%
\begin{tabular}{lcccccccccc}
\toprule
\textbf{Strategy} & \textbf{AR (\%)} & \textbf{ARC (\%)} & \textbf{ASD (\%)} & \textbf{MD (\%)} & \textbf{MLD} & \textbf{IR1} & \textbf{IR2} & \textbf{IR3} & \textbf{Sharpe} & \textbf{ADT} \\
\midrule
\multicolumn{11}{l}{\textbf{Panel A: RL with LSTM and Cash Allowed}} \\
\midrule
Buy \& Hold & 171.0 & 4.57 & 20.41 & 55.80 & 11.345 & 0.2239 & 0.0183 & 0.0074 & 0.3189 & 0.000 \\
Momentum Top-20 & 196.5 & 7.29 & 21.02 & 45.06 & 3.567 & 0.3468 & 0.0561 & 0.1145 & 0.4370 & 35.191 \\
Equal-Weight Monthly & 514.6 & 12.40 & 20.86 & 39.33 & 3.060 & 0.5944 & \textbf{0.1874} & 0.7603 & 0.6630 & 0.261 \\
Markowitz Min Variance & \textbf{665.4} & \textbf{13.17} & 21.76 & 46.23 & 1.048 & \textbf{0.6052} & 0.1724 & 2.1696 & \textbf{0.6772} & \textbf{0.239} \\
LSTM\_1 & 418.2 & 11.22 & 20.96 & 39.09 & 0.690 & 0.5353 & 0.1536 & \textbf{2.4979} & 0.6105 & 23.497 \\
LSTM\_2 & 285.8 & 9.12 & \textbf{18.37} & \textbf{38.27} & \textbf{0.571} & 0.4965 & 0.1183 & 1.8876 & 0.5652 & 18.858 \\
\addlinespace
\multicolumn{11}{l}{\textbf{Panel B: RL with LSTM and Cash Not-Allowed}} \\
\midrule
Buy \& Hold & 171.0 & 4.57 & \textbf{20.41} & 55.80 & 11.345 & 0.2239 & 0.0183 & 0.0074 & 0.3189 & 0.000 \\
Momentum Top-20 & 196.5 & 7.29 & 21.02 & 45.06 & 3.567 & 0.3468 & 0.0561 & 0.1145 & 0.4370 & 35.191 \\
Equal-Weight Monthly & 514.6 & 12.40 & 20.86 & \textbf{39.33} & 3.060 & 0.5944 & \textbf{0.1874} & 0.7603 & 0.6630 & 0.261 \\
Markowitz Min Variance & \textbf{665.4} & \textbf{13.17} & 21.76 & 46.23 & 1.048 & \textbf{0.6052} & 0.1724 & \textbf{2.1696} & \textbf{0.6772} & \textbf{0.239} \\
LSTM\_NC\_1 & 285.7 & 9.11 & 24.15 & 53.07 & \textbf{0.690} & 0.3772 & 0.0648 & 0.8546 & 0.4813 & 20.789 \\
LSTM\_NC\_2 & 291.8 & 9.22 & 23.72 & 49.98 & \textbf{0.690} & 0.3887 & 0.0717 & 0.9570 & 0.4897 & 20.036 \\
\addlinespace
\multicolumn{11}{l}{\textbf{Panel C: RL with Transformers}} \\
\midrule
Buy \& Hold & 171.0 & 4.57 & \textbf{20.41} & 55.80 & 11.345 & 0.2239 & 0.0183 & 0.0074 & 0.3189 & 0.000 \\
Momentum Top-20 & 196.5 & 7.29 & 21.02 & 45.06 & 3.567 & 0.3468 & 0.0561 & 0.1145 & 0.4370 & 35.191 \\
Equal-Weight Monthly & 514.6 & 12.40 & 20.86 & 39.33 & 3.060 & 0.5944 & \textbf{0.1874} & 0.7603 & 0.6630 & 0.261 \\
Markowitz Min Variance & \textbf{665.4} & \textbf{13.17} & 21.76 & 46.23 & 1.048 & \textbf{0.6052} & 0.1724 & \textbf{2.1696} & \textbf{0.6772} & \textbf{0.239} \\
TRANSFORMERS & 283.8 & 9.09 & 21.47 & \textbf{38.83} & \textbf{0.690} & 0.4234 & 0.0991 & 1.3062 & 0.5110 & 20.650 \\
\addlinespace
\bottomrule
\end{tabular}%
}
\end{table}

\vspace{-0.4cm}
\noindent\linespread{0.61}\selectfont
{\scriptsize \textbf{Note}: \textit{All strategies are evaluated on the Nikkei 225, with trading commencing on 2009-04-01. All reported results are rounded to four significant figures. Strategy labels refer to the following configurations --- LSTM\_1: LSTM encoder, flat Dirichlet policy, log-return reward, cash allowed; LSTM\_2: LSTM encoder, hierarchical policy, log-return reward, cash allowed; LSTM\_NC\_1: LSTM encoder, flat Dirichlet policy, benchmark-relative reward, fully invested; LSTM\_NC\_2: LSTM encoder, flat Dirichlet policy, benchmark-relative reward, fully invested; TRANSFORMERS: Transformer encoder, flat Dirichlet policy, log-return reward, cash allowed. AR denotes the total cumulative return over the full evaluation period; ARC denotes the annualized compounded return and is the primary return criterion used throughout this study. IR2 is the primary evaluation metric for risk-adjusted performance.}}

\linespread{1.5}\selectfont

\hypertarget{er_EUSTXX}{%
\subsection{EURO STOXX 50}\label{er_EUSTXX}}

\begin{figure}[H]
\centering

{\scriptsize \textbf\textbf{Panel A: RL with LSTM and Cash Allowed}\par}
\vspace{0.3cm}
\includegraphics[width=\textwidth, height=0.25\textheight, trim={0 0 0 0.65cm}, clip]{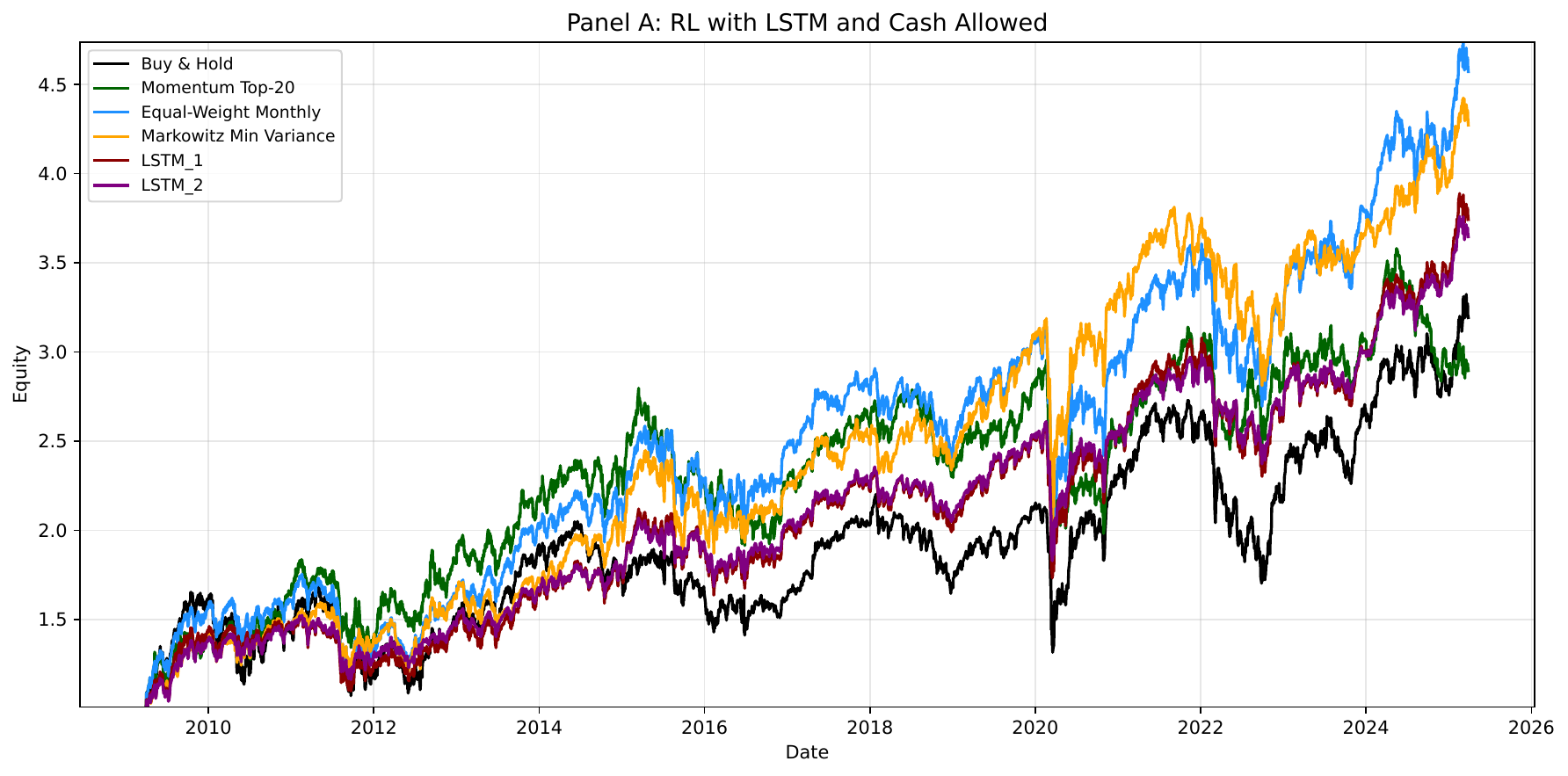}

{\scriptsize \textbf\textbf{Panel B: RL with LSTM and Cash Not-Allowed}\par}
\vspace{0.3cm}
\includegraphics[width=\textwidth, height=0.25\textheight, trim={0 0 0 0.65cm}, clip]{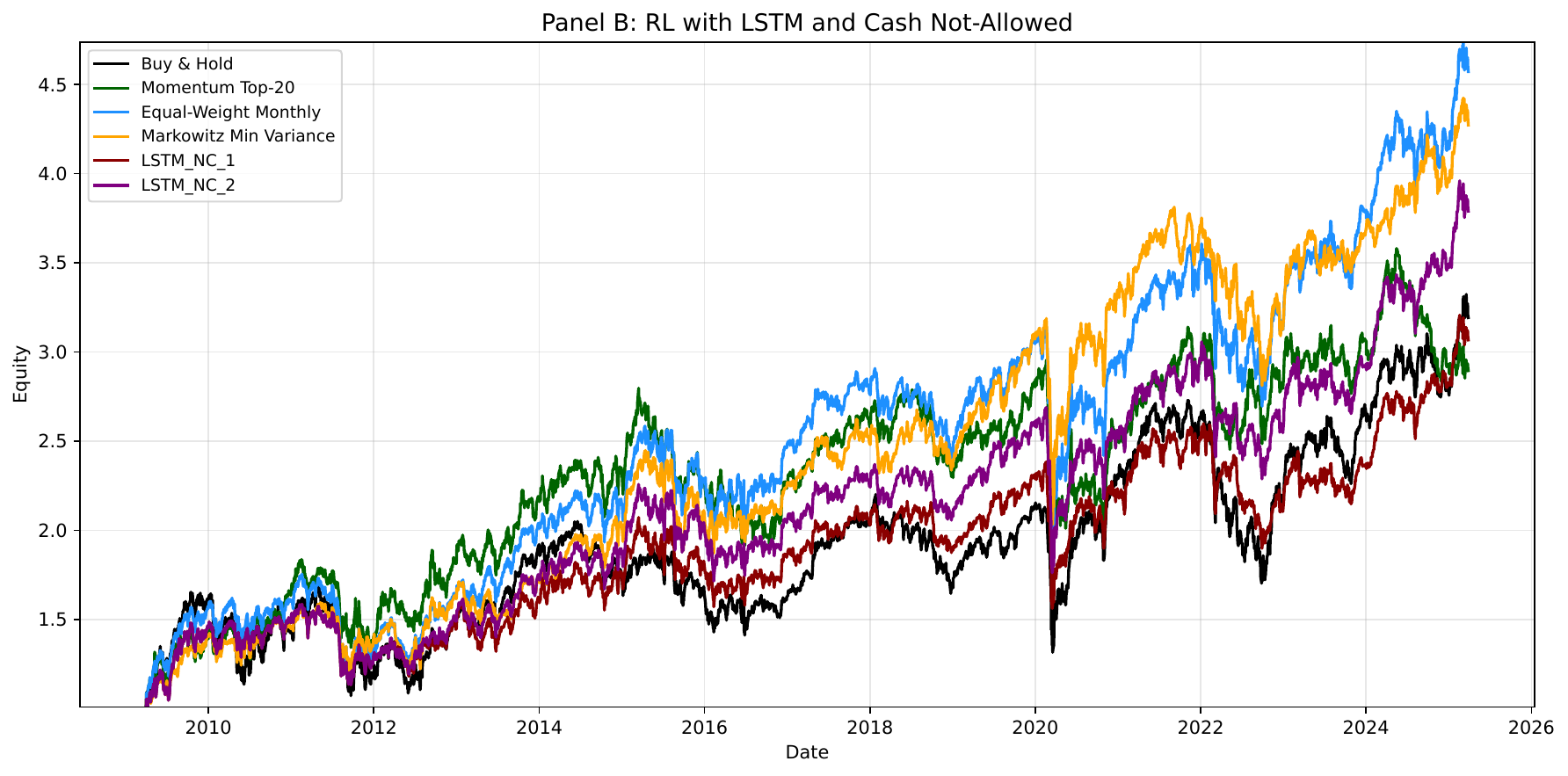}

{\scriptsize \textbf\textbf{Panel C: RL with Transformers}\par}
\vspace{0.3cm}
\includegraphics[width=\textwidth, height=0.25\textheight, trim={0 0 0 0.65cm}, clip]{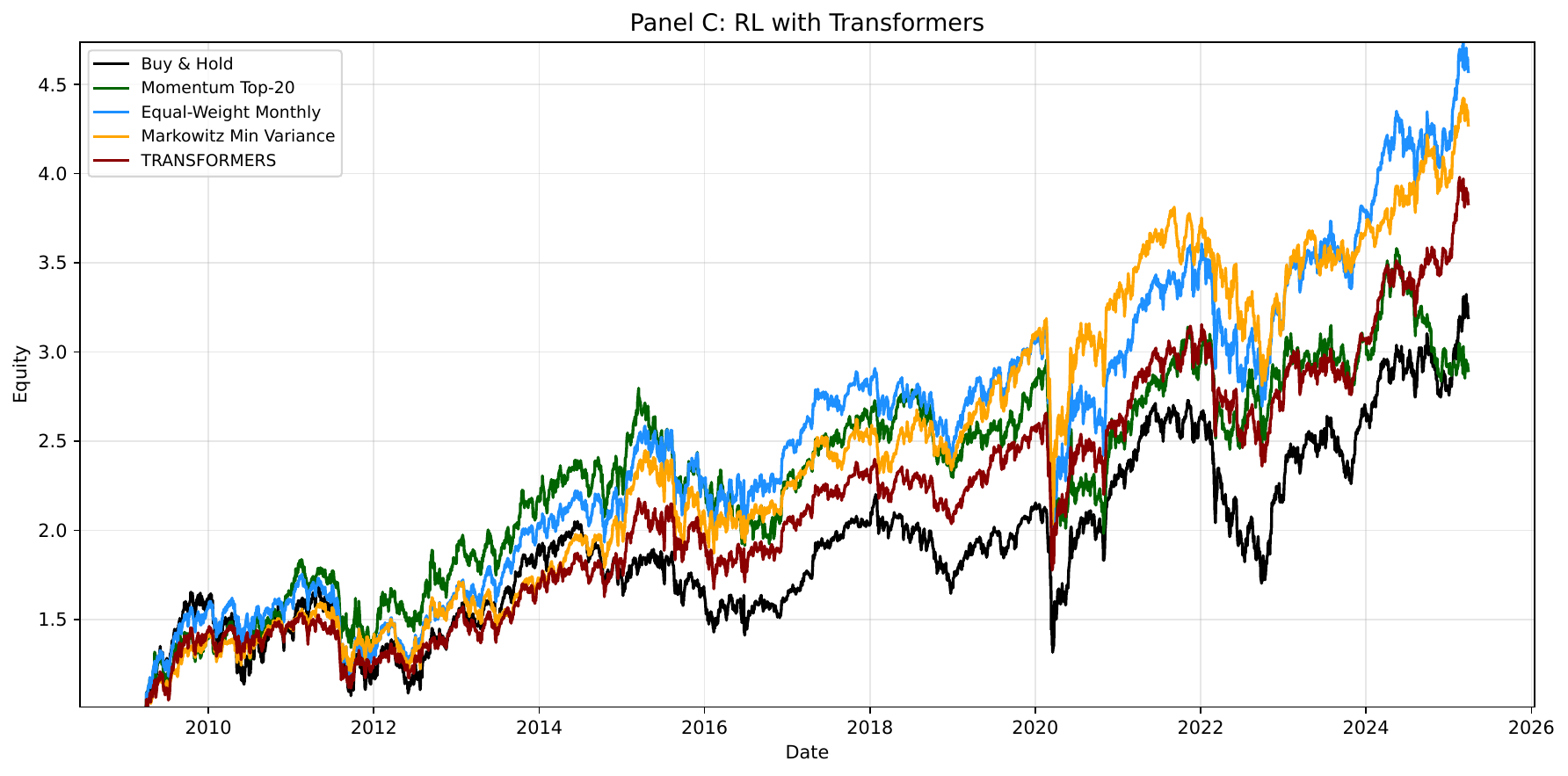}

\caption{Empirical results for the EURO STOXX 50 across the three reinforcement learning configurations.}
\label{fig:eustx_empirical_results}
\end{figure}

Figure~\ref{fig:eustx_empirical_results} presents the equity curves for all strategies across the three configurations for the EURO STOXX 50 equity index. Across all panels, the Equal-Weight Monthly portfolio exhibits the strongest long-term growth, followed by the Markowitz Minimum Variance portfolio. In Panel A, LSTM\_2 tracks closest to the leading benchmarks among RL strategies, exhibiting notably smoother growth than the Buy \& Hold benchmark. In Panel B, the fully invested configurations underperform relative to Panel A but remain above the Buy \& Hold benchmark. Panel C shows the Transformer model achieving competitive performance, closely tracking LSTM\_1.

Table~\ref{tab:perf_metric_eustxx} reports the corresponding performance metrics. The Equal-Weight Monthly portfolio achieves the highest ARC (10.24\%) and AR (384.8\%) across all strategies. Among the RL strategies, LSTM\_2 in Panel A achieves the strongest risk-adjusted performance, with the highest IR2 (0.15) and IR1 (0.52) across all RL configurations, and the lowest volatility and drawdown (ASD = 15.97\%, MD = 29.94\%). Notably, all RL strategies outperform the Buy \& Hold benchmark on IR2, in contrast to the results observed for the NASDAQ-100 and Nikkei 225. The Markowitz Minimum Variance portfolio achieves the lowest MLD (0.413) and ADT (0.172), reflecting its focus on variance minimization.

Overall, the EURO STOXX 50 results present the most favorable environment for reinforcement learning allocation among the three markets studied. All RL strategies outperform the Buy \& Hold benchmark on risk-adjusted metrics, with LSTM\_2 delivering the strongest performance. The relatively lower trend persistence and higher structural uncertainty of the European equity market appear to provide conditions more conducive to active RL-based allocation.

\vspace{-0.4cm}

\linespread{0.61}\selectfont
{\noindent\scriptsize \textbf{Note}: \textit{The figure presents the empirical results for the EURO STOXX 50 under the three model configurations considered in the study. Panel A corresponds to the LSTM-based model with cash allowed (LSTM\_1: flat Dirichlet policy; LSTM\_2: hierarchical policy), Panel B to the LSTM-based model without cash and benchmark-relative reward (LSTM\_NC\_1: flat Dirichlet, top-$k$=10/20; LSTM\_NC\_2: flat Dirichlet, top-$k$=20), and Panel C to the Transformer-based model with flat Dirichlet policy and cash allowed (TRANSFORMERS). Trading begins on 2009-04-01 for the EURO STOXX 50.}}

\linespread{1.5}\selectfont

\newpage

\begin{table}[htbp]
\centering
\scriptsize
\setlength{\tabcolsep}{3pt}
\renewcommand{\arraystretch}{0.95}

\caption{Performance Metrics for EURO STOXX 50 Strategies}
\label{tab:perf_metric_eustxx}
\resizebox{\textwidth}{!}{%
\begin{tabular}{lcccccccccc}
\toprule
\textbf{Strategy} & \textbf{AR (\%)} & \textbf{ARC (\%)} & \textbf{ASD (\%)} & \textbf{MD (\%)} & \textbf{MLD} & \textbf{IR1} & \textbf{IR2} & \textbf{IR3} & \textbf{Sharpe} & \textbf{ADT} \\
\midrule
\multicolumn{11}{l}{\textbf{Panel A: RL with LSTM and Cash Allowed}} \\
\midrule
Buy \& Hold & 238.2 & 7.81 & 23.92 & 39.69 & 1.643 & 0.3265 & 0.0642 & 0.3051 & 0.4333 & 0.000 \\
Momentum Top-20 & 209.0 & 7.22 & 21.07 & 33.37 & 1.548 & 0.3427 & 0.0741 & 0.3465 & 0.4348 & 39.184 \\
Equal-Weight Monthly & \textbf{384.8} & \textbf{10.24} & 20.63 & 39.40 & 1.139 & 0.4964 & 0.1290 & 1.1582 & 0.5738 & 0.859 \\
Markowitz Min Variance & 343.3 & 9.11 & 18.31 & 36.24 & \textbf{0.413} & 0.4975 & 0.1251 & \textbf{2.7640} & 0.5653 & \textbf{0.172} \\
LSTM\_1 & 275.5 & 8.55 & 18.71 & 33.52 & 1.048 & 0.4570 & 0.1166 & 0.9508 & 0.5301 & 10.613 \\
LSTM\_2 & 265.4 & 8.36 & \textbf{15.97} & \textbf{29.94} & 1.060 & \textbf{0.5235} & \textbf{0.1462} & 1.1525 & \textbf{0.5807} & 11.416 \\
\addlinespace
\multicolumn{11}{l}{\textbf{Panel B: RL with LSTM and Cash Not-Allowed}} \\
\midrule
Buy \& Hold & 238.2 & 7.81 & 23.92 & 39.69 & 1.643 & 0.3265 & 0.0642 & 0.3051 & 0.4333 & 0.000 \\
Momentum Top-20 & 209.0 & 7.22 & 21.07 & \textbf{33.37} & 1.548 & 0.3427 & 0.0741 & 0.3465 & 0.4348 & 39.184 \\
Equal-Weight Monthly & \textbf{384.8} & \textbf{10.24} & 20.63 & 39.40 & 1.139 & 0.4964 & \textbf{0.1290} & 1.1582 & \textbf{0.5738} & 0.859 \\
Markowitz Min Variance & 343.3 & 9.11 & \textbf{18.31} & 36.24 & \textbf{0.413} & \textbf{0.4975} & 0.1251 & \textbf{2.7640} & 0.5653 & \textbf{0.172} \\
LSTM\_NC\_1 & 208.3 & 7.23 & 19.87 & 34.49 & 1.131 & 0.3639 & 0.0763 & 0.4885 & 0.4492 & 15.661 \\
LSTM\_NC\_2 & 280.6 & 8.64 & 19.52 & 34.50 & 1.091 & 0.4426 & 0.1108 & 0.8765 & 0.5202 & 12.697 \\
\addlinespace
\multicolumn{11}{l}{\textbf{Panel C: RL with Transformers}} \\
\midrule
Buy \& Hold & 238.2 & 7.81 & 23.92 & 39.69 & 1.643 & 0.3265 & 0.0642 & 0.3051 & 0.4333 & 0.000 \\
Momentum Top-20 & 209.0 & 7.22 & 21.07 & 33.37 & 1.548 & 0.3427 & 0.0741 & 0.3465 & 0.4348 & 39.184 \\
Equal-Weight Monthly & \textbf{384.8} & \textbf{10.24} & 20.63 & 39.40 & 1.139 & 0.4964 & \textbf{0.1290} & 1.1582 & \textbf{0.5738} & 0.859 \\
Markowitz Min Variance & 343.3 & 9.11 & \textbf{18.31} & 36.24 & \textbf{0.413} & \textbf{0.4975} & 0.1251 & \textbf{2.7640} & 0.5653 & \textbf{0.172} \\
TRANSFORMERS & 284.7 & 8.71 & 18.78 & \textbf{33.16} & 1.048 & 0.4638 & 0.1218 & 1.0126 & 0.5367 & 8.418 \\
\addlinespace
\bottomrule
\end{tabular}%
}
\end{table}

\vspace{-0.4cm}
\noindent\linespread{0.65}\selectfont
{\scriptsize \textbf{Note}: \textit{All strategies are evaluated on the EURO STOXX 50, with trading commencing on 2009-04-01. All reported results are rounded to four significant figures. Strategy labels refer to the following configurations --- LSTM\_1: LSTM encoder, flat Dirichlet policy, log-return reward, cash allowed; LSTM\_2: LSTM encoder, hierarchical policy, log-return reward, cash allowed; LSTM\_NC\_1: LSTM encoder, flat Dirichlet policy, benchmark-relative reward, fully invested; LSTM\_NC\_2: LSTM encoder, flat Dirichlet policy, benchmark-relative reward, fully invested; TRANSFORMERS: Transformer encoder, flat Dirichlet policy, log-return reward, cash allowed. AR denotes the total cumulative return over the full evaluation period; ARC denotes the annualized compounded return and is the primary return criterion used throughout this study. IR2 is the primary evaluation metric for risk-adjusted performance.}}

\linespread{1.5}\selectfont

\hypertarget{er_statistical_significance}{%
\subsection{Statistical Significance}\label{er_statistical_significance}}

While several reinforcement learning strategies exhibit competitive performance in terms of risk-adjusted metrics, formal statistical inference is required to determine whether these differences are statistically meaningful. A key challenge in this context is that financial return series typically violate the assumptions underlying standard parametric tests. In particular, daily returns are known to exhibit autocorrelation, heteroskedasticity, and non-normality (e.g., volatility clustering and heavy tails). As a result, conventional paired sample t-tests may lead to biased inference due to underestimated standard errors.

To address these issues, statistical significance is evaluated using two complementary approaches. First, we employ Newey--West heteroskedasticity and autocorrelation consistent (HAC) estimators (\citet{NEWEY_WEST}) to obtain robust standard errors for the mean return differences between each strategy and the benchmark. Second, we apply a stationary block bootstrap procedure (\citet{POLITIS}) to construct empirical distributions of performance measures and assess significance without relying on parametric assumptions. 

In addition to mean return differences, statistical significance is also evaluated for risk-adjusted performance metrics, namely the Sharpe ratio and IR2, using bootstrap-based inference.

The null hypothesis in this setting is defined as:

\begin{equation}
\begin{cases}
H_0: \mu_{d} = \mu_{\text{strategy}} - \mu_{\text{benchmark}} = 0 \\
H_1: \mu_{d} > 0
\end{cases}
\end{equation}

\noindent where $\mu_{\text{strategy}}$ and $\mu_{\text{benchmark}}$ denote the expected returns of the evaluated strategy and the Buy \& Hold benchmark, respectively. The significance level is set at 10\%. In all tests, the Buy \& Hold strategy serves as the primary benchmark, consistent with the evaluation framework described in Section~\ref{sec:benchmarks}.

The results of the HAC-adjusted tests and stationary bootstrap inference are reported in Table~\ref{tab:hac_bootstrap_all}. Across all considered markets, none of the reinforcement learning strategies achieve statistical significance in terms of mean return differences relative to the benchmark. This holds consistently for both HAC-adjusted p-values and bootstrap based inference for Sharpe and IR2 differences. These findings indicate that, despite observable differences in performance metrics, the null hypothesis of equal expected returns cannot be rejected. Consequently, the apparent outperformance of certain strategies should be interpreted with caution, as it may reflect random variation rather than systematic excess returns.

\begin{table}[H]
\centering
\small
\setlength{\tabcolsep}{3pt}
\renewcommand{\arraystretch}{1.1}

\caption{HAC-adjusted and bootstrap-based statistical significance results}
\label{tab:hac_bootstrap_all}

\begin{tabular}{lcccccc}

\toprule
\textbf{Model} &
\textbf{Mean Diff.} &
\textbf{HAC p-val} &
\textbf{$\Delta$ Sharpe} &
\textbf{p-val} &
\textbf{$\Delta$ IR2} &
\textbf{p-val} \\
\midrule

\multicolumn{7}{l}{\textbf{Panel A: NASDAQ-100}} \\
\midrule
LSTM-1      & -0.0000 & 0.6654 & -0.0972 & 0.8322 & -0.0443 & 0.7872 \\
LSTM-2      & -0.0001 & 0.9489 & -0.0838 & 0.7762 & -0.0326 & 0.7502 \\
LSTM-NC-1   &  0.0001 & 0.3129 & -0.0915 & 0.7832 & -0.0986 & 0.7313 \\
LSTM-NC-2   &  0.0000 & 0.4914 & -0.0917 & 0.8062 & -0.0417 & 0.7672 \\
Transformer & -0.0000 & 0.6968 & -0.0975 & 0.8282 & -0.0470 & 0.7882 \\

\addlinespace

\multicolumn{7}{l}{\textbf{Panel B: Nikkei 225}} \\
\midrule
LSTM-1      &  0.0000 & 0.3381 & 0.0566 & 0.3556 & 0.0060 & 0.3636 \\
LSTM-2      & -0.0001 & 0.6949 & 0.0131 & 0.5005 & -0.0288 & 0.4725 \\
LSTM-NC-1   & -0.0000 & 0.5006 & -0.0705 & 0.7403 & -0.0828 & 0.7612 \\
LSTM-NC-2   & -0.0000 & 0.5025 & -0.0624 & 0.7213 & -0.0759 & 0.7273 \\
Transformer & -0.0000 & 0.6034 & -0.0427 & 0.6913 & -0.0488 & 0.6903 \\

\addlinespace

\multicolumn{7}{l}{\textbf{Panel C: EURO STOXX 50}} \\
\midrule
LSTM-1      &  0.0000 & 0.4992 & 0.1144 & 0.1538 & 0.0597 & 0.1349 \\
LSTM-2      & -0.0000 & 0.5840 & 0.1645 & \textbf{0.0719} & 0.0886 & \textbf{0.0619} \\
LSTM-NC-1   & -0.0000 & 0.6263 & 0.0335 & 0.4086 & 0.0205 & 0.3786 \\
LSTM-NC-2   &  0.0000 & 0.4665 & 0.1045 & 0.1928 & 0.0540 & 0.1608 \\
Transformer &  0.0000 & 0.4758 & 0.1209 & 0.1399 & 0.0648 & 0.1229 \\

\bottomrule

\end{tabular}
\end{table}

\vspace{-0.4cm}
\linespread{0.65}\selectfont
{\noindent\scriptsize \textbf{Note}: \textit{HAC p-values are based on Newey--West standard errors for mean return differences. Bootstrap p-values are based on the stationary bootstrap for Sharpe and IR2 differences. Statistical significance at the 10\% level is indicated in bold.}}

\linespread{1.5}\selectfont

To further investigate the presence of abnormal performance, we estimate a regression model of strategy returns on benchmark returns, following a standard approach in empirical finance:

\begin{equation}
R^{\text{strategy}}_t = \alpha + \beta \, R^{\text{benchmark}}_t + \epsilon_t
\end{equation}

\noindent where the intercept $\alpha$ represents abnormal returns relative to the benchmark. To ensure robustness, Newey--West HAC-adjusted standard errors are used to account for autocorrelation and heteroskedasticity in the residuals. Statistical significance is evaluated based on the following hypotheses:

\begin{equation}
\begin{cases}
H_0: \alpha = 0 \\
H_1: \alpha > 0
\end{cases}
\end{equation}

The regression results are presented in Table~\ref{tab:alpha_regression_all}. For the NASDAQ-100 and Nikkei 225 indices, none of the strategies exhibit statistically significant abnormal returns after applying HAC corrections. In contrast, for the EURO STOXX 50 index, several strategies particularly LSTM-based models and the Transformer model achieve positive and statistically significant intercepts at the 10\% level. This suggests the presence of abnormal returns in this market, even after accounting for time-series dependencies in the data.

\begin{table}[H]
\centering
\small
\setlength{\tabcolsep}{3pt}
\renewcommand{\arraystretch}{1.1}

\caption{Regression-based statistical significance results (HAC-adjusted)}
\label{tab:alpha_regression_all}

\begin{tabular}{lccccc}

\toprule
\textbf{Model} & 
\textbf{$\alpha$} & 
\textbf{SE($\alpha$)} & 
\textbf{$t_\alpha$} & 
\textbf{$p_\alpha$} \\
\midrule

\multicolumn{5}{l}{\textbf{Panel A: NASDAQ-100}} \\
\midrule
LSTM-1      & -0.0000 & 0.0001 & -0.06 & 0.5220 \\
LSTM-2      & -0.0000 & 0.0001 & -0.01 & 0.5053 \\
LSTM-NC-1   &  0.0000 & 0.0001 & 0.29 & 0.3860 \\
LSTM-NC-2   &  0.0000 & 0.0001 & 0.04 & 0.4836 \\
Transformer & -0.0000 & 0.0001 & -0.06 & 0.5246 \\

\addlinespace

\multicolumn{5}{l}{\textbf{Panel B: Nikkei 225}} \\
\midrule
LSTM-1      & 0.0001 & 0.0001 & 1.08 & 0.1399 \\
LSTM-2      & 0.0001 & 0.0001 & 0.70 & 0.2427 \\
LSTM-NC-1   & 0.0000 & 0.0001 & 0.11 & 0.4564 \\
LSTM-NC-2   & 0.0000 & 0.0001 & 0.12 & 0.4506 \\
Transformer & 0.0000 & 0.0001 & 0.28 & 0.3899 \\

\addlinespace

\multicolumn{5}{l}{\textbf{Panel C: EURO STOXX 50}} \\
\midrule
LSTM-1      & 0.0002 & 0.0001 & 1.83 & \textbf{0.0333} \\
LSTM-2      & 0.0002 & 0.0001 & 2.26 & \textbf{0.0120} \\
LSTM-NC-1   & 0.0001 & 0.0001 & 1.15 & 0.1243 \\
LSTM-NC-2   & 0.0002 & 0.0001 & 1.73 & \textbf{0.0417} \\
Transformer & 0.0002 & 0.0001 & 1.89 & \textbf{0.0291} \\

\bottomrule

\end{tabular}
\end{table}

\vspace{-0.4cm}
\linespread{0.65}\selectfont
{\noindent\scriptsize \textbf{Note}: \textit{The table reports abnormal returns ($\alpha$) estimated from the regression model with Newey--West HAC-adjusted standard errors. Statistical significance at the 10\% level is indicated in bold.}}

\linespread{1.5}\selectfont

Taken together, the results provide only partial support for the main hypothesis. While no statistically significant outperformance is observed in terms of raw return differences across any market, there is evidence of abnormal performance in the EURO STOXX 50 based on regression analysis. Overall, these findings should be interpreted as exploratory, indicating that reinforcement learning strategies may capture certain market-specific inefficiencies, but do not consistently outperform benchmark strategies in a statistically robust manner across different markets.

\hypertarget{er_Summary}{%
\subsection{Summary}\label{er_Summary}}

Based on the results presented in Tables~\ref{tab:perf_metric_qqq}, 
\ref{tab:perf_metric_nky}, and \ref{tab:perf_metric_eustxx}, the performance of the proposed reinforcement learning strategies differs substantially across the considered equity indices. Using the Modified Information Ratio (IR2) as the primary evaluation criterion, the best-performing reinforcement learning strategy for the NASDAQ-100 is LSTM\_2 with cash allowed (IR2 = 0.46), however the Equal-Weight Monthly portfolio (IR2 = 0.49) and the Buy \& Hold benchmark (IR2 = 0.52) remain superior in this market. For the Nikkei 225, LSTM\_1 with cash allowed achieves the strongest RL performance (IR2 = 0.15), but is outperformed by the Equal-Weight Monthly portfolio (IR2 = 0.19) and the Markowitz Minimum Variance portfolio (IR1 = 0.61). In contrast, for the EURO STOXX 50, all reinforcement learning strategies outperform the Buy \& Hold benchmark on IR2. The strongest result is obtained by LSTM\_2 with cash allowed (IR2 = 0.15), followed by the Equal-Weight Monthly portfolio (IR2 = 0.13) and the Transformer model (IR2 = 0.12).

It is also important to note that the fully invested configurations generally produce higher absolute returns, but at the cost of larger drawdowns and lower risk-adjusted performance. This pattern is particularly visible for the NASDAQ-100, where LSTM\_NC\_1 attains the highest absolute return of 1666.4\%, yet its IR2 falls to 0.39, below that of the cash-allowed LSTM\_2 model. These results suggest that allowing the agent to hold cash improves the balance between return generation and drawdown control.

Furthermore, the HAC-adjusted tests and stationary bootstrap inference reported in Table~\ref{tab:hac_bootstrap_all} indicate that none of the evaluated reinforcement learning strategies achieve statistically significant outperformance relative to the Buy \& Hold benchmark in terms of mean return differences or risk-adjusted performance measures. However, the regression analysis reported in Table~\ref{tab:alpha_regression_all} reveals that several strategies exhibit statistically significant abnormal returns in the EURO STOXX 50, while no such significance is observed for the NASDAQ-100 or Nikkei 225. These findings suggest that while reinforcement learning strategies do not exhibit statistically robust outperformance in terms of raw returns, there is evidence of market-specific abnormal performance in the EURO STOXX 50.

\hypertarget{regimeness}{%
\section{Regime Analysis}\label{regimeness}}

To assess whether aggregate results mask heterogeneous performance across market environments, the evaluation period is decomposed into three macroeconomic regimes: the post-GFC recovery (2009--2013), the secular bull market (2014--2019), and the COVID-19 shock combined with the rate hike cycle (2020--2026). Tables~\ref{tab:subperiod_qqq}, \ref{tab:subperiod_nky}, and \ref{tab:subperiod_eustx} report the corresponding results.

For the \textbf{NASDAQ-100}, the Markowitz Minimum Variance portfolio dominates the post-GFC recovery period across all metrics, followed by RL strategies which outperform Buy \& Hold during this regime. However, in both subsequent periods passive benchmarks dominate, with Buy \& Hold and Equal-Weight Monthly leading on IR2. The persistent upward trend of the NASDAQ-100 systematically favors passive allocation and limits the value of active management across both classical and RL-based approaches.

For the \textbf{Nikkei 225}, the Equal-Weight Monthly and Markowitz portfolios dominate during the post-GFC recovery, while Buy \& Hold delivers the weakest performance across all strategies in this period. The pattern reverses in 2014--2019, where LSTM\_2 achieves the strongest performance across all strategies including the classical benchmarks. In 2020--2026, Markowitz Minimum Variance leads, followed by LSTM-based RL strategies, while Buy \& Hold again delivers the weakest result. This consistent underperformance of passive index exposure on the Japanese market highlights the structural challenges of the Nikkei 225 as a buy-and-hold investment over this period.

For the \textbf{EURO STOXX 50}, the Momentum Top-20 and Markowitz portfolios lead in the first two regimes respectively, with all RL strategies consistently outperforming Buy \& Hold on IR2 across all three periods. In 2020--2026, LSTM\_2 achieves the highest IR2 among all strategies, while Momentum Top-20 collapses to near-zero performance, suggesting that momentum-based selection is particularly fragile in the post-COVID European equity environment.

Overall, the sub-period decomposition reveals that no single strategy type dominates consistently across all markets and regimes. RL strategies add meaningful value relative to Buy \& Hold in several settings, particularly on the Nikkei 225 and EURO STOXX 50, but do not consistently outperform classical benchmarks such as Equal-Weight Monthly and Markowitz Minimum Variance. These findings underscore the importance of evaluating active strategies against a broad set of benchmarks and highlight the regime-dependence of RL portfolio allocation performance.

% ── NASDAQ-100 ──────────────────────────────────────────────────────────────
\begin{table}[H]
\centering
\scriptsize
\setlength{\tabcolsep}{3pt}
\renewcommand{\arraystretch}{0.7}
\caption{Sub-Period Performance Metrics -- NASDAQ-100}
\label{tab:subperiod_qqq}
\resizebox{\textwidth}{!}{%
\begin{tabular}{llccccccccc}
\toprule
\textbf{Period} & \textbf{Strategy} & \textbf{AR (\%)} & \textbf{ARC (\%)} & \textbf{ASD (\%)} & \textbf{MD (\%)} & \textbf{MLD} & \textbf{IR1} & \textbf{IR2} & \textbf{IR3} & \textbf{Sharpe} \\
\midrule

\multicolumn{11}{l}{\textbf{2009--2013: Post-GFC Recovery}} \\
\midrule
& Buy \& Hold          & 178.04 & 23.98 & 18.33 & 16.10 & 0.480 & 1.3082 & 1.9485  & 9736.71   & 1.2735 \\
& Equal-Weight Monthly & 204.06 & 26.41 & 20.04 & 21.29 & 0.766 & 1.3179 & 1.6348  & 5637.73   & 1.2761 \\
& Markowitz Min Var    & \textbf{245.84} & \textbf{29.96} & 18.34 & \textbf{17.03} & \textbf{0.310} & \textbf{1.6336} & \textbf{2.8739}  & \textbf{27826.48}  & \textbf{1.5276} \\
& Momentum Top-20      & 184.26 & 24.54 & 21.41 & 22.40 & 0.972 & 1.1462 & 1.2557  & 3168.66   & 1.1409 \\
& LSTM\_1              & 222.91 & 27.93 & 20.68 & 17.51 & 0.909 & 1.3506 & 2.1543  & 6622.14   & 1.3035 \\
& LSTM\_2              & 175.93 & 23.81 & \textbf{17.40} & 16.24 & 0.742 & 1.3684 & 2.0062  & 6438.49   & 1.3221 \\
& TRANSFORMERS         & 212.99 & 27.09 & 20.16 & 17.31 & 0.913 & 1.3437 & 2.1030  & 6244.44   & 1.2993 \\
\addlinespace

\multicolumn{11}{l}{\textbf{2014--2019: Secular Bull Market}} \\
\midrule
& Buy \& Hold          & \textbf{164.14} & \textbf{17.62} & 16.63 & 22.80 & 0.619 & 1.0595 & 0.8188  & 2332.98  & \textbf{1.0625} \\
& Equal-Weight Monthly & 154.54 & 16.89 & 16.30 & 20.51 & \textbf{0.552} & 1.0362 & 0.8533  & 2613.22  & 1.0427 \\
& Markowitz Min Var    & 103.42 & 12.60 & \textbf{16.31} & 24.46 & \textbf{0.552} & 0.7725  & 0.3980  & 909.05   & 0.8139 \\
& Momentum Top-20      & 57.94  & 7.90  & 17.23 & 27.77 & 1.433 & 0.4585  & 0.1304  & 71.92    & 0.5309 \\
& LSTM\_1              & 122.42 & 14.26 & 17.20 & 23.51 & 1.214 & 0.8291  & 0.5029  & 590.75   & 0.8654 \\
& LSTM\_2              & 115.02 & 13.63 & 14.74 & \textbf{20.37} & 0.579 & 0.9247  & \textbf{0.6187}  & \textbf{1455.26}  & 0.9446 \\
& TRANSFORMERS         & 122.58 & 14.27 & 17.26 & 23.57 & 1.214 & 0.8268  & 0.5006  & 588.84   & 0.8635 \\
\addlinespace

\multicolumn{11}{l}{\textbf{2020--2026: COVID + Rate Hike Cycle}} \\
\midrule
& Buy \& Hold          & \textbf{118.58} & \textbf{15.99} & 25.50 & 36.43 & 1.964 & \textbf{0.6271}  & \textbf{0.2752}  & \textbf{224.02}   & \textbf{0.7164} \\
& Equal-Weight Monthly & 81.62  & 12.00 & 24.73 & 33.68 & 2.075 & 0.4852  & 0.1729  & 99.98    & 0.5870 \\
& Markowitz Min Var    & 80.84  & 11.99 & \textbf{22.66} & \textbf{31.00} & \textbf{0.722} & 0.5291  & 0.2047  & 340.06   & 0.6173 \\
& Momentum Top-20      & 34.53  & 5.67  & 25.96 & 40.74 & 2.254 & 0.2184  & 0.0304   & 7.64     & 0.3495 \\
& LSTM\_1              & 85.86  & 12.39 & 26.46 & 32.34 & 2.258 & 0.4683  & 0.1794  & 98.50    & 0.5825 \\
& LSTM\_2              & 71.89  & 10.73 & 23.25 & 28.77 & 2.262 & 0.4615  & 0.1721  & 81.74    & 0.5640 \\
& TRANSFORMERS         & 87.60  & 12.60 & 26.29 & 32.27 & 2.262 & 0.4793  & 0.1871  & 104.21   & 0.5914 \\

\bottomrule
\end{tabular}}
\end{table}

\vspace{-0.4cm}
\noindent\linespread{0.65}\selectfont
{\scriptsize \textbf{Note}: \textit{Sub-period performance metrics for NASDAQ-100. Buy \& Hold is the Invesco QQQ Trust ETF (QQQ).}}

\linespread{1.5}\selectfont

% ── NIKKEI 225 ──────────────────────────────────────────────────────────────
\begin{table}[H]
\centering
\scriptsize
\setlength{\tabcolsep}{3pt}
\renewcommand{\arraystretch}{0.7}
\caption{Sub-Period Performance Metrics -- Nikkei 225}
\label{tab:subperiod_nky}
\resizebox{\textwidth}{!}{%
\begin{tabular}{llccccccccc}
\toprule
\textbf{Period} & \textbf{Strategy} & \textbf{AR (\%)} & \textbf{ARC (\%)} & \textbf{ASD (\%)} & \textbf{MD (\%)} & \textbf{MLD} & \textbf{IR1} & \textbf{IR2} & \textbf{IR3} & \textbf{Sharpe} \\
\midrule

\multicolumn{11}{l}{\textbf{2009--2013: Post-GFC Recovery}} \\
\midrule
& Buy \& Hold          & 52.43  & 9.35  & 19.21 & 25.29 & 2.131 & 0.4867  & 0.1799  & 78.95     & 0.5724 \\
& Equal-Weight Monthly & \textbf{113.82} & \textbf{17.68} & 23.05 & 28.10 & \textbf{0.520} & \textbf{0.7670}  & \textbf{0.4826}  & \textbf{1642.69}  & \textbf{0.8319} \\
& Markowitz Min Var    & 92.67  & 15.08 & 23.41 & \textbf{27.84} & 0.607 & 0.6442  & 0.3489  & 866.20    & 0.7271 \\
& Momentum Top-20      & 51.88  & 9.39  & 24.12 & 44.75 & 3.560 & 0.3893  & 0.0817   & 21.58     & 0.4973 \\
& LSTM\_1              & 64.43  & 11.32 & 22.39 & 35.80 & 2.925 & 0.5056  & 0.1599  & 61.88     & 0.5942 \\
& LSTM\_2              & 28.95  & 5.58  & \textbf{20.59} & 38.27 & 3.583 & 0.2710  & 0.0395   & 6.16      & 0.3712 \\
& TRANSFORMERS         & 40.32  & 7.56  & 23.28 & 38.82 & 2.925 & 0.3247  & 0.0632   & 16.34     & 0.4327 \\
\addlinespace

\multicolumn{11}{l}{\textbf{2014--2019: Secular Bull Market}} \\
\midrule
& Buy \& Hold          & 16.78  & 2.83  & 15.02 & 25.69 & 1.909 & 0.1884  & 0.0208   & 3.08      & 0.2509 \\
& Equal-Weight Monthly & 49.00  & 7.16  & 19.64 & 27.91 & 1.921 & 0.3646  & 0.0935   & 34.84     & 0.4437 \\
& Markowitz Min Var    & 8.41   & 1.50  & 19.58 & 38.84 & 1.952 & 0.0766   & 0.0030   & 0.23      & 0.1684 \\
& Momentum Top-20      & 20.80  & 3.39  & 19.43 & 35.20 & 2.175 & 0.1745  & 0.0168   & 2.63      & 0.2628 \\
& LSTM\_1              & 62.98  & 8.71  & 19.77 & 24.39 & 1.214 & 0.4406  & 0.1573  & 113.03    & 0.5192 \\
& LSTM\_2              & \textbf{63.55} & \textbf{8.76} & \textbf{16.62} & \textbf{17.41} & \textbf{1.214} & \textbf{0.5271}  & \textbf{0.2652}  & \textbf{191.49}   & \textbf{0.5864} \\
& TRANSFORMERS         & 48.28  & 6.98  & 19.32 & 30.61 & 1.901 & 0.3613  & 0.0824   & 30.29     & 0.4435 \\
\addlinespace

\multicolumn{11}{l}{\textbf{2020--2026: COVID + Rate Hike Cycle}} \\
\midrule
& Buy \& Hold          & 13.77  & 2.998 & 19.03 & 35.92 & 3.417 & 0.1575  & 0.0131   & 1.15      & 0.2298 \\
& Equal-Weight Monthly & 79.70  & 12.48 & 20.20 & 33.59 & 0.675 & 0.6178  & 0.2295  & 424.52    & 0.6759 \\
& Markowitz Min Var    & \textbf{113.38} & \textbf{16.43} & 21.28 & \textbf{28.75} & \textbf{0.683} & \textbf{0.7721}  & \textbf{0.4412}  & \textbf{1063.07}  & \textbf{0.8148} \\
& Momentum Top-20      & 51.16  & 8.74  & 19.80 & 30.45 & 0.687 & 0.4414  & 0.1267  & 161.54    & 0.5125 \\
& LSTM\_1              & 93.36  & 14.12 & \textbf{20.94} & 31.38 & 0.691 & 0.6743  & 0.3034  & 620.58    & 0.7292 \\
& LSTM\_2              & 82.93  & 12.84 & 18.15 & 27.88 & 0.571 & 0.7074  & 0.3258  & 731.43    & 0.7502 \\
& TRANSFORMERS         & 84.45  & 13.06 & 22.12 & 31.32 & 0.691 & 0.5904  & 0.2462  & 465.71    & 0.6597 \\

\bottomrule
\end{tabular}}
\end{table}
\vspace{-0.4cm}
\noindent\linespread{0.65}\selectfont
{\scriptsize \textbf{Note}: \textit{Sub-period performance metrics for Nikkei 225. Buy \& Hold is the iShares MSCI Japan ETF (EWJ).}}

\linespread{1.5}\selectfont

% ── EURO STOXX 50 ───────────────────────────────────────────────────────────
\begin{table}[H]
\centering
\small
\setlength{\tabcolsep}{3pt}
\renewcommand{\arraystretch}{0.7}
\caption{Sub-Period Performance Metrics -- EURO STOXX 50}
\label{tab:subperiod_eustx}
\resizebox{\textwidth}{!}{%
\begin{tabular}{llccccccccc}
\toprule
\textbf{Period} & \textbf{Strategy} & \textbf{AR (\%)} & \textbf{ARC (\%)} & \textbf{ASD (\%)} & \textbf{MD (\%)} & \textbf{MLD} & \textbf{IR1} & \textbf{IR2} & \textbf{IR3} & \textbf{Sharpe} \\
\midrule

\multicolumn{11}{l}{\textbf{2009--2013: Post-GFC Recovery}} \\
\midrule
& Buy \& Hold          & 85.86  & 13.65 & 30.42 & 37.80 & 2.488 & 0.4487  & 0.1620  & 88.89     & 0.5750 \\
& Equal-Weight Monthly & \textbf{101.17} & \textbf{15.50} & 24.28 & 34.71 & 2.262 & 0.6384  & 0.2851  & 195.16    & 0.7184 \\
& Markowitz Min Var    & 77.71  & 12.65 & \textbf{20.24} & \textbf{24.28} & \textbf{1.365} & 0.6250  & 0.3256  & 301.90    & 0.6914 \\
& Momentum Top-20      & 114.98 & 17.13 & 25.31 & 29.02 & 1.591 & \textbf{0.6768}  & \textbf{0.3995}  & \textbf{430.22}   & \textbf{0.7541} \\
& LSTM\_1              & 67.48  & 11.23 & 21.56 & 27.45 & 1.944 & 0.5209  & 0.2131  & 123.06    & 0.6037 \\
& LSTM\_2              & 69.61  & 11.52 & 18.61 & 23.74 & 1.917 & 0.6190  & 0.3004  & 180.71    & 0.6815 \\
& TRANSFORMERS         & 71.77  & 11.81 & 21.68 & 27.40 & 1.917 & 0.5447  & 0.2348  & 144.63    & 0.6257 \\
\addlinespace

\multicolumn{11}{l}{\textbf{2014--2019: Secular Bull Market}} \\
\midrule
& Buy \& Hold          & 11.22  & 1.68  & 17.13 & 30.84 & 3.298 & 0.0981   & 0.0053   & 0.27      & 0.1887 \\
& Equal-Weight Monthly & 51.66  & 7.12  & 17.12 & 25.11 & 1.972 & 0.4159  & 0.1179  & 42.59     & 0.4866 \\
& Markowitz Min Var    & \textbf{77.28} & \textbf{9.92} & \textbf{16.48} & 23.59 & 2.083 & \textbf{0.6019}  & \textbf{0.2531}  & \textbf{120.50}   & \textbf{0.6569} \\
& Momentum Top-20      & 29.84  & 4.42  & 17.67 & 31.38 & 3.286 & 0.2501  & 0.0352   & 4.75      & 0.3319 \\
& LSTM\_1              & 53.47  & 7.32  & 16.22 & 22.50 & 2.135 & 0.4513  & 0.1468  & 50.26     & 0.5163 \\
& LSTM\_2              & 52.15  & 7.16  & 13.82 & \textbf{18.38} & \textbf{1.897} & 0.5181  & 0.2018  & 76.31     & 0.5696 \\
& TRANSFORMERS         & 52.76  & 7.23  & 16.34 & 22.95 & 2.135 & 0.4425  & 0.1394  & 47.29     & 0.5090 \\
\addlinespace

\multicolumn{11}{l}{\textbf{2020--2026: COVID + Rate Hike Cycle}} \\
\midrule
& Buy \& Hold          & 47.37  & 7.66  & 23.91 & 38.95 & 2.175 & 0.3204  & 0.0630   & 22.18     & 0.4256 \\
& Equal-Weight Monthly & 45.36  & 7.43  & 20.68 & 39.40 & 1.163 & 0.3593  & 0.0678   & 43.26     & 0.4443 \\
& Markowitz Min Var    & 39.08  & 6.60  & \textbf{18.59} & 36.24 & \textbf{0.726} & 0.3550  & 0.0647   & 58.82     & 0.4286 \\
& Momentum Top-20      & 0.61   & 0.24  & 20.38 & 35.27 & 1.663 & 0.0118   & 0.0001   & 0.00      & 0.1080 \\
& LSTM\_1              & 46.08  & 7.53  & 18.56 & 33.52 & 1.048 & 0.4057  & 0.0911   & 65.40     & 0.4775 \\
& LSTM\_2              & 41.60  & 6.87  & 15.66 & \textbf{29.94} & 1.060 & 0.4387  & \textbf{0.1007}  & \textbf{65.37}    & \textbf{0.4967} \\
& TRANSFORMERS         & \textbf{46.60} & \textbf{7.60} & 18.55 & 33.16 & 1.048 & \textbf{0.4097}  & 0.0939   & 68.05     & 0.4813 \\

\bottomrule
\end{tabular}}
\end{table}
\vspace{-0.4cm}
\noindent\linespread{0.65}\selectfont
{\scriptsize \textbf{Note}: \textit{Sub-period performance metrics for EURO STOXX 50 strategies across three macroeconomic regimes. Buy \& Hold is the SPDR Euro Stoxx 50 ETF (FEZ). Strategy labels refer to the following configurations --- LSTM\_1: LSTM encoder, flat Dirichlet policy, log-return reward, cash allowed; LSTM\_2: LSTM encoder, hierarchical policy, log-return reward, cash allowed; TRANSFORMERS: Transformer encoder, flat Dirichlet policy, log-return reward, cash allowed. AR denotes total cumulative return over the sub-period; ARC denotes annualized compounded return. IR2 is the primary risk-adjusted evaluation metric. Bold denotes the best value per metric within each period.}}

\linespread{1.5}\selectfont

\newpage

\hypertarget{ensemble}{%
\section{Ensemble Total Fund}\label{ensemble}}

The ensemble framework evaluates whether combining strategy signals across multiple equity markets leads to improved portfolio performance. Cross-asset portfolios are constructed by aggregating out-of-sample daily returns from the NASDAQ-100, Nikkei 225, and EURO STOXX 50 for each reinforcement learning configuration using equal weights. All return series are aligned on a common date index and evaluated over the same out-of-sample period commencing 2009-04-06. The benchmark corresponds to the cross-sectional average of the Buy \& Hold strategies across the three markets.

\begin{figure}[H]
\centering
\includegraphics[width=0.9\linewidth]{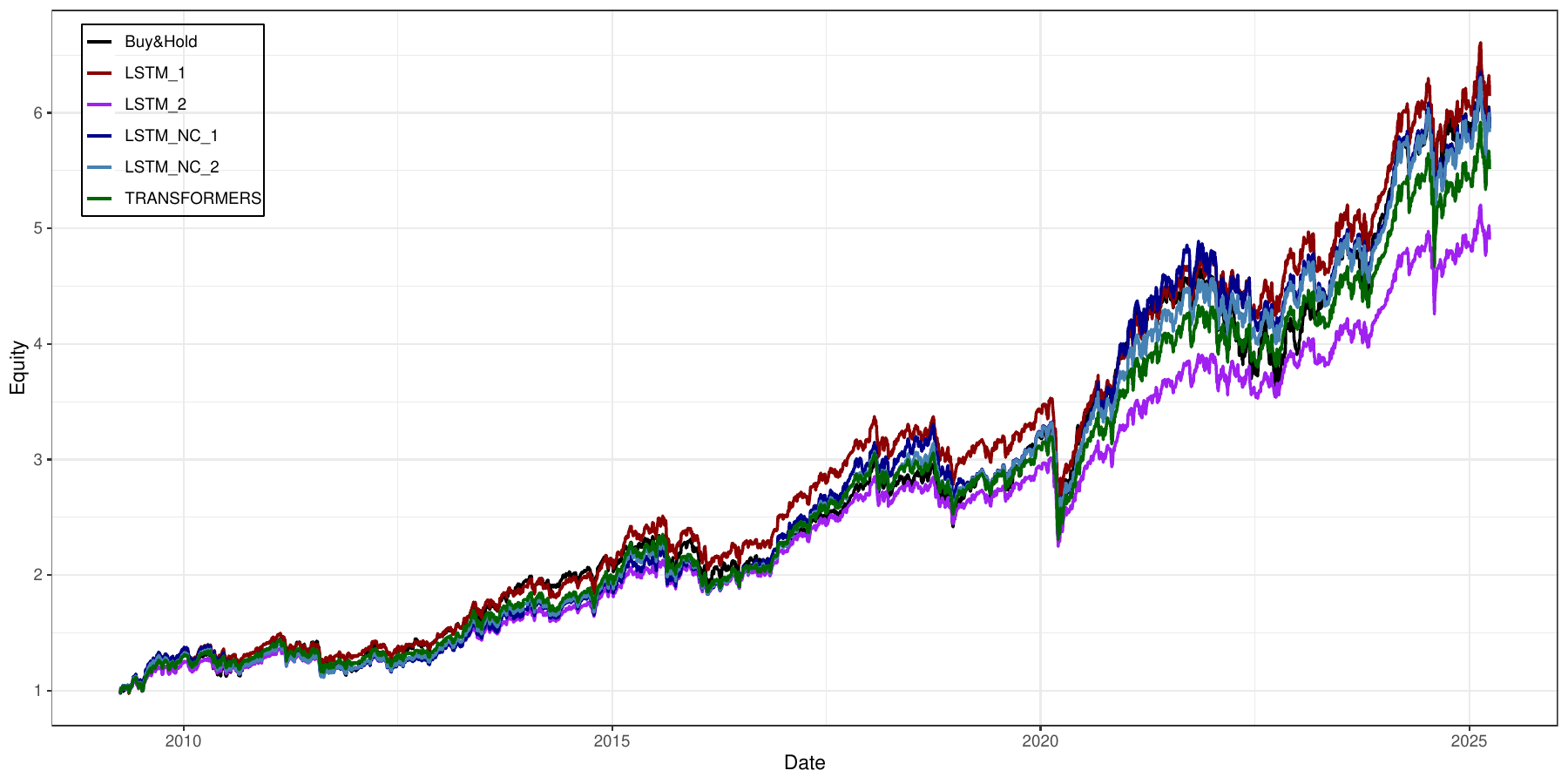}
\caption{Equity curves for cross-asset ensemble portfolios across reinforcement learning strategies.}
\label{fig:ensemble_results}
\end{figure}

Figure~\ref{fig:ensemble_results} presents the corresponding equity curves, expressed as the growth of one unit of capital. All reinforcement learning strategies exhibit broadly similar long-term trajectories, with LSTM\_1 delivering the strongest absolute growth. Table~\ref{tab:perf_metric_ensemble} reports the corresponding performance metrics. Based on the IR2 criterion, LSTM\_1 achieves the highest risk-adjusted performance (IR2 = 0.41), outperforming both the benchmark (IR2 = 0.34) and all alternative configurations. LSTM\_2 achieves the lowest volatility and maximum drawdown among all strategies, reflecting superior risk control, though its lower absolute return results in a slightly lower IR2. The Transformer model delivers competitive risk-adjusted performance with low volatility, while the NC configurations closely track the benchmark.

\vspace{-0.4cm}

\noindent\linespread{0.58}\selectfont 
{\scriptsize \textbf{Note}: \textit{The figure presents the out-of-sample equity curves of cross-asset portfolios constructed as equal-weight combinations of strategy-specific returns across the NASDAQ-100, Nikkei 225, and EURO STOXX 50. Strategy labels refer to the following configurations --- LSTM\_1: LSTM encoder, flat Dirichlet policy, log-return reward, cash allowed; LSTM\_2: LSTM encoder, hierarchical policy, log-return reward, cash allowed; LSTM\_NC\_1: LSTM encoder, flat Dirichlet policy, benchmark-relative reward, fully invested; LSTM\_NC\_2: LSTM encoder, flat Dirichlet policy, benchmark-relative reward, fully invested; TRANSFORMERS: Transformer encoder, flat Dirichlet policy, log-return reward, cash allowed. The benchmark represents the cross-sectional average of the Buy \& Hold strategies across the three markets. All series are aligned on a common date index and expressed as the growth of one unit of capital.}}

\linespread{1.5}\selectfont

\begin{table}[H]
\centering
\small
\setlength{\tabcolsep}{3pt}
\renewcommand{\arraystretch}{1.1}
\caption{Performance Metrics for Ensemble Strategies}
\label{tab:perf_metric_ensemble}
\resizebox{\textwidth}{!}{%
\begin{tabular}{lcccccccccc}
\toprule
\textbf{Strategy} & \textbf{AR (\%)} & \textbf{ARC (\%)} & \textbf{ASD (\%)} & \textbf{MD (\%)} & \textbf{MLD} & \textbf{IR1} & \textbf{IR2} & \textbf{IR3} & \textbf{Sharpe} \\
\midrule
Common Benchmark  & 486.79 & 12.68 & 16.53 & 28.63 & \textbf{1.70} & 0.7671 & 0.3397 & 2.53 & 0.8055 \\
LSTM\_1           & \textbf{514.60} & \textbf{13.03} & 14.72 & 28.16 & 1.79 & 0.8852 & \textbf{0.4096} & \textbf{2.98} & \textbf{0.9062} \\
LSTM\_2           & 390.49 & 11.33 & \textbf{12.75} & \textbf{25.46} & 1.75 & \textbf{0.8886} & 0.3954 & 2.55 & 0.9054 \\
LSTM\_NC\_1       & 487.84 & 12.69 & 16.14 & 28.76 & 1.98 & 0.7862 & 0.3469 & 2.23 & 0.8218 \\
LSTM\_NC\_2       & 483.44 & 12.64 & 15.76 & 28.77 & 1.79 & 0.8020 & 0.3524 & 2.49 & 0.8342 \\
TRANSFORMERS      & 451.32 & 12.21 & 14.77 & 27.92 & 1.77 & 0.8267 & 0.3615 & 2.49 & 0.8541 \\
\bottomrule
\end{tabular}}
\end{table}
\vspace{-0.4cm}
\noindent\linespread{0.65}\selectfont
{\scriptsize \textbf{Note}: \textit{The table reports out-of-sample performance metrics for cross-asset ensemble portfolios constructed as equal-weight combinations of strategy-specific returns across the NASDAQ-100, Nikkei 225, and EURO STOXX 50. Strategy labels refer to the following configurations --- LSTM\_1: LSTM encoder, flat Dirichlet policy, log-return reward, cash allowed; LSTM\_2: LSTM encoder, hierarchical policy, log-return reward, cash allowed; LSTM\_NC\_1: LSTM encoder, flat Dirichlet policy, benchmark-relative reward, fully invested; LSTM\_NC\_2: LSTM encoder, flat Dirichlet policy, benchmark-relative reward, fully invested; TRANSFORMERS: Transformer encoder, flat Dirichlet policy, log-return reward, cash allowed. The benchmark corresponds to the cross-sectional average of the Buy \& Hold strategies across the three markets. AR denotes total cumulative return over the full evaluation period; ARC denotes annualized compounded return and is the primary return criterion. IR2 is the primary risk-adjusted evaluation metric. All reported results are rounded to four significant figures.}}

\linespread{1.5}\selectfont

To assess statistical significance, we apply the same HAC-adjusted tests and stationary bootstrap inference as described in Section~\ref{er_statistical_significance}. The results indicate that none of the ensemble strategies achieve statistically significant outperformance relative to the benchmark in terms of mean returns or risk-adjusted measures.

Regression-based analysis, however, provides partial evidence of abnormal performance. In particular, the \textbf{LSTM\_1} and \textbf{LSTM\_2} ensemble strategies exhibit positive and statistically significant intercepts at the 10\% level, with HAC-adjusted p-values of \emph{0.0496} and \emph{0.0481}, respectively, while no significance is observed for the remaining strategies.

Overall, while the ensemble framework improves performance metrics in economic terms, the statistical evidence for outperformance remains limited. The results should therefore be interpreted as indicative rather than conclusive. The corresponding statistical results are summarized in Table~\ref{tab:ensemble_stat}.

\begin{table}[H]
\centering
\small
\setlength{\tabcolsep}{3pt}
\renewcommand{\arraystretch}{1.1}

\caption{Statistical significance results for ensemble strategies}
\label{tab:ensemble_stat}

\begin{tabular}{lcccccc}

\toprule
\textbf{Model} &
\textbf{Mean Diff.} &
\textbf{HAC p-val} &
\textbf{$\Delta$ Sharpe} &
\textbf{p-val} &
\textbf{$\Delta$ IR2} &
\textbf{p-val} \\
\midrule

\multicolumn{7}{l}{\textbf{Panel A: HAC and Bootstrap Tests}} \\
\midrule
LSTM\_1      &  0.0000 & 0.4933 & 0.1007 & 0.1898 & 0.0705 & 0.2537 \\
LSTM\_2      & -0.0001 & 0.8320 & 0.0999 & 0.1788 & 0.0553 & 0.2458 \\
LSTM\_NC\_DAILY   & -0.0000 & 0.5098 & 0.0163 & 0.4665 & 0.0069 & 0.4925 \\
LSTM\_NC\_DAILY\_2 & -0.0000 & 0.5341 & 0.0287 & 0.4156 & 0.0119 & 0.4605 \\
TRANSFORMERS & -0.0000 & 0.6531 & 0.0486 & 0.3407 & 0.0217 & 0.4026 \\

\addlinespace

\multicolumn{7}{l}{\textbf{Panel B: Regression (HAC-adjusted)}} \\
\midrule
\textbf{Model} & \textbf{$\alpha$} & \textbf{SE($\alpha$)} & \textbf{$t_\alpha$} & \textbf{$p_\alpha$} & \multicolumn{2}{c}{} \\
\midrule
LSTM\_1      & 0.0001 & 0.0001 & 1.65 & \textbf{0.0496} & \multicolumn{2}{c}{} \\
LSTM\_2      & 0.0001 & 0.0001 & 1.66 & \textbf{0.0481} & \multicolumn{2}{c}{} \\
LSTM\_NC\_DAILY   & 0.0001 & 0.0001 & 0.99 & 0.1605 & \multicolumn{2}{c}{} \\
LSTM\_NC\_DAILY\_2 & 0.0001 & 0.0001 & 1.06 & 0.1451 & \multicolumn{2}{c}{} \\
TRANSFORMERS & 0.0001 & 0.0001 & 1.21 & 0.1125 & \multicolumn{2}{c}{} \\

\bottomrule

\end{tabular}
\end{table}

\vspace{-0.4cm}
\noindent\linespread{0.65}\selectfont
{\scriptsize \textbf{Note}: \textit{Panel A reports Newey--West HAC-adjusted p-values for mean return differences and stationary bootstrap p-values for Sharpe and IR2 differences. Panel B reports regression-based abnormal returns ($\alpha$) with HAC-adjusted standard errors. Statistical significance at the 10\% level is indicated in bold.}}

\linespread{1.5}\selectfont

\hypertarget{discussion}{%
\section{Discussion}\label{discussion}}

The objective of this study was to evaluate whether reinforcement learning based portfolio allocation strategies can improve risk-adjusted performance under realistic market conditions. The proposed framework combines SAC-based agents with different architectural and economic configurations, evaluated through a walk-forward optimization procedure against four benchmarks mentioned in Section~\ref{sec:benchmarks}: Buy \& Hold, Equal-Weight Monthly, Markowitz Minimum Variance, and Momentum Top-20.

The empirical results, reported in Tables~\ref{tab:perf_metric_qqq}, \ref{tab:perf_metric_nky}, and \ref{tab:perf_metric_eustxx}, indicate that reinforcement learning strategy effectiveness varies significantly across markets. The strongest results are observed for the EURO STOXX 50, where all RL strategies outperform the Buy \& Hold benchmark in terms of IR2. In contrast, for the NASDAQ-100 and Nikkei 225, passive and classical benchmarks remain competitive or dominant. Notably, the Equal-Weight Monthly and Markowitz Minimum Variance portfolios prove difficult to outperform across multiple markets and regimes, consistent with the findings of \citet{DE_MIGEUL}.

At the aggregate level, the ensemble results in Table~\ref{tab:perf_metric_ensemble} demonstrate that combining strategies across markets leads to improved stability and superior risk-adjusted performance relative to the benchmark. The LSTM\_1 ensemble achieves the highest IR2 (0.41), outperforming the common benchmark (0.34), highlighting the benefits of cross-market diversification.

The results provide partial support for the central hypothesis. While reinforcement learning strategies achieve meaningful performance in certain settings, particularly in the EURO STOXX 50 and in the ensemble framework, they do not consistently outperform passive or classical benchmarks across all markets. The regime analysis presented in Section~\ref{regimeness} further reveals that RL strategies tend to add the most value during periods of elevated uncertainty and lower trend persistence, while passive benchmarks dominate in strongly trending markets such as the NASDAQ-100.

Statistical inference further limits the strength of these conclusions. As shown in Table~\ref{tab:hac_bootstrap_all}, none of the strategies achieve statistically significant excess returns relative to the Buy \& Hold benchmark based on HAC-adjusted tests and stationary bootstrap inference. However, the regression results in Table~\ref{tab:alpha_regression_all} indicate that LSTM\_1, LSTM\_2, and the Transformer model exhibit statistically significant positive intercepts in the EURO STOXX 50, suggesting the presence of abnormal returns in this specific market.

\noindent\textbf{RQ1. How does the choice of reward formulation (absolute vs.\ benchmark-relative) affect the performance and stability of learned strategies?}
\newline The results indicate that reward formulation has a limited impact on performance. Benchmark-relative models (LSTM\_NC) do not consistently outperform absolute-return models across markets. In the NASDAQ-100, LSTM\_2 achieves a higher IR2 than LSTM\_NC\_2 despite using an absolute reward, and a similar pattern holds in the EURO STOXX 50. The experimental configurations were not designed as ceteris paribus experiments, however, as reward formulation co-varies with portfolio constraints across configurations. Consequently, this finding should be interpreted as exploratory rather than causal.

\noindent\textbf{RQ2. What is the impact of different policy structures (flat vs.\ hierarchical allocation) on portfolio performance and diversification?}
\newline The hierarchical policy structure consistently improves risk-adjusted performance. LSTM\_2, which employs a hierarchical Dirichlet policy separating the equity-cash allocation decision from individual asset selection, achieves lower volatility and lower maximum drawdown than LSTM\_1 across all three markets. This suggests that decomposing the allocation decision into sequential stages improves portfolio stability, though the same confounding caveat noted for RQ1 applies here.

\noindent\textbf{RQ3. How do portfolio constraints (fully invested vs.\ flexible exposure) influence trading behavior and risk-adjusted returns?}
\newline Portfolio constraints substantially affect risk-adjusted performance. Cash-allowed configurations consistently achieve lower drawdowns and higher IR2 than their fully invested counterparts, at the cost of somewhat lower absolute returns. This pattern is most visible in the NASDAQ-100, where LSTM\_NC\_1 attains the highest absolute return among all RL strategies yet its IR2 falls well below that of the cash-allowed LSTM\_2. These results suggest that the ability to hold cash provides meaningful downside protection, particularly during volatile market conditions.

\noindent\textbf{RQ4. Does the inclusion of different temporal encoders (LSTM vs.\ Transformer) improve the model's ability to capture market dynamics and enhance performance?}
\newline LSTM-based models outperform the Transformer in risk-adjusted terms across all three markets. While the Transformer delivers competitive absolute returns and achieves the highest AR on the EURO STOXX 50, it does not consistently improve stability relative to LSTM-based configurations.

\noindent\textbf{RQ5. Does combining strategy-specific signals across multiple markets through an ensemble approach improve risk-adjusted performance compared to single-market strategies and the benchmark?}
\newline The ensemble approach consistently improves risk-adjusted performance relative to individual market strategies. All ensemble configurations outperform the common benchmark on IR2, with LSTM\_1 achieving the strongest result. Volatility and maximum drawdown are also reduced relative to single-market deployments, confirming that geographic diversification across the NASDAQ-100, Nikkei 225, and EURO STOXX 50 provides meaningful risk reduction benefits.

\noindent\textbf{RQ6. Does the performance of reinforcement learning strategies 
vary systematically across macroeconomic regimes, and in which market environments do RL strategies add the most value?}
\newline The regime analysis presented in Section~\ref{regimeness} reveals substantial heterogeneity in strategy performance across macroeconomic periods. On the NASDAQ-100, RL strategies outperform Buy \& Hold only during the post-GFC recovery (2009--2013), with passive benchmarks dominating in the subsequent bull market and COVID-era periods. On the Nikkei 225, RL strategies and classical benchmarks consistently outperform the weak Buy \& Hold index exposure, with LSTM\_2 leading across all strategies in 2014--2019. On the EURO STOXX 50, RL strategies deliver consistent risk-adjusted outperformance over Buy \& Hold across all three regimes, though classical benchmarks remain competitive. Overall, RL strategies appear to add the most value in markets characterized by lower trend persistence and greater structural uncertainty, while their advantage diminishes in strongly trending environments where passive strategies are most effective.

The difficulty of outperforming passive and classical benchmarks across all markets and regimes is consistent with the findings of \citet{DE_MIGEUL}, who demonstrate that even sophisticated optimization models struggle to systematically beat naive equal-weight allocation out-of-sample. The ensemble results further reinforce the importance of diversification as a primary mechanism for improving risk-adjusted performance within the proposed framework.

\hypertarget{conclusions}{%
\section{Conclusions}\label{conclusions}}

This study develops and evaluates a unified deep reinforcement learning framework for dynamic portfolio allocation across three major global equity markets: the NASDAQ-100, Nikkei 225, and EURO STOXX 50. The framework combines the Soft Actor--Critic algorithm with a structured walk-forward optimization procedure, realistic market frictions, and a systematic comparison of architectural and economic design choices. The empirical analysis covers the period from 2009 to 2026 and benchmarks the proposed strategies against four reference portfolios: Buy \& Hold, Equal-Weight Monthly, Markowitz Minimum Variance, and Momentum Top-20.

The empirical results demonstrate that reinforcement learning strategies achieve competitive risk-adjusted performance in several settings, most notably in the EURO STOXX 50, where all RL configurations outperform the Buy \& Hold benchmark on IR2, and in the ensemble framework, where LSTM\_1 achieves an IR2 of 0.41 against the common benchmark of 0.34. However, the hypothesis is only partially confirmed: RL strategies do not consistently outperform passive or classical benchmarks across all markets, and none of the strategies achieve statistically significant excess returns relative to Buy \& Hold based on HAC-adjusted inference. The regime analysis further reveals that RL strategies add the most value during periods of elevated uncertainty and lower trend persistence, while passive benchmarks dominate in strongly trending environments such as the NASDAQ-100 during 2014--2019.

This study makes several contributions to the existing literature on deep reinforcement learning for portfolio management. First, it provides a unified multi-market evaluation framework that applies identical walk-forward optimization, adaptive retraining, and performance assessment procedures across three economically distinct equity markets, enabling direct cross-market comparison an aspect largely absent from prior work focused on single-market settings \citep{yang2020deep, JIANG2024101016}. Second, it introduces a hierarchical Dirichlet policy structure that separates the equity-cash allocation decision from individual asset selection, providing a principled mechanism for dynamic cash management within the SAC framework that extends flat policy architectures used in related studies \citep{Hambly_2023}. Third, it proposes an adaptive retraining criterion based on rolling validation Sharpe ratios that selectively updates the model only when performance deteriorates, reducing computational cost while maintaining deployment realism a mechanism not present in standard WFO implementations in the literature. Fourth, it conducts a systematic regime decomposition across three macroeconomic periods, providing empirical evidence on the conditions under which reinforcement learning allocation adds value relative to passive and classical benchmarks, contributing to the broader debate on the regime-dependence of active portfolio management \citep{DE_MIGEUL}.

Despite these contributions, several limitations must be acknowledged. First, none of the reinforcement learning strategies achieve statistically significant outperformance relative to the Buy \& Hold benchmark in two out of three markets, indicating that the observed performance differences may reflect random variation rather than systematic excess returns. Second, the experimental configurations were not designed as ceteris paribus ablation studies, as each configuration varies simultaneously across multiple dimensions including encoder type, reward formulation, and portfolio constraints. Consequently, the answers to RQ1--RQ4 should be interpreted as exploratory rather than causal. Third, the transaction cost assumption of 2 basis points was held fixed throughout training and evaluation; strategies were not retrained under higher cost assumptions, and their robustness to more realistic cost levels remains an open question. Fourth, the walk-forward architecture itself including the choice of training window, validation period, and retraining threshold was not subject to systematic sensitivity analysis, introducing the risk of meta-overfitting as noted by \citet{Bailey2014}. Fifth, the ensemble aggregation employs fixed equal weights across markets, which may not reflect optimal cross-market allocation and limits the adaptability of the ensemble to changing market conditions.

Several directions emerge naturally from this work. The most immediate extension concerns the application of the framework to higher-frequency data and alternative asset classes, including cryptocurrency markets, where microstructure dynamics, liquidity constraints, and volatility regimes differ substantially from traditional equities. High-frequency environments also introduce the opportunity to evaluate intraday allocation policies under more granular transaction cost models. A second direction concerns the enrichment of the state representation through the integration of macroeconomic variables such as yield curve dynamics, credit spreads, and purchasing managers indices alongside sentiment signals derived from financial news using large language model embeddings. Such representations could provide the agent with a richer characterization of the macroeconomic regime, potentially improving performance during structural breaks and crisis periods where price-based features alone may be insufficient. A third direction involves replacing the fixed equal-weight ensemble with a meta-level adaptive aggregation mechanism, where cross-market portfolio weights are learned dynamically based on recent strategy performance and prevailing market conditions. Finally, future work could address the ablation limitation noted above by designing a dedicated ceteris paribus experimental framework in which a single architectural or economic dimension is varied at a time, enabling cleaner attribution of performance differences to individual model components.

\setlength{\bibsep}{2pt}

\bibliography{bibfile1}

\end{document}